\documentclass[11pt]{article}

\usepackage{geometry}
 \usepackage[ruled]{algorithm2e}
 \usepackage{setspace}

\usepackage{appendix}
\usepackage{epsfig}
\usepackage{amsmath,amssymb,amsthm,amsfonts}
\usepackage{tikz}
\usepackage{color}
\usepackage{natbib}
\usepackage{graphicx}
\usepackage{subfigure}
\usepackage{caption}
\usepackage{url}
\usepackage{hyperref}
\usepackage{listings}
\usepackage{enumerate}
\usepackage{wrapfig}
\usepackage{pgfplots}
\usepackage{subfiles}

\numberwithin{equation}{section}

\usepackage{tikz}

\usetikzlibrary{shapes, arrows, calc, positioning,matrix}
\tikzset{
data/.style={circle, draw, text centered, minimum height=2.95em ,minimum width = .5em, inner sep = 2pt},
empty/.style={circle, text centered, minimum height=3em ,minimum width = .55em, inner sep = 2pt},
}

\tikzstyle{state}=[shape=circle,draw=black,fill=white, minimum height=3.0em ]
\tikzstyle{lightedge}=[<-,dotted]
\tikzstyle{mainstate}=[state,thick]
\tikzstyle{mainedge}=[<-,thick]
\tikzstyle{emptynode}=[shape=circle,draw=white,fill=white, minimum height=3.0em ]

\newcommand{\argmax}{\operatornamewithlimits{argmax}}

\begin{document}

 \begin{center}
{\noindent{\LARGE \textbf{Ensemble updating of binary state vectors by maximising the expected number 
of unchanged components} \vspace{1cm} \\} 
{\Large \textsc{Margrethe K. Loe}}\\{\it Department of
    Mathematical Sciences, Norwegian University of Science and
    Technology}\vspace{1cm}\\{\Large \textsc{H\aa kon Tjelmeland}}\\{\it Department of
    Mathematical Sciences, Norwegian University of Science and
    Technology}\vspace{1cm}
}	
\end{center}

\begin{abstract}

In recent years, several ensemble-based filtering methods have been proposed and studied. The main challenge in such procedures is the updating of a prior ensemble to a posterior ensemble at every step of the filtering recursions. In the famous ensemble Kalman filter, the assumption of a linear-Gaussian state space model is introduced in order to overcome this issue, and the prior ensemble is updated with a linear shift closely related to the traditional Kalman filter equations. In the current article, we consider how the ideas underlying the ensemble Kalman filter can be applied when the components of the state vectors are binary variables. While the ensemble Kalman filter relies on Gaussian approximations of the forecast and filtering distributions, we instead use first order Markov chains. To update the prior ensemble, we simulate samples from a distribution constructed such that the expected number of equal components in a prior and posterior state vector is maximised. We demonstrate the performance of our approach in a simulation example inspired by the movement of oil and water in a petroleum reservoir, where also a more na\"{i}ve updating approach is applied for comparison. 
Here, we observe that the Frobenius norm of the difference between the estimated and the true marginal filtering probabilities is reduced to the half with our method compared to the na\"{i}ve approach, indicating that our method is superior. 
Finally, we discuss how our methodology can be generalised from the binary setting  to more complicated situations.

\end{abstract}

\vspace{0.5cm}
\noindent {\it } 
\vspace{-0.1cm}

	\section{Introduction}
	\label{sec:Introduction}
	
A state-space model consists of a latent $\{x^t\}_{t=1}^\infty$ process and an
observed $\{y^t\}_{t=1}^\infty$ process, where $y^t$ is a partial 
observation of $x^t$. More specifically, the $y^t$'s are assumed
to be conditionally independent given the $x^t$ process and $y^t$ only
depends on $x^t$. Estimation of the latent variable at time $t$, $x^t$, given 
all observations up to this time, $y^{1:t} = (y^1,\ldots,y^t)$, is known 
as the filtering or data assimilation problem. In the linear Gaussian 
situation an easy to compute and exact solution is available by the famous 
Kalman filter.
In most non-linear or non-Gaussian situations no computationally feasible
exact solution exists and ensemble methods are therefore frequently adopted. The 
distribution $p(x^t|y^{1:t})$ is then not analytically available, but is
represented by a set of realisations 
$\tilde{x}^{t(1)},\ldots, \tilde{x}^{t(M)}$ from this filtering distribution.
Assuming such an ensemble of realisations to be available for time $t-1$, the 
filtering problem is solved for time $t$ in two steps. First, based on 
the Markov chain model for the $x^t$ process, each $\tilde{x}^{t-1(i)}$ 
is used to simulate a corresponding forecast realisation $x^{t(i)}$, 
which marginally are independent samples from $p(x^{t}|y^{1:t-1})$. 
This is known as the forecast or prediction step.
Second, an update step is performed, where each $x^{t(i)}$ is updated
to take into account the new observation $y^{t}$ and the result
is an updated ensemble $\tilde{x}^{t(1)},\ldots,\tilde{x}^{t(M)}$
which represents the filtering distribution at time 
$t$, $p(x^{t}|y^{1:t})$. The updating step is the difficult one and
the different strategies that have been proposed can be classified into 
two classes, particle filters and ensemble Kalman filters.

In particle filters \citep{book1} each filtering realisation $\tilde{x}^{t(i)}$ comes
with an associated 
weight $\tilde{w}^{t(i)}$, and the pair $(\tilde{w}^{t(i)},\tilde{x}^{t(i)})$
is called a particle. In the forecast step a forecast particle $(w^{t(i)}, x^{t(i)})$
is generated from each filtering particle $(\tilde{w}^{t-1(i)},\tilde{x}^{t-1(i)})$ by
generating $x^{t(i)}$ from $\tilde{x}^{t-1(i)}$ as discussed above and by keeping 
the weight unchanged, i.e. $w^{t(i)}=\tilde{w}^{t-1(i)}$. The updating step
consists of two parts. First the weights are updated by multiplying
each forecast weight $w^{t(i)}$ by the associated likelihood value 
$p(y^{t}|x^{t(i)})$, keeping the $x^t$ component of the particles unchanged.
Thereafter a re-sampling may be performed, where 
$(\tilde{w}^{t(i)},\tilde{x}^{t(i)}),i=1,\ldots,M$ are generated
by sampling the $\tilde{x}^{t(i)}$'s independently from $x^{t(i)},i=1,\ldots,M$
with probabilities proportional to the updated weights, and thereafter setting all 
the new filtering weights $\tilde{w}^{t(i)}$ equal to one. Different criteria can be used to decide
whether or not the re-sampling should be done. The particle filter is very general 
in that it can be formulated for any Markov $x^t$ process and any observation 
distribution $p(y^t|x^t)$. However, when running the particle filter one quite
often ends up with particle depletion, meaning that a significant fraction of 
the particles ends up with negligible weights, which in practice requires
the number of particles to grow exponentially with the dimension of the state
vector $x^t$. To cope with the particle depletion problem various 
modifications of the basic particle filter approach described here have
been proposed, see for example \citet{art15}, \citet{art16} and 
\citet{art17} and references therein.

The ensemble Kalman filter \citep{art2, art3} uses approximations in the update step, and thereby 
produces only an approximate solution to the filtering problem. 
In the update step it starts by using the forecast samples
$x^{t(i)},i=1,\ldots,M$ to estimate a Gaussian approximation to
the forecast distribution $p(x^{t}|y^{1:t-1})$. This is combined with an assumed Gaussian 
observation distribution $p(y^{t}|x^{t})$ to obtain a Gaussian approximation to 
the filtering distribution $p(x^{t}|y^{1:t})$. Based on this Gaussian approximation
the filtering ensemble is generated by sampling $\tilde{x}^{t(i)}$ independently
from Gaussian distributions, where the mean of $\tilde{x}^{t(i)}$ equals
$x^{t(i)}$ plus a shift which depends on the approximate Gaussian filtering distribution.
The associated variance is chosen so that the marginal distribution of the generated
filtering sample $\tilde{x}^{t(i)}$ is equal to the Gaussian approximation to 
$p(x^{t}|x^{1:t})$ when the forecast sample $x^{t(i)}$ is assumed to be distributed
according to the Gaussian approximation to $p(x^{t}|y^{1:t-1})$. The basic
ensemble Kalman filter described here is known to have a tendency to under-estimate 
the variance in the filtering distribution and various remedies have been 
proposed to correct for this, see for example the discussions in \citet{art13, art14} and \citet{art18}. The square root filter \citep{art6, art5} is a special variant of the ensemble Kalman filter
where the update step is deterministic. The filtering ensemble is then generated from the 
forecast ensemble only by adding a shift to each ensemble element. Here the size of 
the shift is chosen to get that the marginal distribution of the filtering 
realisations is equal to the approximated Gaussian filtering distribution.

The Gaussian approximations used in the ensemble Kalman filter update step limits the use
of this filter type to continuous variables, whereas the particle filter setup can be used 
for both continuous and categorical variables. In the literature there exists a few attempts
to use the ensemble Kalman filter setup also for categorical variables, see in 
particular \citet{art4}. The strategy then used for the update step
is first to map the categorical variables over to continuous variables, perform 
the update step as before in the continuous space, and finally map the updated continuous variables
back to corresponding categorical variables. Our goal in this article is to study how 
the basic ensemble Kalman filter idea can be used for categorical variables without having to map 
the categorical variables over to a continuous space. As discussed above the update step
is the difficult one in ensemble filtering methods. The basic ensemble Kalman filter update
starts by estimating a Gaussian approximation to the forecast distribution 
$p(x^{t}|y^{1:t-1})$. More generally one may use another parametric class than the Gaussian. 
For categorical variables the simplest alternative is to consider a first-order Markov chain,
which is what we focus on in this article. Having a computationally feasible 
approximation for the forecast distribution we can find a corresponding 
approximate filtering distribution. Given the forecast ensemble the question then 
is from which distribution to simulate the filtering ensemble to obtain that 
the filtering realisations marginally is distributed according to the given 
approximate filtering distribution, corresponding to the property for  
the standard ensemble Kalman filter. In this article we develop in detail 
one possible way to do this when the elements of the state vector are binary, 
the approximate forecast distribution is a first-order Markov chain, and the 
observation distribution has a specifically simple form.

The article has the following layout. First, in Section \ref{sec:Preliminaries} we specify the
general state-space model and the associated filtering problem and present the ensemble Kalman filter. 
Next, in Section \ref{sec:A general ensemble updating framework} we formulate the general 
ensemble updating strategy briefly discussed above. Then, in Section 
\ref{sec:Ensemble updating of binary state vectors}, we limit the attention to a situation where the elements of 
the state vector are binary and discuss how the update step can be performed
in this case. After that, we present two numerical experiments with simulated data in Section \ref{sec:Numerical}. 
Finally, in Section \ref{sec:Closing remarks} we give some final remarks and briefly discuss how our
setup can be generalised to a situation with more than two classes and an assumed higher-order Markov chain
model for the forecast distribution.

	\section{Preliminaries}
	\label{sec:Preliminaries}

In this section, we review some basic theoretical aspects of ensemble-based filtering methods. The material presented 
should provide the reader with the necessary background for understanding our proposed approach and it
also establishes some of the notations used throughout the article.

\subsection{Review of the filtering problem}

\begin{figure}
\centering
\begin{tikzpicture}
			[
			->,
			>=stealth',
			auto,node distance=3cm,
			thin,
			main node/.style={circle, draw, }
			]
			% states
			\node[state] (s1) at (0,0) {$ x^1$};
			\node[state] (s2) at (2,0) {$ x^2$}
			edge [<-] (s1);
			\node[draw=none,fill=none] (s3) at (4,0){$\cdots$}
			edge [<-]  (s2);
			\node[state] (stt) at (6,0) {$ x^{t-1}$}
			edge [<-] (s3);
			\node[state] (st) at (8,0) {$ x^{t}$} 
			edge [<-] (stt);
			% observations
			\node[state] (y1) at (0,-2) {$ y^1$}
			edge [<-] (s1);
			\node[state] (y2) at (2,-2) {$ y^2$}
			edge [<-] (s2);
			\node[state] (ytt) at (6,-2) {$ y^{t-1}$}
			edge [<-] (stt);
			\node[state] (ytt) at (8,-2) {$ y^{t}$}
				edge [<-] (st);
			\node[emptynode] (e) at (10,0) {$\cdots$}
				edge[<-] (st);
			\end{tikzpicture}
			\caption{Illustration of the model behind the filtering problem. 
	}
	\label{fig:Model}
			\end{figure}
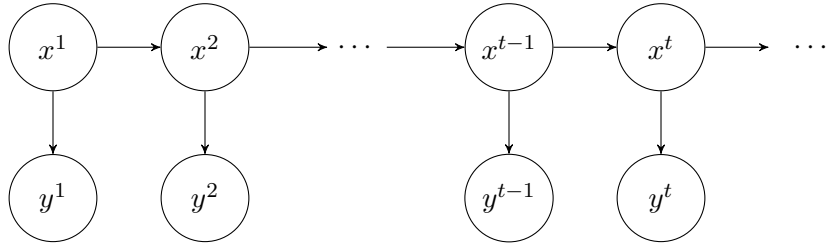

The  filtering problem in statistics can be nicely illustrated with a graphical model; see Figure \ref{fig:Model}. Here, $\{x^{t}\}_{t=1}^{\infty}$ represents a time series of unobserved states and  $\{y^{t}\}_{t=1}^{\infty}$ a corresponding time series of observations. Each state $x^t$ is $n$-dimensional and can take on values in a state space $\Omega_X$, while each observation $y^t$ is $k$-dimensional and can take on values in a state space $\Omega_Y$.   The series of unobserved  states, called  the  state process, constitutes a first order Markov chain with initial state $x^1$, initial distribution $p(x^1)$, and transition probabilities $p(x^{t} | x^{t-1}), t > 1$. For each state $x^t, t \geq 1$, there is a corresponding observation $y^t$. The observations are assumed conditionally independent given the state process, with $y^t$ depending on $\{x^t\}_{t=1}^{\infty}$ only through $x^t$,  according to some likelihood model $p(y^t|x^t)$. To summarise, the model is specified by
\begin{equation}
\begin{aligned}
& x^1  \sim p(x^{1}), \\
& x^t |   x^{t-1}   \sim p(x^{t} | x^{t-1}), \hspace{2mm}   t> 1,  \\ 
&  y^t | x^t  \sim p(y^t | x^t), \hspace{10mm}  t \geq 1 .
\end{aligned}
\label{eq:model}
\end{equation}

The objective of the filtering problem is,  for each $t$,  to compute the so-called filtering distribution, $p(x^t | y^{1:t})$, that is, the distribution of $x^t$ given all observations up to this time, $y^{1:t} = (y^1, \dots, y^t)$. 
   Because of the assumptions about the state and observation processes, it can be shown (see e.g. \cite{chapter1}) that the series of filtering distributions can be computed recursively according to the following equations:
  \begin{subequations}
 \begin{equation}
  \text{i)} \hspace{2mm} p(x^t | y^{1:t-1} )
 = \int_{\Omega_X} p(x^t | x^{t-1}) p(x^{t-1} | y^{1:t-1}) \text{d}x^{t-1}, 
 \label{eq:pred}
 \end{equation}
 \begin{equation}
  \text{ii)} \hspace{2mm} p(x^t  | y^{1:t} ) = \frac{ p(x^t | y^{1:t-1} ) p(y^t | x^t)} {\displaystyle\int_{\Omega_X} p(x^t | y^{1:t-1} ) p(y^t|x^t) \text{d} x^t}.
  \label{eq:filt}
 \end{equation}
 \label{eq:test}
 \end{subequations}
As one can see, the recursions evolve as a two-step process, each iteration consisting of i) a prediction step, and ii) an update step. In the prediction, or forecast step, one computes the predictive, or forecast, distribution $p(x^t | y^{1:t-1})$, while in the update step,  one  computes the filtering distribution $p(x^t | y^{1:t})$ by conditioning the predictive distribution on the incoming observation $y^t$ through application of  Bayes' rule. 
The update step can be formulated as a standard Bayesian inference problem, with $p(x^t | y^{1:t-1})$ becoming the prior, $p(y^t | x^t)$  the likelihood, and $p(x^t | y^{1:t})$  the posterior. 

 There are two important special cases where the analytical solutions to the filtering recursions, \eqref{eq:pred} and \eqref{eq:filt}, can be computed exactly. 
 The first case is the hidden Markov model (HMM). Here,  the state space $\Omega_X$ consists  of a finite number of states, and the integrals in \eqref{eq:pred} and \eqref{eq:filt} reduce to finite sums. 
 If the number of states in $\Omega_X$ is large, however, the summations become computer-intensive, rendering the filtering recursions \textit{computationally} intractable. 
 The second case is the linear Gaussian state space model, which can be formulated as follows:
  $$ x^1 \sim \mathcal N (x^1 | \mu^1, Q^1),$$
  $$ x^t | x^{t-1} = A^t x^{t-1}  + \epsilon^t, \hspace{5mm} \epsilon^t \sim \mathcal N_n(\epsilon | 0, Q^t),$$
  $$ y^t | x^{t} = H^t x^{t}  + \omega^t, \hspace{5mm} \omega^t \sim \mathcal N_k(\omega | 0, R^t), $$
  where $A^t \in \mathbb R^{n \times n}$ and $H^t \in \mathbb R^{k \times n}$ are non-random linear operators, $Q^t \in \mathbb R^{n \times n}$ and $R^t \in  \mathbb R^{k \times k}$ are covariance matrices, and $x^1, \epsilon^1, \epsilon^2, \dots,\omega^1, \omega^2, \dots$ are all independent.
In this case, the sequence of predictive and filtering distributions are Gaussian, and the filtering recursions lead to the famous Kalman filter \citep{art1}.

In general, we are unable to  evaluate the integrals in \eqref{eq:pred} and \eqref{eq:filt}, leaving the filtering recursions intractable. Approximate solutions therefore become necessary.  The most common approach in this regard is the class of ensemble-based methods, where  a set of samples, called an ensemble,  is used to empirically represent the sequence of predictive and  filtering distributions. 
Starting from an initial ensemble $\{  x^{1(1)}, \dots, x^{1(M)} \}$ of $M$ independent realisations from the Markov chain initial model $p(x^1)$, the idea is to advance this ensemble forward in time according to the model dynamics. 
As the original filtering recursions, the propagation of the ensemble alternate between an update step and a prediction step. 
Specifically, suppose at time $t \geq 1$ that an ensemble $\{x^{t(1)}, \dots, x^{t(M)}\}$ of independent realisations from the forecast distribution $p(x^t | y^{1:t-1})$ is available. We  then want to update this forecast ensemble by conditioning on the incoming observation $y^t$ in order to obtain an updated, or posterior, ensemble $\{ \tilde x^{t(1)}, \dots,  \tilde x^{t(M)} \}$ with independent realisations from the filtering distribution $p(x^t | y^{1:t})$. If we are able to carry out this updating, we proceed and propagate the updated ensemble $\{ \tilde x^{t(1)}, \dots,  \tilde x^{t(M)} \}$ one step forward in time by using the Markov chain prior model to simulate $x^{t+1(i)} |  \tilde x^{t(i)} \sim p(x^{t+1} | \tilde x^{t(i)})$ for each $i$. This produces a new forecast ensemble for time $t+1$ with independent realisations from the forecast distribution $p(x^{t+1} | y^{1:t})$. However, while we are often able to simulate from the Markov forward model $p(x^t | x^{t-1})$, there is no straightforward way for carrying out the update of the prior ensemble $\{ \tilde x^{t(1)}, \dots,  \tilde x^{t(M)} \}$ to a posterior ensemble $\{ \tilde x^{t(1)}, \dots,  \tilde x^{t(M)} \}$. Therefore, ensemble methods require approximations in the update step. Consequently, the assumption we make at the beginning of each time step $t$, i.e. that $ x^{t(1)}, \dots,  x^{t(M)}$ are exact and independent realisations from $ p(x^{t} | y^{1:t-1})$, holds only approximately, except in the initial time step.

In the remains of this article, we focus primarily on the challenging updating of a prior ensemble $\{ x^{t(1)}, \dots,  x^{t(M)} \}$ into a posterior ensemble $\{ \tilde x^{t(1)}, \dots,  \tilde x^{t(M)} \}$ at a specific time step $t$. 
We refer to this task as the \textit{ensemble updating} problem. 
For simplicity, we omit the time superscript $t$ and the $y^{1:t-1}$ from the notations in the remaining sections, as these quantities are fixed throughout. That is, we write $x$ instead of $x^t$, $p(x)$ instead of $p(x^t|y^{1:t-1})$, $p(x| y)$ instead of $p(x^t | y^{1:t})$, and so on.

\subsection{The ensemble Kalman filter}
 
The ensemble Kalman filter (EnKF), first introduced in the geophysics literature by \cite{art2}, is an approximate ensemble-based method that relies on Gaussian approximations to overcome the difficult updating of the prior ensemble.
The updating is done in terms of a linear shift of each ensemble member, closely related to the traditional Kalman filter equations. The literature on the EnKF is extensive, but some basic references include \cite{art2} and  \cite{art3}. Here, we only provide a brief presentation. For simplicity reasons, we restrict the focus to linear, Gaussian observational models; that is, the relationship between $y$ and $x$ is  assumed linear with additive zero-mean Gaussian noise, 
\begin{equation}
y | x = Hx + \epsilon, \hspace{0.5cm} \epsilon \sim \mathcal N_k(\epsilon; 0, R).
\label{eq:Gauss-linear}
\end{equation}

There exist two main classes of EnKFs, stochastic filters and deterministic, or so-called square root filters, differing in whether the updating of the ensemble is carried out stochastically or deterministically. The stochastic EnKF is probably the most common version, and we begin our below presentation of the EnKF by focusing on this method.

Consider first a linear Gaussian state space model as introduced in the previous section. 
Under this linear Gaussian model, it follows from the Kalman filter recursions that the current predictive, or prior model $p(x)$ is a Gaussian, $\mathcal N_n(x; \mu, \Sigma)$, with analytically tractable mean $\mu$ and covariance $\Sigma$.
Furthermore, the current filtering, or posterior distribution $p(x|y)$ is a Gaussian, $\mathcal N_n(x; \tilde \mu, \tilde \Sigma)$, with mean $\tilde \mu$ and covariance $\tilde \Sigma$ analytically available from the Kalman filter update formulas as
$$
\tilde \mu = \mu + K (y-H\mu),
$$
$$
\tilde \Sigma = (I-KH)\Sigma,
$$
where  $K = \Sigma H' (H\Sigma H' + R)^{-1}$ is the Kalman gain. 
The stochastic EnKF update 
is based on the following fact:
If $x \sim \mathcal N_n(x; \mu, \Sigma)$ and $\epsilon \sim \mathcal N_k(\epsilon ; 0,R)$  are independent random samples, then
\begin{equation}
\tilde x = x + K  (y - Hx + \epsilon),
\label{eq:eq3}
\end{equation}
is a random sample from $\mathcal N_n(x; \tilde \mu, \tilde \Sigma)$. 
The verification of this result is straightforward. Clearly, under the assumption  that the prior ensemble   $\{x^{(1)}, \dots, x^{(M)}\}$ contains independent samples from the Gaussian $\mathcal N_n(x; \mu, \Sigma)$, one theoretically valid way to obtain the updated ensemble is to plug each $x^{(i)}$ into \eqref{eq:eq3}. 
However, for computational reasons, the prior covariance matrix $\Sigma$ and hence the Kalman gain $K$ can be intractable for high-dimensional systems. To circumvent this issue, stochastic EnKF passes each prior sample  $x^{(i)}$  through a linear shift identical to  \eqref{eq:eq3}, but with the true Kalman gain $K$ replaced with an empirical estimate $\hat K$ inferred from the prior ensemble,
\begin{equation}
\tilde x^{(i)} = 
x^{(i)} + \hat K (y - Hx^{(i)} + \epsilon^{(i)}), \hspace{5mm} i = 1, \dots, M.
\label{eq:stochEnKF}
\end{equation}
In the EnKF literature,  each term $Hx^{(i)} - \epsilon^{(i)}$ is typically referred to as a ''perturbed'' observation. Under the Gaussian-linear assumptions, the update \eqref{eq:stochEnKF} produces approximate samples from the Gaussian posterior model $\mathcal N_n(x; \tilde \mu, \tilde \Sigma)$ that corresponds to the Gaussian prior model $\mathcal N_n (x; \mu, \Sigma)$. The update is consistent in the sense that as the ensemble size goes to infinity, the distribution of the updated samples approaches the Gaussian $\mathcal N_n(x; \tilde \mu, \tilde \Sigma)$, that is, the solution of the Kalman filter. 

Although the EnKF update is based on Gaussian assumptions about the predictive and filtering distributions, it can be applied even if these assumptions are not met. Naturally, bias is in this case introduced, and the updated samples will not converge in distribution to the true posterior $p(x|y)$. However, since the update is a linear combination of the $x^{(i)}$'s, non-Gaussian properties present in the true prior and posterior models can, to some extent, be captured.

Deterministic EnKFs instead use a non-random linear transformation to update the ensemble. In the following, let $\hat \mu$ and $\hat \Sigma$ denote the sample mean and sample covariance, respectively,  of the prior ensemble. 
Further, let
$
\hat {\tilde \mu} 
$
and
$
\hat {\tilde \Sigma}
$
denote the mean and covariance, respectively, of the Gaussian posterior model $\mathcal N_n(x; \hat {\tilde{ \mu}}, \hat {\tilde{ \Sigma}} )$ corresponding to the  Gaussian prior approximation $\mathcal N_n(x; \hat \mu, \hat \Sigma )$. 
Generally, the update equation  of a square root EnKF can be written as
\begin{eqnarray*}
\tilde x^{(i)} = \hat{\mu} + \hat K (y - H\hat \mu) +  B (x^{(i)} - \hat \mu ), & i = 1, \dots, M,
\label{eq:EnSRF}
\end{eqnarray*}
where  $ B \in \mathbb R^{n  \times n }$ is a solution to the quadratic  matrix equation
\begin{equation*}
B \hat \Sigma  B' = (I - \hat K H)\hat \Sigma.
\label{eq:EnSRF B}
\end{equation*}
Note that $B$ is not unique except in the univariate case. This gives rise to a variety  of square root algorithms; see \cite{art5} for further explanation. As such, several square root formulations have been proposed in the literature, including, but not limited to, \cite{art6}, \cite{art7}, and \cite{art8}. 
The non-random update of square root EnKF ensures that the sample mean and sample covariance of the posterior ensemble equal $\hat {\tilde \mu} $ and  $\hat {\tilde \Sigma}$ \textit{exactly}.  
This is different from stochastic EnKFs where, under linear-Gaussian assumptions, the sample mean and sample covariance of the posterior ensemble only equal $\hat {\tilde \mu} $ and  $\hat {\tilde \Sigma}$ in expectation.

	\section{A general ensemble updating framework}
	\label{sec:A general ensemble updating framework}

In this section, we present a general framework for ensemble updating.  Both the EnKF and the update strategy for binary vectors proposed in this article can be viewed as special cases within the framework.

 \subsection{The framework}

For convenience, let us first quickly review the ensemble updating problem. 
Starting out, we have a prior ensemble,  $\{x^{(1)}, \dots, x^{(M)} \}$, which we assume contain  independent realisations from a prior model $p(x)$. 
The prior model $p(x)$ is typically intractable in this context, either computationally or analytically, or both. 
Given an incoming observation $y$ and a corresponding likelihood model $p(y|x)$ the goal is to update the prior ensemble according to Bayes' rule in order to obtain a posterior ensemble, $\{\tilde x^{(1)}, \dots,\tilde  x^{(M)} \}$, with independent realisations from the posterior model $p(x|y)$. However, carrying out this update exactly is generally unfeasible and  approximate strategies are required.

Conceptually, the proposed framework is quite simple. It involves three main steps as follows. 
First, we replace the intractable model, $p(x|y) \propto p(x) p(y|x)$, with a simpler model,  $f(x|y) \propto f(x) p(y|x)$, using information from the prior ensemble to construct the approximate prior model $f(x)$. 
Throughout, we will refer to the model $f(x |y) \propto f(x) p(y|x)$ as the \textit{assumed} model. Notice that  the likelihood model $p(y|x)$ has not been replaced; for simplicity, we assume that this model is already tractable. 
Second, we put forward a distribution conditional on $x$ and $y$,  denoted $q(\tilde x | x,y)$, obeying the following property:
 \begin{equation}
f(\tilde x|y) = \int_{\Omega_X} f(x) q(\tilde x | x,y) \text{d} x.
\label{eq:constr}
\end{equation}
Third, we update the prior ensemble by generating samples from this conditional distribution,
\begin{eqnarray}
\tilde x^{(i)} \sim q(\tilde x | x^{(i)}, y), & i = 1, \dots, M.
\label{eq:update}
\end{eqnarray}

To understand the framework, one must  note that under the assumption that the assumed model is the correct one, the prior samples  have distribution $f(x)$ and the updated samples should have distribution $f(x|y)$. 
 If one is able to compute and sample from $f(x|y)$, one straightforward way to obtain the updated samples is to sample directly from $f(x|y)$. 
However,  since the assumed model is not really the correct one, that is probably not the best way to proceed. The prior ensemble contains valuable information about the true model that may not have been captured by the assumed model, and by straightforward simulation from $f(x|y)$  this information is lost.
To capture more information from the prior ensemble, it is probably advantageous to simulate conditionally on the prior samples. This is why we introduce the  conditional distribution $q(\tilde x | x,y)$. 
The criterion in \eqref{eq:constr} ensures that the marginal distribution of the updated samples $\tilde x^{(1)}, \dots, \tilde x^{(M)}$ produced by $q(\tilde x | x, y)$ still would be $f(x|y)$ given that the assumed model is the correct one. 
However, since the assumed model is not the correct one, the marginal distribution of the updated samples  is not $f(x|y)$, but some other distribution, hopefully one closer to the true posterior  model $p(x|y)$. 

There are two especially important things about the proposed framework that must be taken care of in a practical application. First, we need to select an assumed prior $f(x)$
which, combined with the likelihood model $p(y|x)$, returns a tractable posterior $f(x|y)$. 
Second, we need to construct the updating distribution $q(\tilde x | x,y)$. Clearly, there can be many, or infinitely many different $q(\tilde x | x,y)$ which all fulfil the constraint in \eqref{eq:constr}.
Below, we consider the proposed framework in two specific situations. The first case corresponds to the EnKF where  $f(x)$, $p(y|x)$ and $q(\tilde x | x,y)$ are all Gaussians. The second case is the developed method of this article where $f(x)$ and $p(y|x)$ constitute a hidden Markov model with binary states and $q(\tilde x | x,y)$ represents a transition matrix.

\subsection{The EnKF as a special case} 
\label{sec:enkf}
In the EnKF, the assumed prior model $f(x)$ is a Gaussian distribution. 
Combined with a Gaussian-linear likelihood $p(y|x)$, this leads to a Gaussian assumed posterior $f(x|y)$. 
The conditional distribution $q(\tilde x | x,y)$, which arises from the EnKF linear update, takes a different form depending on whether the filter is stochastic or deterministic. 
In stochastic EnKF, the linear update \eqref{eq:stochEnKF} yields a Gaussian $q(\tilde x | x,y)$ with mean equal to $ x + \hat K(y - Hx)$ and covariance equal to $\hat K R \hat K'$, that is
$$
q(\widetilde x | x,y) = \mathcal{N} \left (\tilde x ; x + \hat K(y - Hx), \hat K R \hat K'  \right ).
$$
In square root EnKF, the case is a bit different. Because the update now is deterministic, $q (\tilde x | x, y)$ has zero covariance and becomes a degenerate Gaussian distribution, or a delta function located at the value to which $x$ is moved, that is
$$
q(\tilde x | x,y) = \delta  \left (\tilde x ; \hat \mu + \hat K (y - H \hat \mu) + B(x^{} - \hat \mu ) \right ).
$$

\subsection{The proposed method  as a special case}
\label{sec:markov}
Suppose $x=(x_1, \dots, x_n)$ is a vector of $n$ binary variables, $ x_i \in \{0,1\}$, and that $x$ is spatially arranged along a line. A possible assumed prior model for $x$ is then a first order Markov chain, 
$$
f(x ) = f(x_1) f(x_2 | x_1) \cdots f(x_n | x_{n-1}).
$$
Furthermore, suppose that for each variable $x_i$ there is a corresponding observation, $y_i$, so that $y = (y_1, \dots, y_n)$, and suppose that the $y_i$'s are conditionally independent given $x$, with $y_i$ depending on $x$ only through $x_i$,
$$
p(y|x) = p(y_1 | x_1) \cdots p(y_n | x_n).
$$
This combination of $f(x)$ and $p(y|x)$ constitutes a hidden Markov model, as introduced in Section 2. It follows that the corresponding assumed posterior $f(x|y)$ is also a first order Markov chain for which all quantities of interest are possible to compute.

Now, since $\Omega_X = \{0,1\}^n$ is a discrete sample space, we rewrite the constraint in \eqref{eq:constr} as a sum,
\begin{equation}
f(\tilde x | y) = \sum_{x \in \Omega_X} f(x) q(\tilde x |x,y). 
\label{eq:sum constr}
\end{equation}
As in the general case, there can be many, or infinitely many, valid solutions of $q(\tilde x |x,y)$. 
Because of the discrete context, 
$q(\tilde x |x,y)$ now represents a transition matrix. Brute force, this transition matrix involves $2^n(2^n-1)$ parameters, and the constraint \eqref{eq:sum constr} leads to a system of $2^n-1$ linear equations on these parameters. 
Even for moderate $n$, solving such a problem is too complicated.
In the next section, we propose a simplifying strategy where we impose Markovian properties on $q(\tilde x |x,y)$, formulate an optimality criterion for $q(\tilde x |x,y)$, and use dynamic programming to construct the optimal solution.

	\section{Ensemble updating of binary state vectors}
	\label{sec:Ensemble updating of binary state vectors}

This section continues on the situation introduced in Section \ref{sec:markov}. In particular, we focus on the construction of 
$q(\tilde x | x,y)$. 
We start out in Section \ref{sec: Tittel ukjent} proposing an optimality criterion  and enforce Markovian properties on  $q(\tilde x | x,y)$. 
Thereafter,  in Section \ref{sec: Construction}, we present a dynamic programming (DP) algorithm that constructs the  optimal solution in a backward-forward recursive manner. 
Finally, in Section \ref{sec:linear} we take a closer look at some more technical aspects of the backward recursion of this DP algorithm.

\subsection{Choice of optimality criterion}
\label{sec: Tittel ukjent}

As mentioned in the previous section, there are infinitely many valid solutions of $q(\tilde x | x, y)$. For us, however,
it is sufficient with \emph{one} solution, preferably an \emph{optimal} solution, $q^{*} (\tilde x | x, y) $, with respect 
to some criterion. In this regard, an optimality criterion needs to be specified.  To specify this criterion, we argue 
that in order for  $q(\tilde x | x, y)$   to retain information from the prior ensemble and capture important properties of the true prior and posterior 
model, it should not make unnecessary changes to the prior samples. That is, as we update each prior sample $x^{(i)}$, 
we should take new information from the incoming observation $y$ into account and,  to a certain extent, push $x^{(i)}$ 
towards $y$, but the adjustment we make should be minimal. With this in mind, we propose to define the optimal solution 
$q^{*} (\tilde x | x, y) $ as the one that maximises the expected number of variables, or components, of $x$ that remain unchanged after the update to $\tilde x$. Mathematically, 
\begin{equation}
q^{*} (\tilde x | x, y) = \argmax_{q(\tilde x | x, y)} \text{E}_{\pi} \left [ \sum_{i=1}^n 1(x_i = \tilde x_i )\right ],
\label{eq:optimal q}
\end{equation}
where the subscript $\pi$ indicates that the expectation is taken over the following joint distribution,
\begin{equation}
\pi (\tilde x, x | y) = f(x) q(\tilde x | x, y),
\label{eq:pi}
\end{equation}
that is, the joint distribution of $x$ and $\tilde x$ given that the assumed models $f(x)$ and $f(x|y)$ are correct.

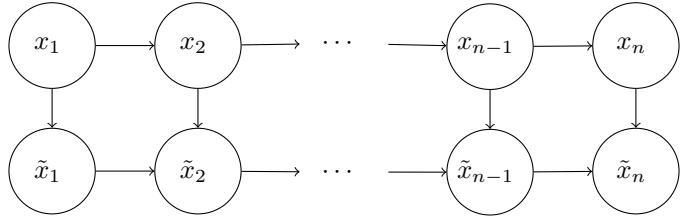
\begin{figure}
	\centering
	\begin{tikzpicture}[
	align = flush center,
	font = \small]
	\matrix [
	matrix of nodes,
	column sep = 2em,
	row sep = 1.3em, 
	nodes = {solid},
	] (m)
	{  
		  |[data]|$x_1$ & |[data]|$x_2$ &  |[empty]| $\cdots$ &  |[data]|$x_{n-1}$ &  |[data]|$x_n$  \\
		 |[data]|$\tilde x_1$ &  |[data]|$\tilde x_2$ & |[empty]| $\cdots$ &  |[data]|$\tilde x_{n-1}$  &  |[data]|$\tilde x_n$  \\
		\\[.7em]
	}; 
	\draw[->](m-1-1) to (m-1-2);
	\draw[->](m-1-2) to (m-1-3);
	\draw[->](m-1-3) to (m-1-4);
	\draw[dotted](m-1-4);
	\draw[->](m-1-4) to (m-1-5);
	
	\draw[->](m-1-1) to (m-2-1);
	\draw[->](m-1-2) to (m-2-2);
	\draw[->](m-1-4) to (m-2-4);
	\draw[->](m-1-5) to (m-2-5);
	
	\draw[->](m-2-1) to (m-2-2);
	\draw[->](m-2-2) to (m-2-3);
	\draw[->](m-2-3) to (m-2-4);
	\draw[dotted](m-1-4);
	\draw[->](m-2-4) to (m-2-5);
	
	\end{tikzpicture}
	\caption{Graphical illustration of dependencies between variables in $x$ and $\tilde x$.}
	\label{fig:q}
\end{figure}

Brute force, the construction of $q(\tilde x | x, y)$ involves the specification of $2^n(2^n-1) = \mathcal O(4^n)$ parameters. 
Clearly, this is intolerable even for moderate $n$. To reduce the number of parameters, we 
propose to enforce Markovian properties on $\pi( \tilde x, x|y)$  as illustrated graphically in Figure \ref{fig:q}. Then, $q(\tilde x|x,y)$ 
can be factorised as
\begin{equation}
q(\tilde x|x,y) = q(\tilde x_1 | x_1, y) q(\tilde x_2 | \tilde x_1, x_2, y) q(\tilde x_3 | \tilde x_2, x_3, y) \cdots q(\tilde x_n | \tilde x_{n-1}, x_n, y).
\label{eq:factorized q}
\end{equation}
Consequently, the number of parameters reduces to $2 + 4(n-1) = \mathcal O (n)$,
namely two parameters for the first factor $ q(\tilde x_1 | x_1, y)$, and four parameters for each $q(\tilde x_{k} | \tilde x_{k-1}, x_{k}, y)$, $k=2, \dots, n$. Another, and just as important consequence of the 
Markovian structure, is that the optimal solution $q^{*} (\tilde x | x, y)$ can be efficiently computed using dynamic programming. Following \eqref{eq:factorized q}, the optimal solution can also be factorised as
\begin{equation}
q^*(\tilde x|x,y) = q^*(\tilde x_1 | x_1, y) q^*(\tilde x_2 | \tilde x_1, x_2, y) q^*(\tilde x_3 | \tilde x_2, x_3, y) \cdots q^*(\tilde x_n | \tilde x_{n-1}, x_n, y).
\label{eq:factorized q optimal}
\end{equation}
The next section presents a DP algorithm where the $n$ factors in \eqref{eq:factorized q optimal}
 are constructed recursively.

% -- -- -- -- --  -- -- -- -- --  -- -- -- -- --  -- -- -- -- --  -- -- -- -- --  -- -- -- -- --  -- -- -- -- --  -- -- -- -- -- %

%  -- -- -- -- --  -- -- -- -- --  -- -- -- -- --  -- -- -- -- --  -- -- -- -- --  -- -- -- -- --  -- -- -- -- -- -- -- -- -- --  -- -- -- -- --  -- -- -- -- -- %

\subsection{Dynamic programming}
\label{sec: Construction}

Here, we describe a DP algorithm for constructing the optimal solution $q^*(\tilde x | x, y)$ introduced in the previous section. The algorithm involves a backward recursion followed by a forward recursion. The main challenge is the backward recursion and the details therein are a bit technical. For simplicity,  this section provides an overall description of the algorithm, while the  more technical aspects of the backward recursion are considered separately in Section \ref{sec:linear}. Throughout, we use the notation $\pi(\tilde x_{i:j}, x_{k:l}  | y)$, $1 \leq i \leq j \leq n$, $1\leq k \leq l \leq n$, to denote the joint distribution of $\tilde x_{i:j} = (\tilde x_i, \dots, \tilde x_j)$ and $ x_{k:l} = (x_k, \dots, x_l)$ 
assuming $x$ is distributed according to $f(x)$ and $\tilde x$ is simulated using $q(\tilde x | x, y)$.
 Furthermore, we introduce the following simplifying notations:
%\begin{eqnarray*}
%\pi_1 &=& \pi(x_1| y), \\
%q_ 1&=& q(\tilde x_1 | x_{1}, y), \\
%\pi_k &=& \pi(\tilde x_{k-1}, x_{k} | y), \\
%q_ k &=& q(\tilde x_k | \tilde x_{k-1}, x_{k}, y),
%\end{eqnarray*}
%for $ 2 \leq k \leq n.$

\begin{eqnarray*}
\pi_k &=& \begin{cases} \pi(x_1 | y), & k = 1, \\  \pi(\tilde x_{k-1}, x_{k} | y), & 2 \leq k \leq n, \end{cases} 
\\
q_ k &=&  \begin{cases} q(\tilde x_1 | x_{1}, y), & k =1, \\ q(\tilde x_k | \tilde x_{k-1}, x_{k}, y), & 2 \leq k \leq n, \end{cases} 
\end{eqnarray*}

The backward recursion of the DP algorithm involves recursive computation of the quantities
\begin{eqnarray}
\max_{q_{k:n}} \text{E}_{\pi} \left [ \sum_{i=k}^{n} 1(x_i= \tilde x_i)  \right ]
\label{eq:partial max}
\end{eqnarray}
for $k=n, n-1, \dots, 1$. 
In words,  \eqref{eq:partial max}  represents the largest possible contribution of  the partial expectation $\text{E}_{\pi} \left [ \sum_{i=k}^{n} 1(x_i= \tilde x_i)  \right ]$ to the full expectation $\text{E}_{\pi} \left [ \sum_{i=1}^{n} 1(x_i= \tilde x_i)  \right ]$ that can be obtained for a fixed $\pi(\tilde x_{1:k-1}, x_{1:k} | y)$.
The recursion uses the fact that, for $k\geq2$, the Markovian structure of $\pi (x, \tilde x | y) $  yields
\begin{eqnarray}
\notag
\hspace{-7mm} \max_{q_{(k-1):n}} \text{E}_{\pi} \left [ \sum_{i=k-1}^{n} 1(x_i= \tilde x_i)  \right ]
& = & \max_{q_{(k-1):n}} \text{E}_{\pi} \left [ 1(x_{k-1}=\tilde x_{k-1}) +  \sum_{i=k}^{n} 1(x_i = \tilde x_{i}) \right ] \\  
%& = & \max_{q^{k-1}} \left ( \max_{q^{k:n}} \text{E}_{\pi} \left [ 1(x_{k-1}=\tilde x_{k-1}) +  \sum_{i=k}^{n} 1(x_i = \tilde x_{i}) \right ] \right ) \\
& = & \max_{q_{k-1}} \left [ \text{E}_{\pi} \left [ 1(x_{k-1}=\tilde x_{k-1}) \right ] +  \max_{q_{k:n}} \text{E}_{\pi} \left [ \sum_{i=k}^{n} 1(x_i = \tilde x_{i}) \right ] \right ]
\label{eq:max recursion}
\end{eqnarray}
suggesting that the full maximum value in \eqref{eq:optimal q} can be computed recursively by recursive maximisation over $q_n$, $q_{n-1}$, $\dots$, $q_1$. 

An essential aspect of the backward recursion are the distributions $\pi_1, \dots, \pi_n$. 
At each step $k$, we compute    \eqref{eq:partial max}  as a function of  $\pi_k$.
%which in turn is a function of $\pi_{k-1}$ and $q_{k-1}$. 
Essentially,  each $\pi_k$, $k\geq 2$,   consists of four numbers, or parameters, one for each possible configuration of the pair $(\tilde x_{k-1}, x_{k})$. 
However, one parameter is lost since $\pi(\tilde x_{k-1}, x_{k} | y)$ is a probability distribution so that the four numbers necessarily sum to one. Another two parameters are lost due to the constraint \eqref{eq:constr} which here entails that $\pi(\tilde x_{k-1} | y) = f(\tilde x_{k-1} | y)$
and $\pi( x_{k} | y) = f(x_{k})$. Thereby, only one parameter, call it $t_k$, remains. 
This parameter $t_k$ is free to vary within an interval $[t_k^{\min}, t_k^{\max}]$, where the bounds $t_k^{\min}$ and $t_k^{\max}$ are determined by the probabilistic nature of $\pi_k$. 
An example parametrisation is to set $t_k = \pi(\tilde x_{k-1} = 0, x_{k} = 0 | y)$, which is the approach taken in this work. Below, the notation  $\pi_{t_k}(\tilde x_{k-1}, x_k | y)$ will, when appropriate, be used instead of $\pi(\tilde x_{k-1}, x_k | y)$,  in order to express the dependence on $t_k$ more explicitly. 
The chosen parameter $t_k$ leads to a parametrisation of $\pi_k$ as follows,
\begin{equation}
\begin{aligned}
\pi_{t_k}(\tilde x_{k-1} = 0, x_{k} = 0 | y) &= t_k, 
\\
\pi_{t_k}(\tilde x_{k-1} = 0, x_{k} = 1 | y) &= f(\tilde x_{k-1}=0|y) - t_k,
\\
\pi_{t_k}(\tilde x_{k-1} = 1, x_{k} = 0 | y) &= f( x_{k}=0) - t_k,
\\
\pi_{t_k}(\tilde x_{k-1} = 1, x_{k} = 1 | y) &= 1- f( x_{k}=0) - f(\tilde x_{k-1}=0 | y)+ t_k,
\end{aligned}
\label{eq:parametrization}
\end{equation}
and the bounds of the interval $[t_k^{\min}, t_k^{\max}]$ are given as 
\begin{equation}
t_k^{\min} = \max \Big \{  0,  f(x_k=0) + f(x_{k-1} = 0|y) -1 \Big \},
\label{eq:tmin}
\end{equation}
\begin{equation}
t_k^{\max} = \min \Big \{f(x_k=0), f(x_{k-1} = 0 | y)  \Big \}.
\label{eq:tmax}
\end{equation}
For $k=1$, the situation is a bit different, since there is only one variable, $x_1$, involved in $\pi_1$. In fact, due to the criterion \eqref{eq:constr}, we have $\pi(x_1 | y) = f(x_1)$. Consequently, $t_1$ is not a parameter free to vary within a certain range, but a fixed number, for example either $ f(x_1 = 0)$ or $f(x_1 = 1)$. 
Here, we take $t_1 =  f(x_1 = 0)$. 

Apart from the parametrisation of $\pi_k$, an essential feature of each $\pi_k,$ for  $k\geq2,$ is its dependence on $\pi_{k-1}$ and $q_{k-1}$. 
This connection is due to the particular structure of $\pi(\tilde x, x|y)$ cf. Figure \ref{fig:q}. Generally, for $k\geq 3$, we know that $\pi_k$, or $\pi(\tilde x_{k-1}, x_{k} | y)$, can be deduced from the  joint distribution $\pi (\tilde x_{k-2}, \tilde x_{k-1}, x_{k-1}, x_{k} | y)$ by summing out the variables $\tilde x_{k-2}$ and $x_{k-1}$,
\begin{eqnarray}
\pi(\tilde x_{k-1},x_{k} | y) = \sum_{\tilde x_{k-2}} \sum_{x_{k-1}}  \pi (\tilde x_{k-2}, \tilde x_{k-1}, x_{k-1}, x_{k} | y) , 
% \sum_{\tilde x_{k-2}} \sum_{x_{k-1}} \pi (\tilde x_{k-2}, x_{k-1}| y) q(\tilde x_{k-1} | \tilde x_{k-2}, x_{k-1}, y) f(x_{k} | x_{k-1}).
\label{eq:pi k}
\end{eqnarray}
where
$\pi (\tilde x_{k-2}, \tilde x_{k-1}, x_{k-1}, x_{k} | y)$ can be written in the particular form
$$
\pi (\tilde x_{k-2}, \tilde x_{k-1}, x_{k-1}, x_{k} | y) = \pi (\tilde x_{k-2}, x_{k-1}| y) q(\tilde x_{k-1} | \tilde x_{k-2}, x_{k-1}, y) f(x_{k} | x_{k-1}).
$$
Similarly, for the special case $k=2$, we can compute $\pi(\tilde x_{1}, x_{2} | y)$ by summing out $x_2$ from $\pi(\tilde x_{1}, x_1, x_{2} | y)$,
\begin{eqnarray}
\pi(\tilde x_{1}, x_{2} | y) &=& \sum \limits_{x_1} \pi(\tilde x_{1}, x_1, x_{2} | y), 
% \\
%& = & \sum \limits_{x_2} f(x_1) q(\tilde x_1 | x_1, y) f(x_2 | x_1) 
\label{eq:pi 2}
\end{eqnarray}
where  $\pi(\tilde x_{1}, x_1, x_{2} | y)$ can be factorised as 
\begin{eqnarray}
\pi(\tilde x_{1}, x_1, x_{2} | y)= f(x_1) q(\tilde x_1 | x_1, y) f(x_2 | x_1) . 
\label{eq:pi 2 2}
\end{eqnarray}
Plugging in $\tilde x_{k-1} = 0$ and $x_{k} = 0$ in \eqref{eq:pi k},  and  using that  $\pi_{k-1}$ is parametrised by $t_{k-1}$, we obtain a formula for $t_{k}$ in terms of $t_{k-1}$ and $q_{k-1}$,  $k\geq 3$.
Likewise, plugging in  $\tilde x_{1} = 0$ and $x_{2} = 0$ in \eqref{eq:pi 2}, and using that $f(x_1 = 0) = t_1$, we obtain a formula for $t_2$ in terms of $t_1$ and $q_1$. 
To  express the dependence of $t_k$ on $t_{k-1}$ and $q_{k-1}$, $k\geq 2$, we will use the notation
$$
t_k = t_{k}(t_{k-1}, q_{k-1}).
$$

In some of the following equations, it will be necessary to explicitly express that  \eqref{eq:partial max} 
is  a function of $t_k$. 
For this reason, we now define
$$
\text{E}_{k:n}^*(t_k) =  \max_{q_{k:n}} \text{E}_{\pi} \left [ \sum_{i=k}^{n} 1(x_i= \tilde x_i)  \right ].
$$
  Similarly, we need a notation for the argument of the maximum in  \eqref{eq:max recursion}  as a function of $t_k$:
  $$
  q^*_{t_k}(\tilde x_k | \tilde x_{k-1}, x_{k}, y) = \argmax_{q_k} \left [ \text{E}_{\pi} \left [ 1(x_{k}=\tilde x_{k}) \right ] +  \max_{q_{(k+1):n}} \text{E}_{\pi} \left [ \sum_{i=k}^{n} 1(x_i = \tilde x_{i}) \right ] \right ], \quad 2 \leq k \leq n,
  $$
   $$
  q^*_{t_1}(\tilde x_1 | x_{1}, y) = \argmax_{q_1} \left [ \text{E}_{\pi} \left [ 1(x_{1}=\tilde x_{1}) \right ] +  \max_{q_{2:n}} \text{E}_{\pi} \left [ \sum_{i=1}^{n} 1(x_i = \tilde x_{i}) \right ] \right ].
  $$
  If $q^*_{t_k}(\tilde x_k | \tilde x_{k-1}, x_{k}, y)$ and $q^*_{t_1}(\tilde x_1 | x_{1}, y)$ are discussed in a context where the specific values of the involved variables are not important, simpler notations are preferable. In this regard, we also introduce 
  $$
  q^*_k(t_k) = \begin{cases}q^*_{t_k}(\tilde x_k | \tilde x_{k-1}, x_{k}, y), & 2\leq k \leq n, \\ q^*_{t_1}(\tilde x_1 |  x_{1}, y), & k = 1.  \end{cases}
  $$
Also, we need a notation for $\text{E}_{\pi}[1(x_k = \tilde x_k)]$ indicating that this is a function of both $t_k$ and $q_k$,
 $$
 \text{E}_{k} (t_{k}, q_{k}) = \text{E}_{\pi} [1(x_{k}= \tilde x_{k})].
 $$

 The backward recursion computes $\text{E}_{k:n}^*(t_k)$ recursively for $k=n, n-1, ..., 1$, each step  performing a maximisation over $q_k$ as a function of  the parameter $t_k$. 
The recursion is initialised by 
\begin{eqnarray}
\text{E}^*_{n}(t_{n}) = \max_{q_{n}} \Big [ \text{E}_n(t_n, q_n) \Big ],
\label{eq:max backward n}
\end{eqnarray}
\begin{eqnarray}
q_n^*(t_n) =  \argmax_{q_{n}} \Big [ \text{E}_n(t_n, q_n)   ] \Big ].
%  q^*_{t_n}(\tilde x_n | \tilde x_{n-1}, x_{n}, y)= \argmax_{q_{n}} \Big [ \text{E}_n(t_n, q_n)   ] \Big ],
\label{eq:argmax backward n}
\end{eqnarray}
Then, for $k=n-1, n-2, \dots, 1$, the recursion proceeds according to
 \begin{eqnarray}
\text{E}^*_{k:n}(t_{k}) = \max_{q_{k}} \Big [ \text{E}_{k} \left (t_{k}, q_{k}\right ) +  \text{E}^*_{(k+1):n}(t_{k+1}(t_{k}, q_{k}) ) \Big ],
\label{eq:E star}
\end{eqnarray}
\begin{equation}
q^*_k(t_k) = \argmax_{q_{k}} \Big [ \text{E}_{k} \left (t_{k}, q_{k}\right ) +  \text{E}^*_{(k+1):n}(t_{k+1}(t_{k}, q_{k}) ) \Big ],
\label{eq:q k star}
\end{equation}
%\begin{subequations}
%\begin{equation}
%  q^*_{t_k}(\tilde x_k | \tilde x_{k-1}, x_{k}, y) = \argmax_{q_{k}} \Big [ \text{E}_{k} \left (t_{k}, q_{k}\right ) +  \text{E}^*_{(k+1):n}(t_{k+1}(t_{k}, q_{k}) ) \Big ],
%  \label{eq:q star}
%\end{equation}
%\begin{equation}
%  q^*_{t_1}(\tilde x_1 | x_{1}, y) = \argmax_{q_{1}} \Big [ \text{E}_{1} \left (t_{1}, q_{1}\right ) +  \text{E}^*_{2:n}(t_{2}(t_{1}, q_{1}) ) \Big ].
%  \label{eq:q star 1}
%\end{equation}
%\end{subequations}
Note that at the final step of the backward recursion, where $k=1$, we compute $\text{E}^*_{1:n}(t_1)$ and $q^*_{1}(t_1)$. Now, since we have one specific value for $t_1$, we also obtain one specific value for $\text{E}^*_{1:n}(t_1)$ and corresponding specific values for $q^*_{1}(t_1)$. This completes the backward recursion.

After the backward recursion, the forward recursion can proceed. Here, we recursively compute the specific values for $t_2, t_3, \dots, t_n$ corresponding to  $\text{E}^*_{1:n}(t_1)$ and $q^*_{1}(t_1) $. Hence,
we recursively obtain the optimal values $q^*(\tilde x_2 | \tilde x_1, x_2, y)$, $q^*(\tilde x_3 | \tilde x_2, x_3, y)$, $\dots$, $q^*(\tilde x_{n} | \tilde x_{n-1}, x_{n}, y)$ in \eqref{eq:factorized q optimal}. 
The forward recursion is initialised by 
$$
t_1^* = t_1,
$$
$$
q^*(\tilde x_1 | x_1, y) = q^*_{t_1^*}(\tilde x_1 | x_{1}, y).
$$
Then, for $k = 2, 3, \dots, n$, the recursion proceeds according to
\begin{eqnarray}
t_k^* = t_{k}(t_{k-1}^*, q^*_{k-1}(t_{k-1}^*)),
\label{eq:t k forward}
\end{eqnarray}
$$
q^*(\tilde x_k | \tilde x_{k-1}, x_k, y) 
= q^*_{t_k^*}(\tilde x_k | \tilde x_{k-1}, x_k, y),
$$
%where for simplicity of writing in \eqref{eq:t k forward}, we have used the shorthand notations $q^*_1$ for  $q^*_{t_1^*}(\tilde x_1 | x_1, y)$ and $q^*_k$ for  $q^*_{t_{k}^*}(\tilde x_{k} | \tilde x_{k-1}, x_{k}, y)$, $k\geq2$. 
When the forward recursion terminates, the optimal solution $q^*(\tilde x | x, y)$ is readily available.

%% ------ -------  ------ -------  ------ -------  ------ -------  ------ -------  ------ -------  ------ -------  %% 

\subsection{Parametric, piecewise linear programming}
\label{sec:linear}
In this section, we look further into the backward recursion of the DP algorithm described in the previous section. As we shall see, each step of this recursion involves the set-up of an optimisation problem that we refer to as a 'parametric, piecewise linear program'; namely,  an optimisation problem with a piecewise linear objective function subject to a set of linear constraints, which at each step $k$ is solved as a function of the parameter $t_k$.
For simplicity of writing in following equations, we introduce the notations:
\begin{subequations}
\begin{align}
\label{eq:q notation}
{}& q^{ij}_{k} = q(\tilde x_{k} = 0 |  \tilde x_{k-1} = i, x_{k} = j, y), 
\\
\begin{split}
\label{eq: q1 notation}
{} & q_1^i = q(\tilde x_1 = 0 | x_1 = i, y)
\end{split}
\\
\begin{split}
\label{eq:f notation}
{}& f^{ij}_{k} = f(x_{k-1} = i, x_{k} = j | y),
\end{split}
\\
\begin{split}
\label{eq:pi k notation}
{}& 
\pi^{ij}_k(t_k) = \pi_{t_k}(\tilde x_{k-1} = i, x_{k} = j | y),
\end{split}
\\
\begin{split}
 {}& q^{*ij}_{k}(t_k) = q^*_{t_k}(\tilde x_{k} = 0 |  \tilde x_{k-1} = i, x_{k} = j, y),
\end{split}
\\
\begin{split}
 {}& \rho_{k-1}^{i | j} = f(x_k = i | x_{k-1} = j),
\end{split}
\end{align}
\end{subequations}
for $i,j \in \{0,1\}$ and $k\geq 2$.

%
%\begin{subequations}
%%
%\begin{equation}
%\pi^{ij}_k(t_k) = \pi_{t_k}(\tilde x_{k-1} = i, x_{k} = j | y),
%\end{equation}
%%
%\begin{equation}
%f^{ij}_{k} = f(x_{k-1} = i, x_{k} = j | y), 
%\end{equation}
%%
%\begin{equation}
%q^{ij}_{k} = q(\tilde x_{k} = 0 |  \tilde x_{k-1} = i, x_{k} = j, y),
%\end{equation}
%%
%\begin{equation}
% q^{*ij}_{k}(t_k) = q^*_{t_k}(\tilde x_{k} = 0 |  \tilde x_{k-1} = i, x_{k} = j, y),
%\end{equation}
%%
%\begin{equation}
%\rho_{k-1}^{ij} = f(x_k = j | x_{k-1} = i),
%\end{equation}
%%
%\end{subequations}
%or $i,j \in \{0,1\}$ and $k\geq 2$.
%
%\begin{equation}
%\begin{aligned}
%\pi^{ij}_k(t_k) = \pi_{t_k}(\tilde x_{k-1} = i, x_{k} = j | y), \\
%f^{ij}_{k} = f(x_{k-1} = i, x_{k} = j | y), \\
%q^{ij}_{k} = q(\tilde x_{k} = 0 |  \tilde x_{k-1} = i, x_{k} = j, y), \\
%q^{*ij}_{k}(t_k) = q^*_{t_k}(\tilde x_{k} = 0 |  \tilde x_{k-1} = i, x_{k} = j, y), \\
%\rho_{k-1}^{ij} = f(x_k = j | x_{k-1} = i),
%\end{aligned}
%\label{eq:notations}
%\end{equation}
%for $i,j \in \{0,1\}$ and $k\geq 2$.

Reconsider the initial step of the backward recursion. Here,  the aim is to compute $E^*_n(t_n)$  in \eqref{eq:max backward n} and $q_{n}^*(t_n)$ in \eqref{eq:argmax backward n}.
The objective function at this step, $\text{E}_n(t_n, q_n)$, can be computed as 
\begin{eqnarray*}
E_n(t_n, q_n) &=& \pi^{00}_{n}(t_n) q_{n}^{00} +  \pi^{01}_{n}(t_n)(1- q_{n}^{01}) + \pi^{10}_{n}(t_n)q_{n}^{10} + \pi^{11}_{n}(t_n)(1- q_{n}^{11}) .
\end{eqnarray*}
Since $\pi^{01}_{n}(t_n) + \pi^{11}_{n}(t_n) = f(x_n=1)$, we can, after rearranging the terms,  rewrite this function as
\begin{eqnarray}
E_n(t_n, q_n) &=& \pi^{00}_{n}(t_n) q_{n}^{00} -  \pi^{01}_{n}(t_n)q_{n}^{01} + \pi^{10}_{n}(t_n)q_{n}^{10} - \pi^{11}_{n}(t_n)q_{n}^{11} + f(x_n=1).
\label{eq:objective n}
\end{eqnarray}
As a function of the parameter $t_n \in [t_n^{\min}, t_n^{\max}]$, we are interested in computing the solution of $q_n$ which maximises \eqref{eq:objective n}.
In this regard, one needs to take the constraint \eqref{eq:constr} into account. Specifically, the constraint entails at this step that
$$
 \pi(\tilde x_{n-1}, \tilde x_{n} | y ) = f(\tilde x_{n-1}, \tilde x_{n} | y)
$$
for all $\tilde x_{n-1}, \tilde x_{n} \in \{0,1\}$. 
Hence, using that $\pi(\tilde x_{n-1}, \tilde x_{n} , x_{n}| y)$ 
$=\pi(\tilde x_{n-1}, x_{n} | y)  q(\tilde x_{n} | \tilde x_{n-1}, x_{n}, y)$,
and that $\pi(\tilde x_{n-1}, \tilde x_{n} | y)$ follows by summing out $x_{n}$ from $\pi(\tilde x_{n-1}, \tilde x_{n} , x_{n}| y)$, we see that $q_n$ must fulfil
\begin{eqnarray*}
f(\tilde x_{n-1},\tilde x_{n} |  y) 
&=& {\sum_{x_{n}} \pi(\tilde x_{n-1}, x_{n} | y) q(\tilde x_{n} | \tilde x_{n-1}, x_{n}, y)}.
\end{eqnarray*}
This requirement leads to four linear equations of which two are linearly independent, one where we set $\tilde x_{n-1}=0$ and one where we set  $\tilde x_{n-1} = 1$. Using the notations in \eqref{eq:q notation}-\eqref{eq:pi k notation}, the two linearly independent equations can be written as
\begin{eqnarray}
f_{n}^{00} = \pi_{n}^{00}(t_n) q_{n}^{00} + \pi_{n}^{01}(t_n) q_{n}^{01}, \label{eq: equality constr 1} \\
f_{n}^{10} = \pi_{n}^{10}(t_n) q_{n}^{10} + \pi_{n}^{11}(t_n) q_{n}^{11}. \label{eq: equality constr 2}
\label{eq: equality constr}
\end{eqnarray}
Additionally, we know that $q_n^{00}, q_n^{01}, q_n^{10},$ and $q_n^{11}$  can only take values within the interval $[0,1]$, 
\begin{equation}
0 \leq q_n^{ij} \leq 1, \quad \text{ for all } i,j \in \{0,1\}. 
\label{eq:inequality constr}
\end{equation}
To summarise, we want, as a function of the parameter $t_n \in [t_n^{\min}, t_n^{\max}]$, to compute the solutions of $q_n^{00}, q_n^{01}, q_n^{10}$, and $q_n^{11}$ which maximise the function \eqref{eq:objective n} subject to the constraints \eqref{eq: equality constr 1}-\eqref{eq:inequality constr}. For any fixed $t_n $, this is a maximisation problem where both the objective function and all the constraints are linear in  $q_n^{00}, q_n^{01}, q_n^{10}$, and $q_n^{11}$. As such, the maximisation problem can, for a given value of $t_n$, be formulated as a linear program and solved accordingly. 
In Appendix A, we show that the optimal solutions  $q_n^{*00}(t_n)$, $q_n^{*01}(t_n)$, $q_n^{*10}(t_n)$, and $q_n^{*11}(t_n)$ are piecewise-defined functions of $t_n$ and easy to compute analytically. Furthermore, we show that the corresponding function $E^*_n(t_n)$, obtained by inserting $q_n^{*00}(t_n)$, $q_n^{*01}(t_n)$, $q_n^{*10}(t_n)$, and $q_n^{*11}(t_n)$ into \eqref{eq:objective n}, is a continuous piecewise linear (CPL)  function of $t_n$.

Next, consider the intermediate steps of the backward recursion, that is $k = n-1, n-2, \dots, 2.$ At each such step, 
the aim is to compute 
$E^*_{k:n}(t_{k})$ in \eqref{eq:E star} and $q_{{k}}^*(t_k)$ in \eqref{eq:q k star}. 
The objective function at each step reads
\begin{equation}
E_{k:n}(t_{k}, q_{k}) = E_{k}(t_{k}, q_{k}) + E^*_{(k+1):n}(t_{k+1}(t_{k}, q_{k})),
\label{eq:objective k}
\end{equation}
which is to be maximised with respect to $q_k$. 
Here, the first term, $E_{k}(t_{k}, q_{k})$, can 
be computed as 
\begin{eqnarray}
E_{k}(t_{k}, q_{k}) &=& \pi^{00}_{k}(t_{k}) q_{k}^{00} -  \pi^{01}_{k}(t_{k})q_{k}^{01} + \pi^{10}_{k}(t_{k})q_{k}^{10}   - \pi^{11}_{k}(t_{k})q_{k}^{11} + f(x_{k}=1).
\label{eq:E k}
\end{eqnarray}
The second term, $E^*_{(k+1):n}(t_{k+1}(t_{k}, q_{k}))$,  is a CPL function of $t_{k+1}$.
For $k = n-1$, this result is immediate, since we know from the first iteration that  $E_n^*(t_n)$ is CPL. For $k<n-1$, the result is explained in Appendix A. 
Since $t_{k+1}(t_k, q_k)$ is linear in $q_k$, it follows that $E^*_{k+1}(t_{k+1}(t_k, q_k))$ is CPL in $q_k$ for any given $t_k\in [t_k^{\min}, t_k^{\max}]$. Hence, the objective function \eqref{eq:objective k} is also CPL in $q_k$ for any $t_k \in [t_k^{\min}, t_k^{\max}]$. 
As in the first backward step, we have the following equality and inequality constraints for $q_k$:
\begin{subequations}
\label{eq:constr k both}
\begin{equation}
f_{k}^{00} = \pi_{k}^{00}(t_{k}) q_{k}^{00} + \pi_{k}^{01}(t_{k}) q_{k}^{01},
\label{eq:constr k 1}
\end{equation}
\begin{equation}
f_{k}^{10} = \pi_{k}^{10}(t_{k}) q_{k}^{10} + \pi_{k}^{11}(t_{k}) q_{k}^{11},
\label{eq:constr k 2}
\end{equation}
\end{subequations}
and
\begin{equation}
0 \leq q_{k}^{00}, q_{k}^{01}, q_{k}^{10}, q_{k}^{11} \leq 1.
\label{eq:ineq constr k}
\end{equation}
Additionally, we need to incorporate constraints ensuring that $q_k$ and $t_k$ return a value $t_{k+1}$ within the interval $[t_{k+1}^{\min}, t_{k+1}^{\max}]$, with $t_{k+1}^{\min}$ and $t_{k+1}^{\max}$ given by \eqref{eq:tmin} and \eqref{eq:tmax}, respectively. That is, we require
\begin{equation}
t_{k+1}^{\min} \leq t_{k+1} (t_{k}, q_{k}) \leq t_{k+1}^{\max}.
\label{eq:ineq addition}
\end{equation}
From  \eqref{eq:pi k},  the formula $t_{k+1} (t_{k}, q_{k})$ follows  as 
\begin{eqnarray}
t_{k+1} (t_{k}, q_{k})
= \pi_{k}^{00}(t_{k}) q_{k}^{00} \rho_{k}^{0|0}  + \pi_{k}^{01}(t_{k}) q_{k}^{01} \rho_{k}^{0|1} 
+  \pi_{k}^{10}(t_{k}) q_{k}^{10} \rho_{k}^{0|0}  + \pi_{k}^{11}(t_{k}) q_{k}^{11} \rho_{k}^{0|1}.
\label{eq:t k}
\end{eqnarray}
Clearly, for any fixed $t_k \in [t_k^{\min}, t_k^{\max}]$, all the constraints \eqref{eq:constr k both}-\eqref{eq:ineq addition} are  linear in $q_k$. However, the objective function in \eqref{eq:objective k}  is only piecewise linear. As such, we are not faced with a standard linear program, but a
'piecewise linear program'. 
Piecewise linear programs are a well-studied field of linear optimisation and several techniques for solving such problems have been proposed and studied, see for instance \cite{art10, art11, art12}. The most straightforward approach is to solve the standard linear program corresponding to each line segment of the objective function separately, and afterwards compare the solutions and store the overall optimum. 
This technique 
can be inefficient and is not recommended if the number of pieces of the objective function is relatively large.
However, in our case,  the objective functions normally consist of only a few pieces. For example, in the simulation experiment of Section 5.2, where a model $q(\tilde x  | x, y)$ was constructed as much as 1,000 times, the largest number of intervals observed was {10} and the average number of intervals was {4.35}.  
Therefore, we consider the straightforward approach as a convenient method  for solving our piecewise linear programs, although we are aware that more elegant strategies exist and might have their advantages. Further details of our solution are presented below.

First, some new notations needs to be introduced.  For each $2\leq k\leq n$, we let  $M_{k}$ denote the number of pieces, or intervals, of $E^*_{k:n}(t_{k})$, and we let  $t_{k}^{B(j)},$ $ j = 1, \dots, M_{k}+1$, denote the corresponding breakpoints. Note that for the first and last breakpoints, we have $t_{k}^{B(1)} = t_{k}^{\min}$ and $t_{k}^{B(M_{k}+1)} = t_{k}^{\max}$.
Further, we let 
${I}_{k}^{(j)} = [t_{k}^{B(j)}, t_{k}^{B(j+1)} ] \subseteq [t_{k}^{\min}, t_{k}^{\max}]$ denote interval no. $j $, and $\mathcal S_{k}  = \{1, 2, \dots, M_{k}\} $ the set of interval indices. 
For each $j \in \mathcal S_{k}$,  $E^*_{k:n}(t_{k})$ is defined by a linear function, which we denote by $E^{*(j)}_{k}(t_{k})$, whose intercept and slope we denote by  $a_{k}^{(j)}$ and $b_{k}^{(j)}$, respectively.

Each 
 linear piece, $E^{*(j)}_{k+1}(t_{k+1})$,  of the piecewise linear function $E^*_{(k+1):n}(t_{k+1})$ leads to a standard parametric linear program.
Specifically, if $E_{(k+1):n}^{*} (t_{k+1} (t_{k}, q_{k} ))$ in \eqref{eq:objective k} is replaced  by $E_{(k+1):n}^{*(j)} (t_{k+1} (t_{k}, q_{k} ))$, we obtain an objective function
 \begin{equation}
 E_{k:n}^{(j)} (t_{k}, q_{k}) = E_{k} (t_{k}, q_{k}) + E_{(k+1):n}^{*(j)} (t_{k+1} (t_{k}, q_{k} )),
 \label{eq:objective k j}
 \end{equation}
 which is linear, not piecewise linear, as a function of $q_k$. The corresponding constraints for $q_k$ are given in \eqref{eq:constr k both} and \eqref{eq:ineq constr k}, but instead of \eqref{eq:ineq addition}, we require that $t_k$ and $q_k$ return a value $t_{k+1}(t_{k}, q_{k})$ within the interval $I_{k+1}^{(j)}$, 
 \begin{eqnarray}
t_{k+1}^{B(j)} \leq t_{k+1} (t_{k}, q_{k}) \leq t_{k+1}^{B(j+1)}.
\label{eq: tk constr}
\end{eqnarray}
 Using  \eqref{eq:E k},
  \eqref{eq:t k}, and that $E^{*(j)}_{k+1}(t_{k+1}) = a_{k+1}^{(j)} + b_{k+1}^{(j)} t_{k+1}$,  we can for each $j\in \mathcal S_{k+1}$  rewrite
   \eqref{eq:objective k j}
   as 
\begin{eqnarray}
E_{k:n}^{(j)} (t_{k}, q_{k}) = \beta_{k}^{00(j)} (t_{k}) q_{k}^{00} + \beta_{k}^{01(j)}(t_{k}) q_{n-1}^{01}   + \beta_{k}^{10(j)} (t_{k}) q_{k}^{10}  + \beta_{k}^{11(j)} (t_{k}) q_{k}^{11} + \alpha_{k}^{(j)},
\label{eq:k rewritten}
\end{eqnarray}
where 
$$
\beta_{k}^{00(j)}(t_{k}) = \left ( b_{k+1}^{(j)}\rho_{k}^{0|0} + 1 \right)  \pi_{k}^{00} (t_{k}),
$$
$$
\beta_{k}^{01(j)}(t_{k}) = \left ( b_{k+1}^{(j)}\rho_{k}^{0|1} - 1 \right ) \pi_{k}^{01} (t_{k}), 
$$
$$
\beta_{k}^{10(j)}(t_{k}) =\left ( b_{k+1}^{(j)}\rho_{k}^{0|0} +1 \right) \pi_{k}^{10}(t_{k}),
$$
$$
\beta_{k}^{11(j)}(t_{k})= \left( b_{k+1}^{(j)}\rho_{k}^{0|1} - 1\right) \pi_{k}^{11}(t_{k}), 
$$
and
$$
\alpha_{k}^{(j)}= a_{k+1}^{(j)} + f(x_{k}=1).
$$
%\begin{eqnarray*}
%\beta_{k}^{00(j)}(t_{k})& =& \left ( b_{k+1}^{(j)}\rho_{k}^{00} + 1 \right)  \pi_{k}^{00} (t_{k}), \\
%\beta_{k}^{01(j)}(t_{k})& =& \left ( b_{k+1}^{(j)}\rho_{k}^{10} - 1 \right ) \pi_{k}^{01} (t_{k}), \\
%\beta_{k}^{10(j)}(t_{k})& =&\left ( b_{k+1}^{(j)}\rho_{k}^{00} +1 \right) \pi_{k}^{10}(t_{k}), \\
%\beta_{k}^{11(j)}(t_{k})& =& \left( b_{k+1}^{(j)}\rho_{k}^{10} - 1\right) \pi_{k}^{11}(t_{k}), \\
%\alpha_{k}^{(j)} &=& a_{k+1}^{(j)} + f(x_{k}=1).
%\end{eqnarray*}
 To summarize, we obtain for each $j \in \mathcal S_{k+1}$ a standard parametric linear program, with the objective function given in \eqref{eq:k rewritten} and the constraints given in \eqref{eq:constr k both}, \eqref{eq:ineq constr k}, and \eqref{eq: tk constr}. 
Solving the parametric linear program corresponding to each $j \in \mathcal S_{k+1}$, yields the following quantities:
 \begin{equation}
 \widetilde E^{(j)}_{k:n} (t_k) = \max\limits_{q_k} E_{k:n}^{(j)} (t_k, q_k),
 \label{eq:tilde E}
 \end{equation}
  \begin{equation}
 \widetilde q_{k}^{(j)}(t_k)   = \argmax_{q_k} E_{k:n}^{(j)} (t_k, q_k).
 \label{eq:tilde q}
 \end{equation}
 The overall maximum value $E_{k:n}^*(t_k)$ and corresponding argument $q_{k}^{*}  (t_k) $ are then available as
 \begin{equation*}
 E_{k:n}^*(t_k) = E_{k:n}^{j^*_{k+1}(t_k)} (t_k)
 \end{equation*}
 and
 \begin{equation*}
 q_k^*(t_k) = \widetilde q_k^{(j^*_{k+1}(t_k))}(t_k)
 \end{equation*}
 where
 $$
 j_{k+1}^*(t_k) = \argmax_{j\in \mathcal S_{k+1}} \widetilde E_{k:n}^{(j)} (t_k).
 $$
 As previously mentioned, and as shown in Appendix A,  $E_{k:n}^*(t_k)$ is a CPL function of $t_k$. As such, $E_{k:n}^*(t_k)$ is fully specified by its breakpoints and corresponding function values.  
The breakpoints of $E_{k:n}^*(t_k)$ can be computed beforehand.  
 Thereby, we can
obtain $E_{k:n}^*(t_k)$ for all values of  $t_k$ quite efficiently, as we only need to solve our parametric, piecewise linear program at the breakpoints of $E_{k:n}^*(t_k)$.

Finally, consider the last step of the backward recursion, $k=1$. Here, the aim is to compute $q^*_{t_1}(\tilde x_1 | x_1,y)$ and $E_{1:n}^*(t_1)$. Essentially, this step proceeds in the same fashion as the intermediate steps, but some technicalities are a bit different since there are only two variables involved in $q_1$, namely $q_1^0 = q(\tilde x_1 = 0 | x_1 = 0, y)$ and $q_1^1 = q(\tilde x_1 = 0 | x_1 = 1, y)$. Also, $t_1$ is not a parameter free to vary within a certain range, but a fixed number, namely $t_1 = f(x_1 = 0)$, meaning that we obtain specific values for $q^*_{t_1}(\tilde x_1 | x_1,y)$ and $E_{1:n}^*(t_1)$. 
The function we want to maximise at this final backward step, with respect to $q_1$, is
\begin{equation}
E_{1:n} (t_1, q_1) = E_1 (t_1, q_1) + E^*_{2:n} ( t_2 (t_1, q_1) ),
\label{eq:objective 1}
\end{equation}
where now, recalling that $\pi(x_1 | y) = f(x_1)$, the first term, $E_1 (t_1, q_1)$, can be written as
\begin{equation}
E_1(t_1, q_1) = t_1 q_1^0  + (1-t_1) (1-q_1^1), 
\label{eq:E1}
\end{equation}
Again, as in the intermediate steps, we have a piecewise linear, not a linear, objective function. 
To determine the constraints for $q_1$, we note that the 
requirement  \eqref{eq:constr} for $q(\tilde x | x, y)$ entails that
$$
f(\tilde x_1 | y) = \pi(\tilde x_1 | y).
$$
Thereby, since $t_1 = f(x_1 = 0)$ and using that  $f(\tilde x_1 | y) = \sum_{x_1} \pi(\tilde x_1, x_1 | y)$ and  $ \pi(\tilde x_1, x_1 | y) = f(x_1 ) q(\tilde x_1 | x_1 , y)$, we see that the following requirement must be met by $q(\tilde x_1 | x_1, y)$:
\begin{equation}
f(\tilde x_1 |y ) =t_1 q(\tilde x_1 | x_1 = 0, y) + (1-t_1)q(\tilde x_1 | x_1 = 1, y).
\label{eq:constr last eq}
\end{equation}
Additionally, we have the inequality constraints
\begin{equation}
0 \leq q_1^0, q_1^1 \leq 1.
\label{eq:constr last ineq}
\end{equation}
So, we are faced with a piecewise linear program, with the piecewise linear objective function \eqref{eq:objective 1} and the linear constraints \eqref{eq:constr last eq} and \eqref{eq:constr last ineq}. 
Again, we proceed by iterating through each linear piece of $E^*_{2:n} ( t_2 (t_1, q_1) )$, solving the standard linear program corresponding to each piece separately. 
That is, for each $j\in \mathcal S_2$, we replace $E^*_{2:n} ( t_2 (t_1, q_1) )$ in \eqref{eq:objective 1} by $E^{*(j)}_{2:n} ( t_2 (t_1, q_1) )$ and consider instead the objective function 
\begin{equation}
E_{1:n}^{(j)} (t_1, q_1) = E_1 (t_1, q_1) + E^{*(j)}_{2:n} ( t_2 (t_1, q_1) ),
\label{eq:objective 1 j}
\end{equation}
which is linear, not piecewise linear, as a function of $q_1$. 
As we did for each subproblem $j \in \mathcal S_{k+1}$ in every intermediate backward iteration, we must for each subproblem $j\in \mathcal S_2$ incorporate the inequality constraints 
 \begin{eqnarray}
t_{2}^{B(j)} \leq t_{2} (t_{1}, q_{1}) \leq t_{2}^{B(j+1)},
\label{eq: t1 constr}
\end{eqnarray} 
where now the formula $t_{2} (t_{1}, q_{1})$ follows from  \eqref{eq:pi 2} and \eqref{eq:pi 2 2} as
\begin{equation}
t_2(t_1, q_1) = t_1 q_1^0 \rho_1^{0|0} + (1-t_1) q_1^1 \rho_1^{0|1}.
\label{eq:t2}
\end{equation}
Using \eqref{eq:E1}, \eqref{eq:t2}, and that $E^{*(j)}_{2:n} ( t_2) = a_2^{(j)} + b_2^{(j)}t_2 $, we can rewrite the function in \eqref{eq:objective 1 j}
as
\begin{equation}
E_{1:n}^{(j)} (t_1, q_1) = \beta_1^{0(j)}(t_1) q_1^{0} + \beta_1^{1(j)}(t_1) q_1^1 + \alpha_1^{(j)}(t_1),
\label{eq:objective 1 j rewritten}
\end{equation}
where
$$
\beta_1^{0(j)}(t_1) = t_1 (1 + b_2^{(j)} \rho_1^{0|0}),
$$
$$
\beta_1^{1(j)} (t_1)= (1-t_1) (1 + b_2^{(j)} \rho_1^{0|1}),
$$
$$
\alpha_1^{(j)}(t_1) = 1  - t_1 + a_2^{(j)}.
$$
To summarize, we obtain for each $j\in \mathcal S_2$ a standard linear program, where the aim is to maximise the  objective function \eqref{eq:objective 1 j rewritten} with respect to $q_1$ subject to the constraints \eqref{eq:constr last eq}, \eqref{eq:constr last ineq} and \eqref{eq: t1 constr}. This program is solved  for $t_1 = f(x_1 = 0)$. Analogously to  \eqref{eq:tilde E} and \eqref{eq:tilde q}, let
$$
\widetilde E^{(j)}_{1:n}(t_1) = \max_{q_1} E_{1:n}^{(j)} (t_1, q_1), 
$$
$$
\widetilde q_1^{(j)}(t_1) =  \argmax_{q_1} E_{1:n}^{(j)} (t_1, q_1).
$$
Ultimately, we obtain
$$
E_{1:n}^*(t_1) = \widetilde E_{1:n}^{(j_2^*)} (t_1)
$$
and
$$
q_1^*(t_1) = \widetilde q_{1}^{(j_2^*)} (t_1)
$$
where
 $$
 j^*_{2} = \argmax_{j \in \mathcal S_{2}} \left [ \widetilde E_{1:n}^{(j)} (t_1) \right ].
 $$

	\section{Numerical experiments}
	\label{sec:Numerical}

In this section, we present two empirical studies with simulated data.  
In Section \ref{sec:toy example}, we present a toy example where the assumed prior $f(x)$ is a given stationary Markov chain of length $n = 4$. 
Here, we focus on the  construction of $q(\tilde x | x, y)$ for this assumed prior model, not on the application of it in an ensemble-based context. 
In Section \ref{sec:advanced example}, we consider a higher-dimensional and ensemble-based example, inspired by the movement, or flow, of water and oil in a petroleum reservoir. 

\subsection{Toy example}
\label{sec:toy example}

\begin{figure}
\subfigure[]{\includegraphics[width=5.05cm]{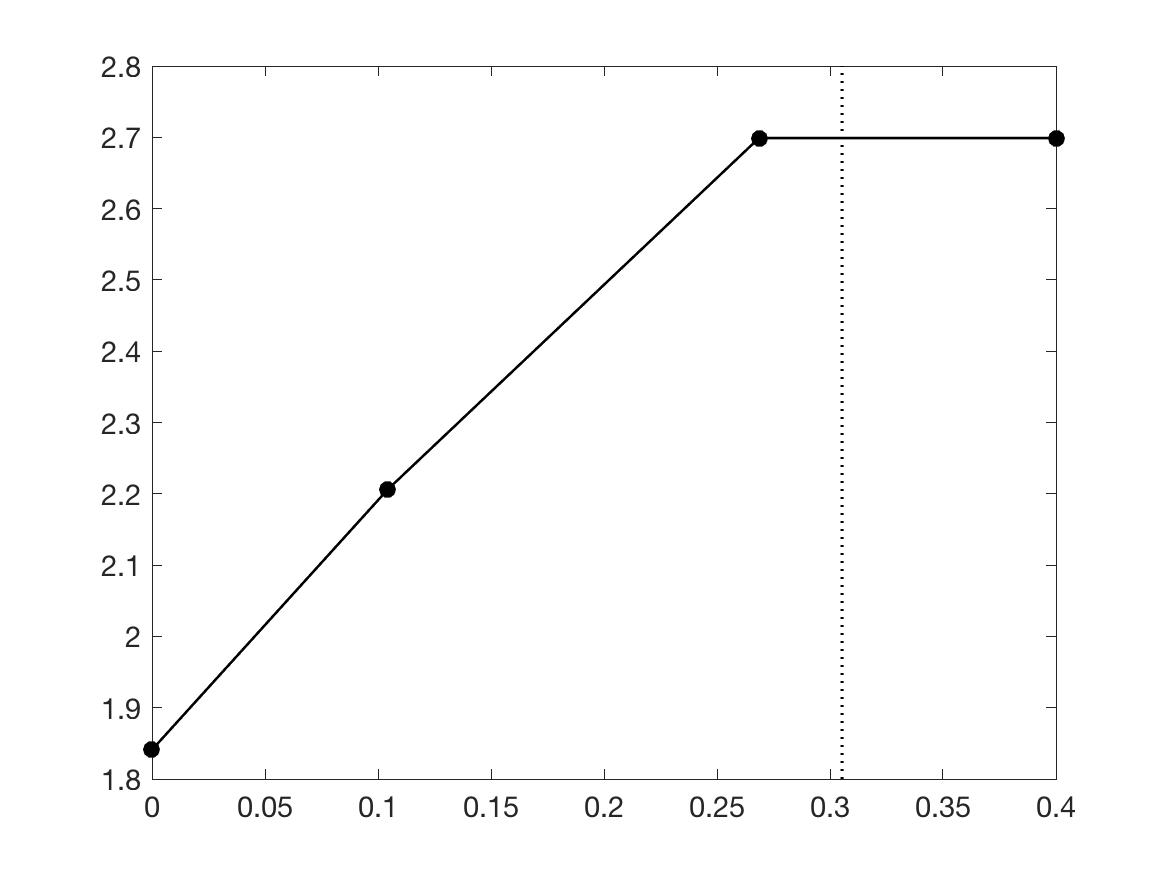}} \hfill
\subfigure[]{\includegraphics[width=5.05cm]{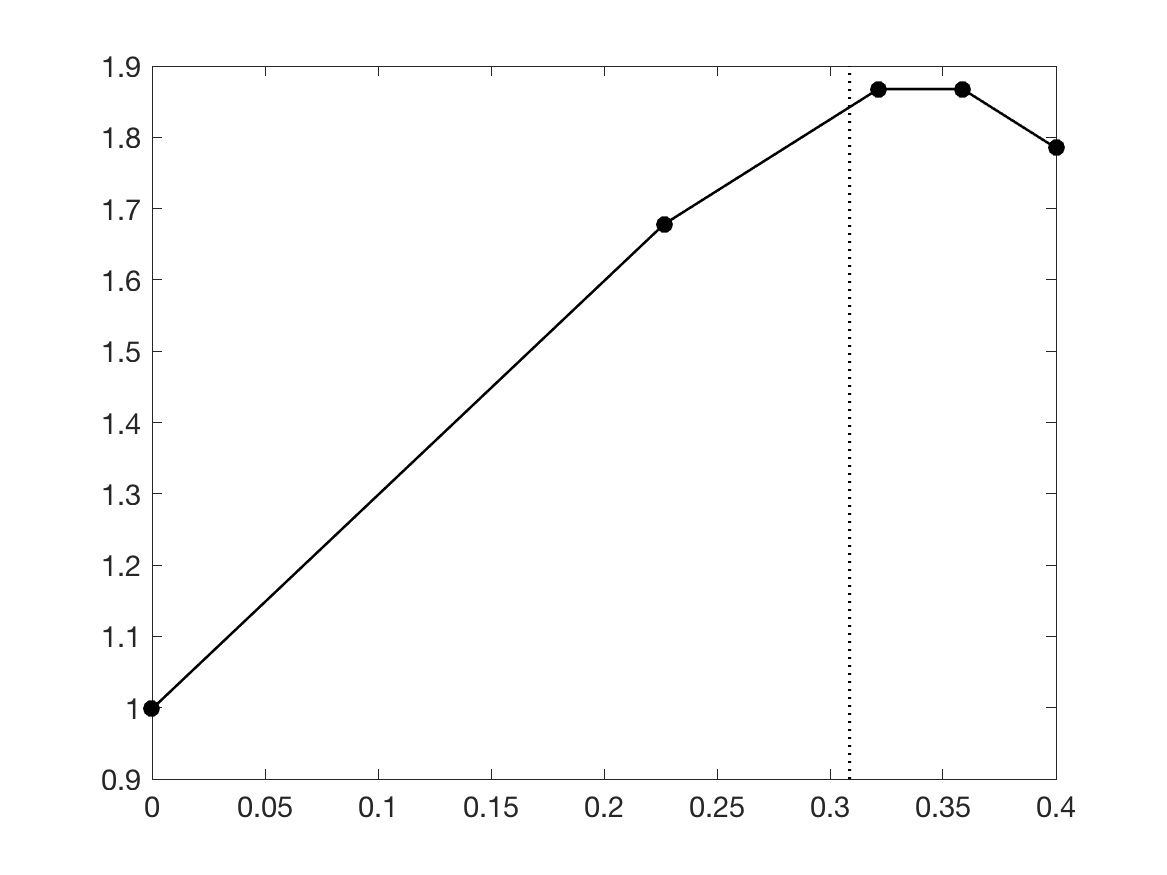}} \hfill
\subfigure[]{\includegraphics[width=5.05cm]{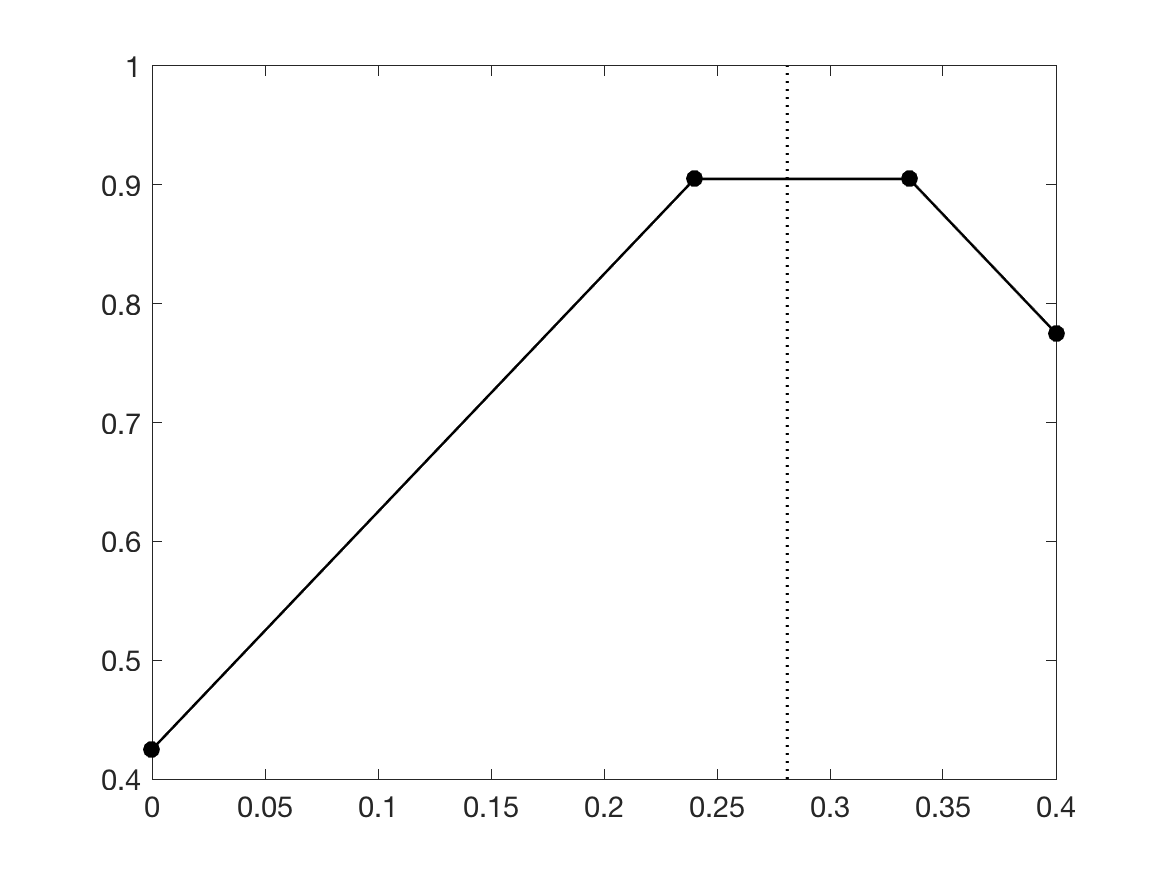}}
\subfigure[]{\includegraphics[width=5.05cm]{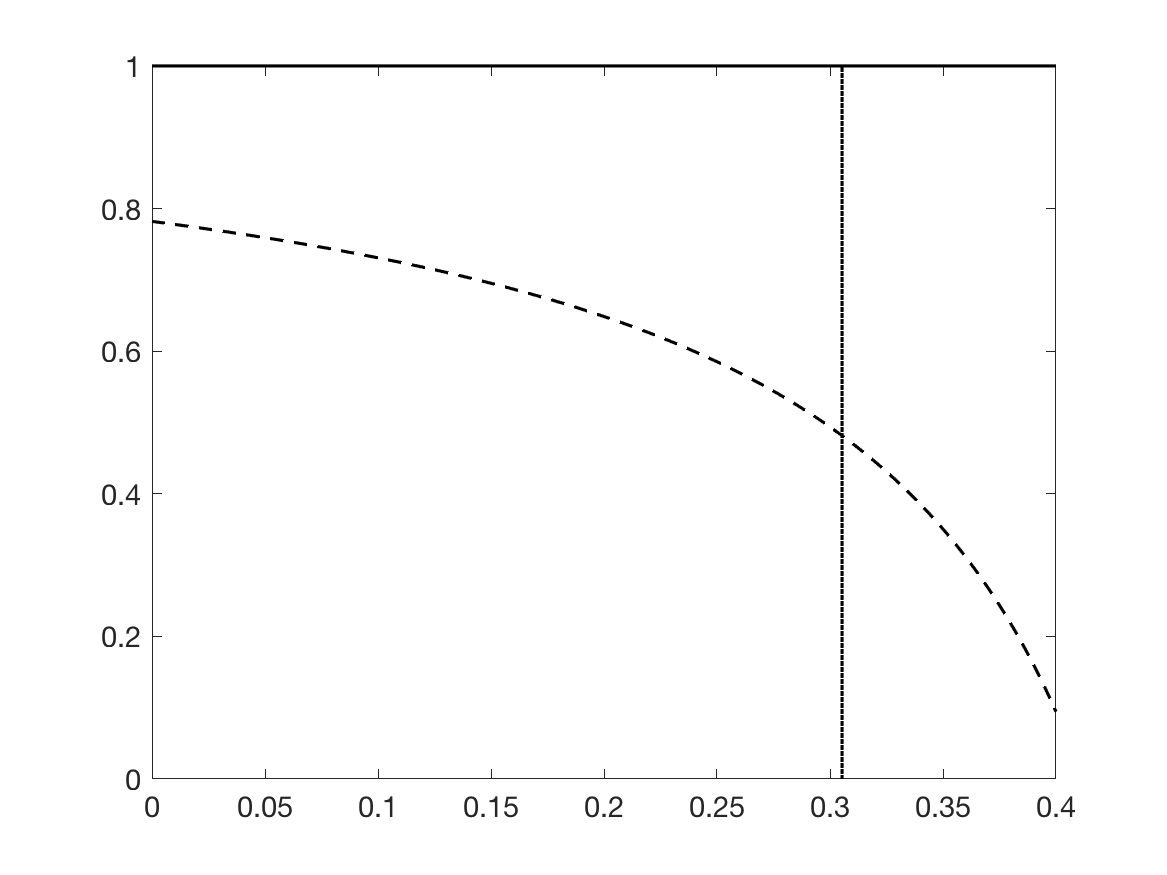}} \hfill
\subfigure[]{\includegraphics[width=5.05cm]{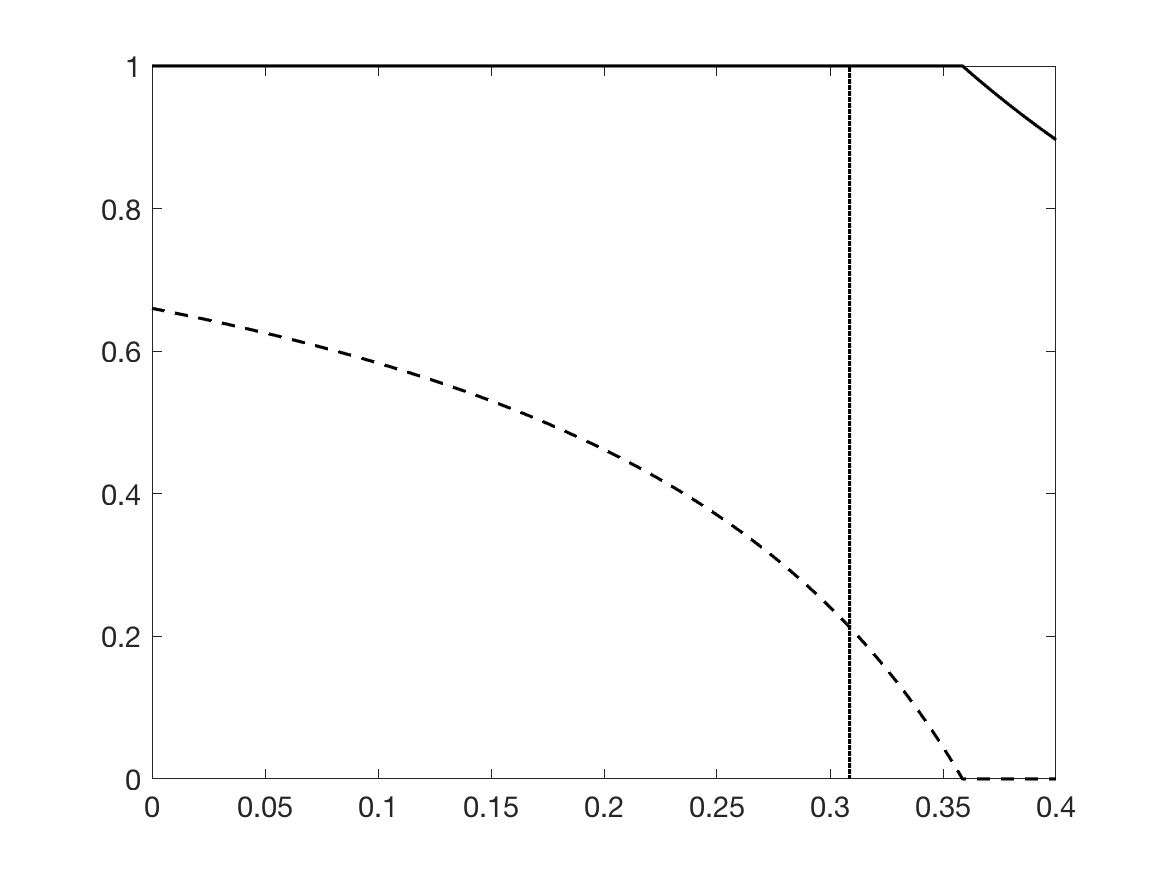}} \hfill
\subfigure[]{\includegraphics[width=5.05cm]{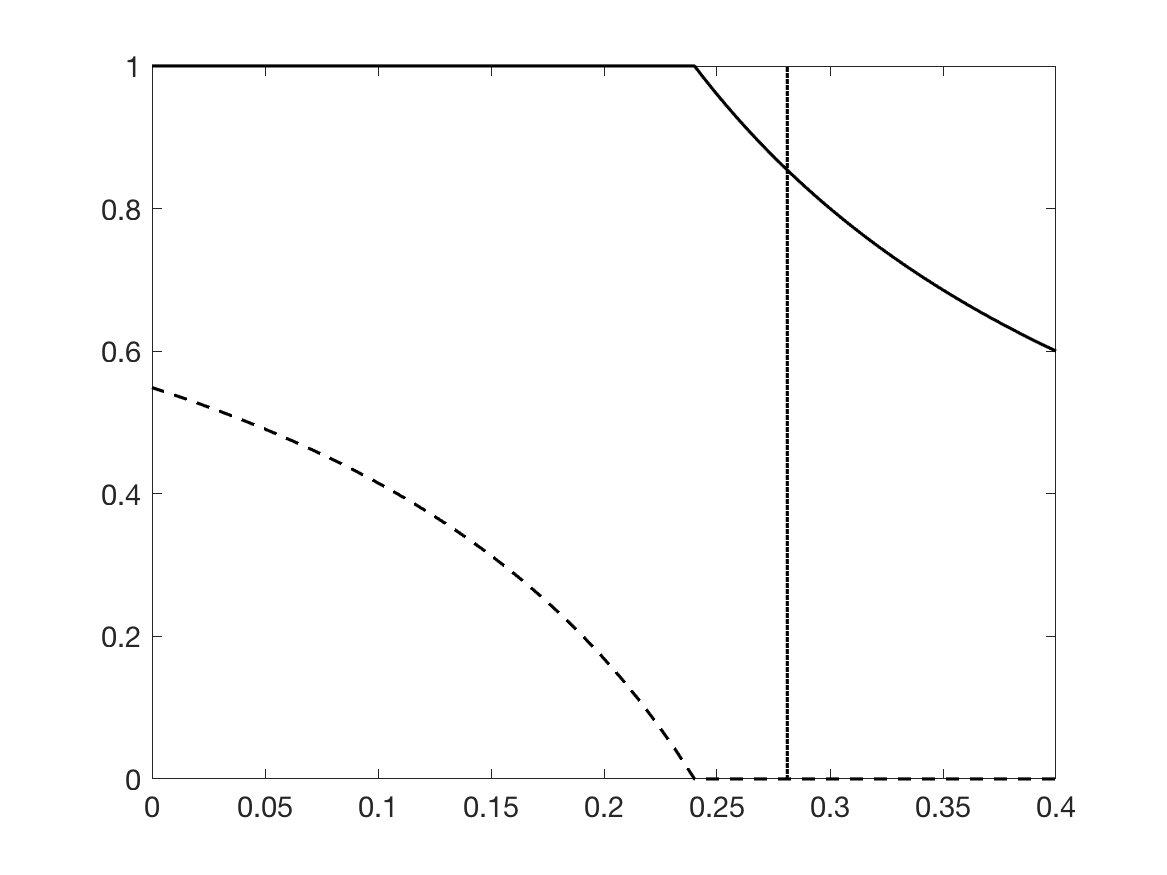}}
\subfigure[]{\includegraphics[width=5.05cm]{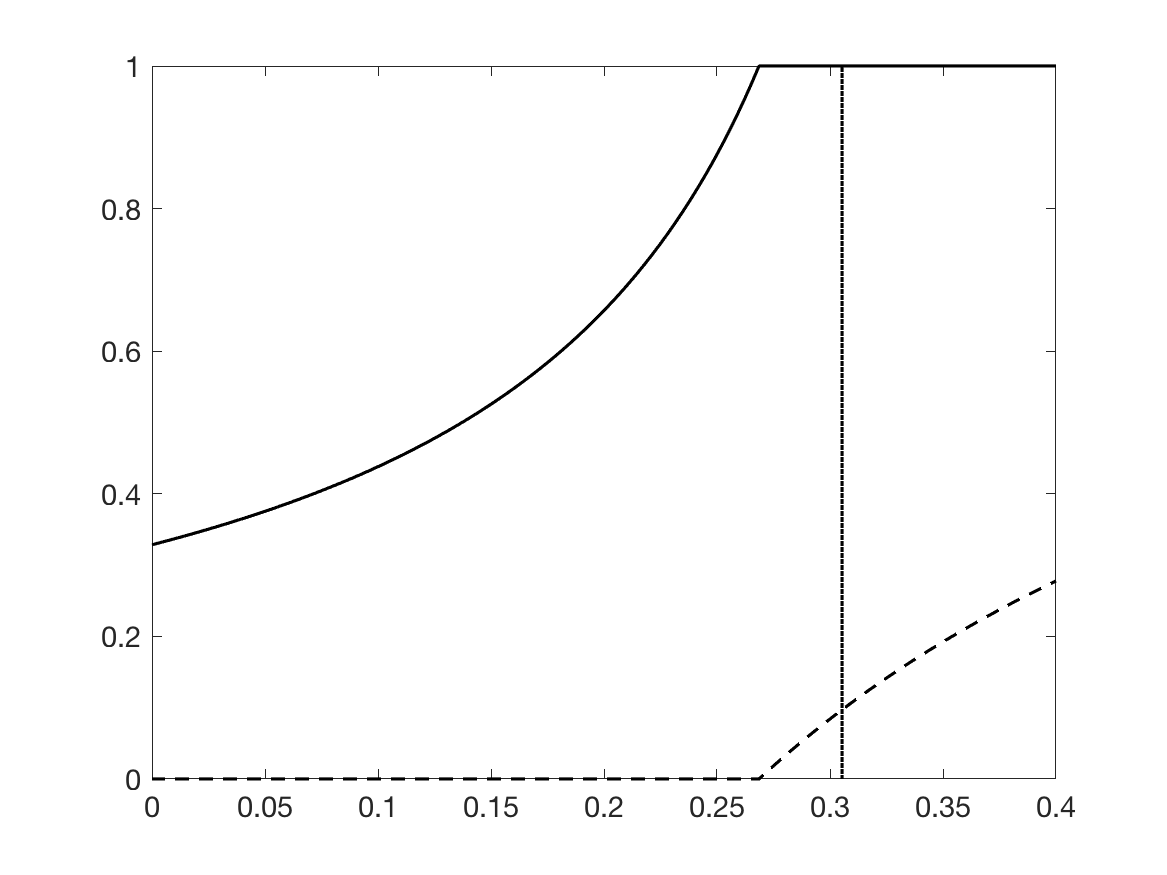}} \hfill
\subfigure[]{\includegraphics[width=5.05cm]{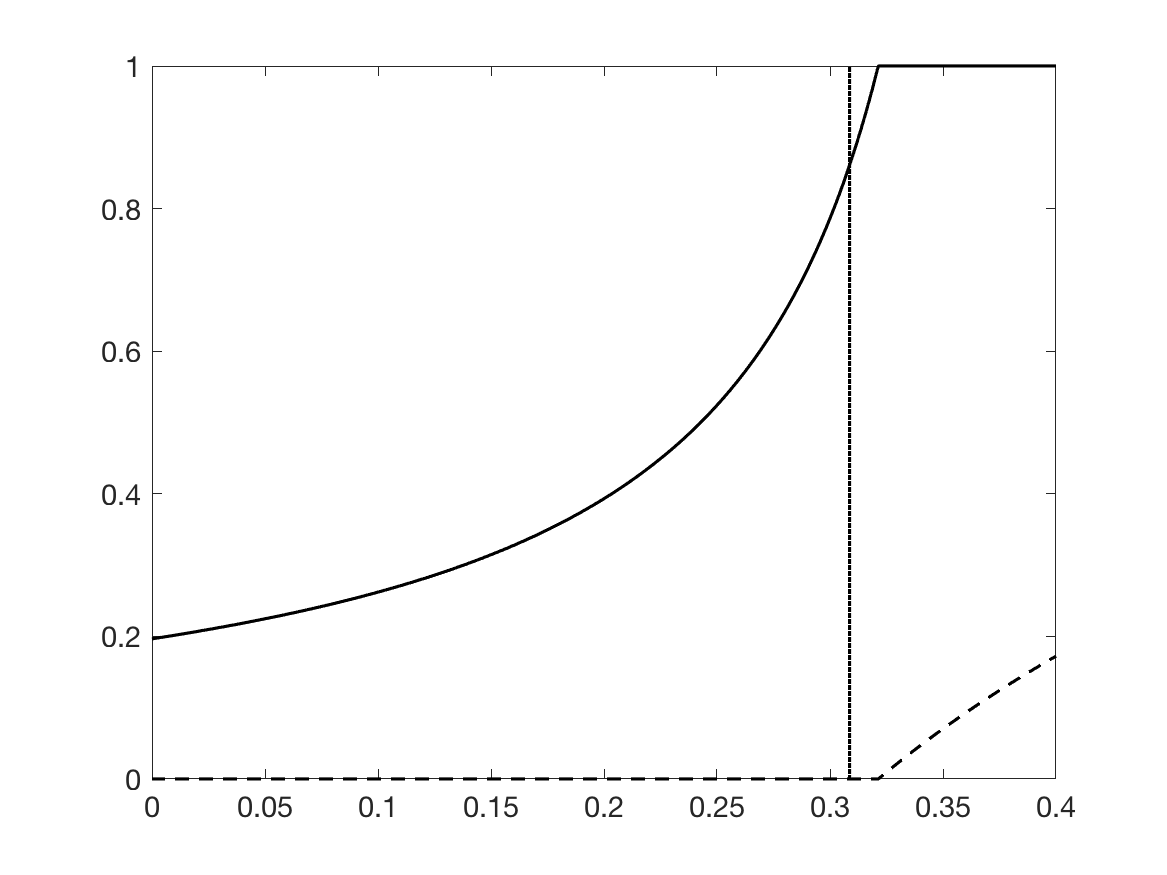}} \hfill
\subfigure[]{\includegraphics[width=5.05cm]{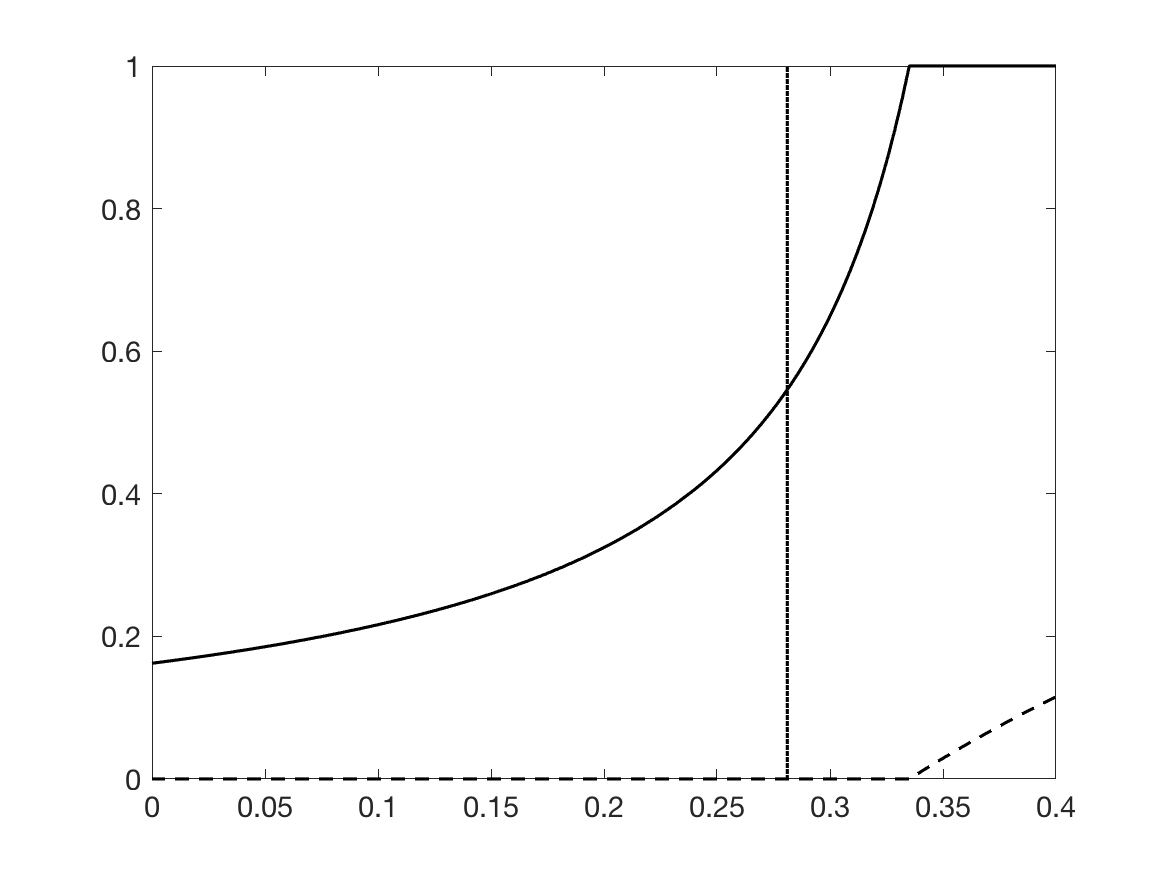}} 
\caption{Results from the toy example. Figures (a), (b) and (c) present $E_{k:4}^*(t_k)$ for $k = 2, 3$ and 4, respectively, with the breakpoints highlighted as black dots. 
Figures (d), (e) and (f) present $q_{k}^{*00}(t_k)$ (solid) and $q_{k}^{*01}(t_k)$ (dashed) for  for $k = 2, 3$ and 4, respectively. Figures (g), (h) and (i) present $q_{k}^{*10}(t_k)$ (solid) and $q_{k}^{*11}(t_k)$ (dashed) for $k = 2, 3$ and 4, respectively.  The vertical dotted line in each figure (a)-(i) represents the corresponding optimal parameter value $t_k^*$. }
\label{fig:toy example}
\end{figure}

Suppose the assumed prior $f(x)$ is a Markov chain of length $n = 4$ with homogenous transition probabilities $f(x_k = 0 | x_{k-1} = 0) = 0.7$ and $f(x_k = 1 | x_{k-1} = 1) = 0.8$ for $k\geq2$, and initial distribution $f(x_1)$ equal to the associated limiting distribution. Thus, the Markov chain $f(x)$  is stationary with marginal probabilities $f(x_k = 0) = 0.40$ for each $k=1,2,3,4$.  Further, suppose every factor $f(y_i|x_i)$ of the likelihood model $f(y|x)$ is a Gaussian distribution with mean $x_i$ and standard deviation $\sigma = 2$, and consider the observation vector $y = (-0.681, -1.585, 0.007, 3.103)$. The corresponding posterior Markov chain model $f(x|y)$ then have the transition probabilities 
\begin{equation}
\begin{aligned}
 f(x_2 = 0 | x_1 = 0, y) = 0.7821,  \quad f(x_2 = 1 | x_1 = 1, y) = 0.7223, \\
  f(x_3 = 0 | x_2 = 0, y) = 0.6600,  \quad f(x_3 = 1 | x_2 = 1, y) = 0.8278, \\
 f(x_4 = 0 | x_3 = 0, y) = 0.5490,  \quad f(x_4 = 1 | x_3 = 1, y) = 0.8846, 
 \label{eq: toy example posterior transition}
 \end{aligned}
\end{equation}
and marginal distributions 
\begin{equation}
\begin{aligned}
f(x_1 = 0 | y) = 0.526779, \\
f(x_2 = 0 | y) = 0.543379, \\
f(x_3 = 0 | y) = 0.437279, \\
f(x_4 = 0 | y) = 0.304977.
\label{eq: toy example posterior marginal}
\end{aligned}
\end{equation}
Based on the prior model $f(x)$ and the posterior model $f(x|y)$, we can construct $q^*(\tilde x | x, y)$ as discussed above. For this simple example, this involves computing fourteen quantities, namely  $q_1^{*0}(t_1^*) = q^*(\tilde x_1 = 0 | x_1 = 0, y)$, $q_1^{*1}(t_1^*) = q^*(\tilde x_1 = 0 | x_1 = 1, y)$, $q_k^{*ij}(t_k^*) = q^*(\tilde x_k = 0 | \tilde x_{k-1} = i, x_k=j, y)$, for $k = 2, 3, 4$, and $i,j = 0,1$.  
As described in the previous section, the construction of $q(\tilde x | x, y)$ involves a backward recursion and a forward recursion. In the backward recursion, we compute  $E_{k:n}^*(t_k)$ and $q_k^{*00}(t_k) $, for $k = 4, 3, 2$. The results for these quantities are presented in Figure \ref{fig:toy example}. 
In the forward recursion, we  start out computing the optimal solution of the first factor, $q^*(\tilde x_1  | x_1 , y)$, and then compute  the remaining optimal parameter values $t_2^*$, $t_3^*$ and $t_4^*$ and corresponding optimal solutions $q_k^{*ij}(t_k^*)$, $k = 2, 3, 4$, $i,j = 0,1$. The results from the forward recursion are given in Table \ref{tab: toy example}. 
\begin{table}
\centering
\subtable[]
{\begin{tabular}{ c | c }
$k$ & 1 \\
\hline 
$t_k^*$ & 0.400000 \\
$q_{k}^{*0}(t_k^*)$ & 1.000000 \\
$q_{k}^{*1}(t_k^*)$ & 0.211299 \\
\end{tabular}}
\hspace{1.5cm}
\subtable[ ]
{\begin{tabular}{c |  c c c}
\ $k$ & 2 & 3 & 4 \\
\hline 
$t_k^*$ & 0.305356 & 0.308676 & 0.281108 \\
$q_k^{*00}(t_k^*)$ & 1.000000 & 1.000000 & 0.853968 \\
$q_k^{*01}(t_k^*)$ & 0.481489 & 0.212926 & 0.000000 \\
$q_k^{*10}(t_k^*)$ & 1.000000 & 0.860986 & 0.546043 \\
$q_k^{*11}(t_k^*)$ &  0.097118 & 0.000000 & 0.000000
\end{tabular}}
\caption{Results for the optimal solution $q^*(\tilde x | x, y)$ of the toy example, in (a) for the first factor $q^*(\tilde x_1 | x_1, y)$, and in (b) for the remaining factors $q^*(\tilde x_k | x_k, \tilde x_{k-1}, y)$. }
\label{tab: toy example}
\end{table}

Taking a closer look at the results for the optimal solution $q^*(\tilde x | x, y)$, we see that many of the probabilities $q_k^{*ij}(t_k^*)$ are either zero or one. 
This feature can be formally explained mathematically (see Appendix A), but is also quite an intuitive result, and has to do with how the probabilities of the prior model $f(x)$ differ from the corresponding probabilities of the posterior model $f(x | y)$. 
Often, if $f(x_k = 0 ) < f(x_k = 0|y)$, we obtain $q_k^{*00}(t_k^*) = 1$ and $q_k^{*10}(t_k^*) = 1$, while $q_k^{*01}(t_k^*)$ and $q_k^{*11}(t_k^*)$ take values somewhere between zero and one. Thus, if we have a prior sample $x$ with $x_k = 0$, the update of $x$ to $\tilde x$ is always such that $\tilde x_k = 0$. 
Specifically, in our toy example, this is the case for $k = 2$;  that is, we have $f( x_2 = 0 ) < f( x_2 = 0 | y)$, and obtained $q_2^{*00}(t_2^*) = 1$ and $q_2^{*10}(t_2^*) = 1$. 
Likewise, if $f(x_k = 0 ) > f(x_k = 0|y)$, we often obtain $q_k^{*01}(t_k^*) = 0$ and $q_k^{*11}(t_k^*) = 0$, while $q_k^{*00}(t_k^*)$ and $q_k^{*10}(t_k^*)$ take values somewhere between zero and one. Thus, if we have a prior sample $x$ with $x_k = 1$, the update of $x$ to $\tilde x$ is always such that $\tilde x_k = 1$. 
In our toy example, this is the case for $k = 4$; that is, we have $f( x_4 = 0 ) > f( x_4 = 0 | y)$, and obtained $q_4^{*01}(t_4^*) = 0$ and $q_4^{*10}(t_4^*) = 0$. 
However, the model $q(\tilde x | x, y)$ is not only constructed so that the marginal probabilities in \eqref{eq: toy example posterior marginal} are fulfilled, but also so that the posterior transition probabilities in \eqref{eq: toy example posterior transition} are reproduced. 
In our toy example, we see for example that for $k=3$ we obtained $q_3^{*10}(t_3^*) <1$ even if $f( x_3 = 0 ) < f( x_3 = 0 | y)$. Instead, we observe another deterministic term, namely $q_3^{*11}(t_3^*) = 0$.

\subsection{Ensemble-based, higher-dimensional example with simulated data}
\label{sec:advanced example}

Until now, we have focused on the ensemble updating problem at a specific time step of the filtering recursions. However, in a practical application, one is interested in the filtering problem as a whole and needs to cope with the ensemble updating problem sequentially for $t = 1, 2, \dots, T$.
We now address this issue and investigate the application of our proposed approach in this context. 
That is, we reconsider the situation with an unobserved Markov process, $\{x^t\}_{t=1}^T$, and a corresponding time series of observations, $\{y^t\}_{t=1}^{T}$, and at every time step $t = 1, \dots, T$, we construct a distribution $q(\tilde x^t | x^t, y^{1:t})$ in order to update the prior ensemble  $\mathcal X^t =  \{x^{t(1)},x^{t(2)}, \dots, x^{t(M)} \}$ to a posterior ensemble $\widetilde {\mathcal X}^t = \{\tilde x^{t(1)}, \tilde x^{t(2)}, \dots, \tilde x^{t(M)} \}$. 
Below, we first present the set-up of our simulation example in Section \ref{sec:set-up}, and thereafter study the performance of our approach in Sections \ref{sec:numerical marginal} and \ref{sec:numerical joint}.

\subsubsection{Specification of simulation example}
\label{sec:set-up}
To construct a simulation example we must first define the $\{ x^t\}_{t=1}^T$  Markov chain. We set $T=100$ and let
$x^t=(x_1^t,\ldots,x_n^t)$ be an $n=400$ dimensional vector of binary variables $x_i^t\in\{0,1\}$ for each
$t=1,\ldots,T$. To simplify the specification of the transition probabilities $p(x^t|x^{t-1})$ 
we make two Markov assumptions. First, conditioned on $x^{t-1}$ we assume the elements in $x^t$
to be a Markov chain so that
\begin{equation*}
  p(x^t|x^{t-1}) = p(x_1^t|x^{t-1}) \prod_{i=2}^n p(x_i^t|x_{i-1}^t,x^{t-1}).
\end{equation*}
The second Markov assumption we make is that
\begin{equation*}
  p(x_i^t|x_{i-1}^t,x^{t-1}) = p(x_i^t|x_{i-1}^t,x_{i-1}^{t-1},x_i^{t-1},x_{i+1}^{t-1}),
\end{equation*}
for $i=2,\ldots,n-1$, i.e. the value in element $i$ at time $t$ only depends on the values in elements $i-1$, $i$ and $i+1$ at
the previous time step. For $i=1$ and $i=n$ we make the corresponding Markov assumptions
\begin{equation*}
  p(x_1^t|x_1^{t-1},x_2^{t-1}) \mbox{~~~and~~~}
  p(x_n^t|x_{n-1}^t,x_{n-1}^{t-1},x_n^{t-1}).
\end{equation*}
To specify the $x^t$ Markov process we thereby need to specify $p(x_i^t|x_{i-1}^t,x_{i-1}^{t-1},x_i^{t-1},x_{i+1}^{t-1})$
for $t=2,\ldots,T$ and $i=2,\ldots,n$ and the corresponding probabilities for $t=1$ and for $i=1$ and $n$.

To get a reasonable test for how our proposed ensemble updating procedure works we want an $\{x^t\}_{t=1}^{T}$ process
with a quite strong dependence between $x^{t-1}$ and $x^t$, also when conditioning on observed data. Moreover,
conditioned on $y^{1:t}$, the elements in $x^t$ should not be first order Markov so that the true model differ from the
{\em assumed} Markov model defined in Sections 3.1 and 3.3. In the following we first discuss the choice of
$p(x_i^t|x_{i-1}^t,x_{i-1}^{t-1},x_i^{t-1},x_{i+1}^{t-1})$ for $t=2,\ldots,T$ and $i=2,\ldots,n$ and thereafter
specify how these are modified for $t=1$ and for $i=1$ and $n$. When specifying the probabilities we are inspired by 
the process of how water comes through to an oil producing well in a petroleum reservoir, but without claiming
our model to be a very realistic model for this situation. Thereby $t$ represents time and $i$ is
location in the well. We let $x_i^t=0$ represent the presence of oil at location or node $i$ at time $t$ and
correspondingly $x_i^t=1$ represents the presence of water. In the start we assume oil is present in the whole well,
but as time goes by more and more water is present and at time $t=T$ water has become the dominating fluid in the well.
Whenever $x_i^{t-1}=1$ we therefore want $x_i^t=1$ with very high probability, especially if also $x_{i-1}^t=1$.
If $x_i^{t-1}=0$ we correspondingly want a high probability for $x_i^t=0$ unless $x_{i-1}^t=1$ and $x_{i-1}^{t-1}=x_{i+1}^{t-1}=1$.
Trying different sets of parameter values according to these rules we found that the values
specified in Table \ref{tab:MM} gave realisations consistent with the requirements discussed above.
\begin{table}
  \begin{center}
  \caption{\label{tab:MM}Probabilities defining the true model $p(x^t|x^{t-1})$ used to simulate a true chain $\{x^t\}_{t=1}^T$.}
  \begin{tabular}{ccc|c|c}
    $x_{i-1}^{t-1}$ & $x_i^{t-1}$ & $x_{i+1}^{t-1}$ & $P(x_i^t=1|x_{i-1}^t=1,x_{i-1:i+1}^{t-1})$ & $P(x_i^t=1|x_{i-1}^t=0,x_{i-1:i+1}^{t-1})$ \\ \hline
    0 & 0 & 0 & 0.0100 & 0.0050 \\
    1 & 0 & 0 & 0.0400 & 0.0100 \\
    0 & 1 & 0 & 0.9999 & 0.9800 \\
    1 & 1 & 0 & 0.9999 & 0.9900 \\
    0 & 0 & 1 & 0.0400 & 0.0400 \\
    1 & 0 & 1 & 0.9800 & 0.0400 \\
    0 & 1 & 1 & 0.9999 & 0.9800 \\
    1 & 1 & 1 & 0.9999 & 0.9800 
  \end{tabular}
  \end{center}
\end{table}
One realisation from this model is shown in Figure \ref{fig:Example 2, Figure 1}(a),
\begin{figure}
\hspace{-1cm}
\begin{tikzpicture}

	\node[inner sep=0pt,label=below:\footnotesize (a) $\{x^t\}_{t=1}^{100}$] (russell) at (1,0) {\includegraphics[width=.30\textwidth]{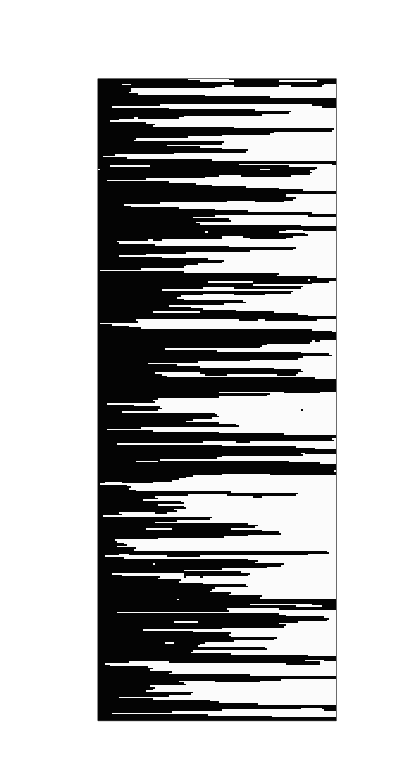}};
	\node[inner sep=0pt,label=below:\footnotesize (b) $\left  \{ \hat p_c(x^t_i | y^{1:t}) \right \}_{t=1}^{100}$] (russell2) at (5,0) {\includegraphics[width=.30\textwidth]{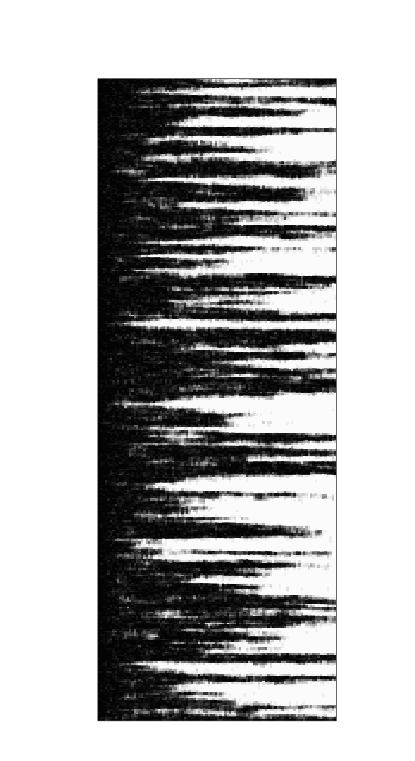}};
	\node[inner sep=0pt,label=below:\footnotesize (c) $\left  \{ \hat p_q(x^t_i | y^{1:t}) \right \}_{t=1}^{100}$] (russell3) at (9,0) {\includegraphics[width=.30\textwidth]{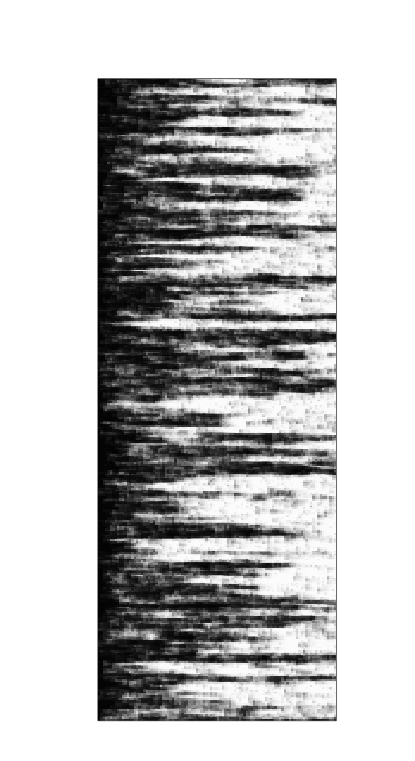}};
	\node[inner sep=0pt,label=below:\footnotesize (d) $\left  \{ \hat p_a(x^t_i | y^{1:t}) \right \}_{t=1}^{100}$] (russell4) at (13,0) {\includegraphics[width=.30\textwidth]{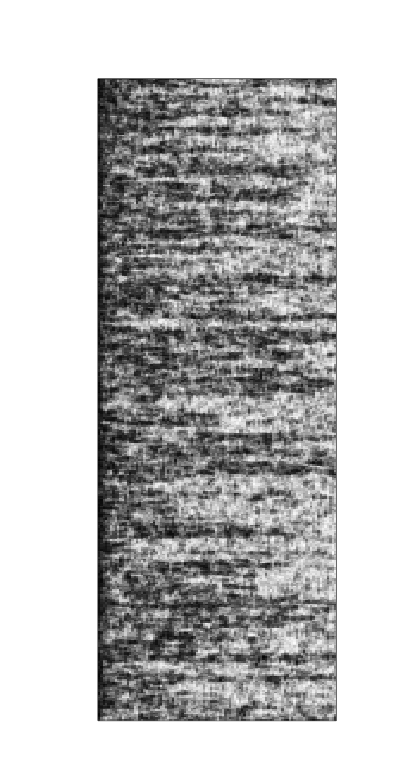}};
	
\draw[->] (-1.5,4) -- (-1.5,3);
\draw[->] (-1.5,4) -- (-0.5,4);
%\draw[->] (russell) -- (russell2);
\node (t) at (-1, 4.25) {$t$};
\node (t) at (-1.75, 3.55) {$i$};

\end{tikzpicture}
\caption{Results from the numerical experiment of Section \ref{sec:advanced example}.  Grayscale images of the true simulated $x^t$ process in (a), and estimates of the marginal probabilities $p(x^t_i =1 | y^{1:t})$ in (b), (c) and (d), where black and white correspond to the values zero and one, respectively.  }
\label{fig:Example 2, Figure 1}
\end{figure}
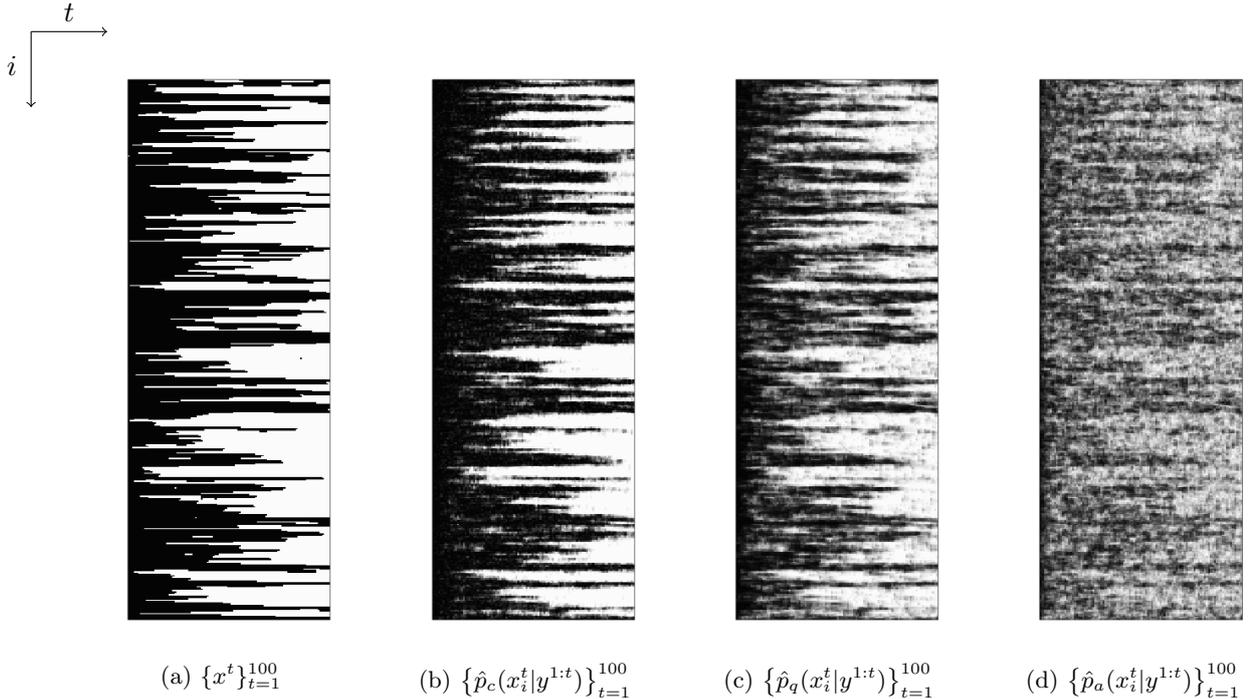
 where black and white represent $0$ (oil) and $1$ (water),
respectively. The corresponding probabilities when $t=1$ and for $i=1$ and $n$ we simply defined from the values in Table
\ref{tab:MM} by defining all values lying outside the $\{(i,t):i=1,\ldots,n;t=1,\ldots,T\}$ lattice to be zero.
In particular this implies that at time $t=0$, which is outside the lattice, oil is present in the whole well.
In the following we consider the realisation shown in Figure \ref{fig:Example 2, Figure 1}(a) to be the (unknown) true $x^t$ process.

The next step in specifying the simulation example is to specify an observational process. For this we simply
assume one scalar observation $y_i^t$ for each node $i$ at each time $t$, and assume the elements
in $y^t=(y_1^t,\ldots,y_n^t)$ to be conditionally independent given $x^t$. 
Furthermore, we let $y_i^t$ be
Gaussian with  mean $x_i^t$ and variance $\sigma^2$. As we  want the dependence between $x^{t-1}$
and $x^t$ to be quite strong also when conditioning on the observations, we need to choose the variance $\sigma^2$
reasonably large, so we set $\sigma^2 = 2^2$. 
Given the true $x^t$ process shown in Figure \ref{fig:Example 2, Figure 1} we simulate $y_i^t$ values from the specified Gaussian distribution, and in the following consider these
values as observations. An image of these observations is not included, since the variance
$\sigma^2$ is so high that such an image is not very informative.

Pretending that the $\{x^t\}_{t=1}^T$ process is unknown and that we only have the observations $\{y^t\}_{t=1}^T$
available, our aim with this simulation study is to evaluate how well our proposed ensemble based filtering procedure
is able to capture the properties of the correct filtering distributions $p(x^t|y^{1:t}),t=1,\ldots,T$. To do
so we first need to evaluate the properties of the correct filtering distribution. It is possible to
get samples from $p(x^t|y^{1:t})$ by simulating from $p(x^{1:t}|y^{1:t})$ with a
Metropolis--Hastings algorithm, but to a very high
computational cost as a separate Metropolis--Hastings run must be performed for each value of $t$. Nevertheless,
we do this to get the optimal solution of the filtering problems to which we can compare the results of
our proposed ensemble based filtering procedure. In our algorithm for simulating from 
$p(x^{1:t}|y^{1:t})$ we combine single site Gibbs updates of each element in $x^{1:t}$ with a
one-block Metropolis--Hastings update of all elements in $x^{1:t}$. To get a reasonable acceptance rate for the one-block
proposals we adopt the approximation procedure introduced in \citet{art158} to obtain a partially ordered
Markov model \citep{art119} approximation to $p(x^{1:t}|y^{1:t})$, propose potential new values for $x^{1:t}$
from this approximate posterior, and accept or reject the proposed values according to the usual Metropolis--Hastings acceptance probability.
For each value of $t$ we run the Metropolis--Hastings algorithm for a large number of iterations and discard a burn-in period.
From the generated realisations we can then estimate the properties of $p(x^t|y^{1:t})$. In particular we can estimate the
marginal probabilities $p(x_i^t=1|y^{1:t})$ as the fraction of realisations in which the simulated $x_i^t=1$. We denote these
estimates of the correct filtering probabilities by $\widehat{p}_c(x_i^t=1|y^{1:t})$. In Figure \ref{fig:Example 2, Figure 1}(b) all these estimates are
visualised as a grayscale image, where black and white correspond to $\widehat{p}_c(x_i^t=1|y^{1:t})$ equal to zero and one,
respectively. It is important to note that Figure \ref{fig:Example 2, Figure 1}(b) is not showing the solution of the smoothing problem, but the solution
of many filtering problems put together as one image.

Now, we are ready to run our proposed ensemble-based filtering method. The input to our algorithm include the simulated observations $\{y^t\}_{t=1}^{T}$, the observational model $p(y^t | x^t)$, the initial model $p(x^1)$ for the $\{x^t\}_{t=1}^T$ process, and an ensemble size $M$. Since in real-world situations it is typically necessary  to choose $M$ much smaller than $n$ for computational reasons, we chose to set $M=20$.
A problem, however, when the ensemble size is this small, is that results may vary a lot from run to run.  
To quantify this between-run variability, we therefore 
reran our proposed approach a total of $B=1,000$ times, each time with a new initial ensemble of $M=20$ realisations from the initial model $p(x^1)$. 
At each time step $t$ we thus achieved a total of $M B = 20,000$ posterior samples of the state vector $x^t$, which can be used to construct an estimate, denoted $\hat p_q (x^t | y^{1:t})$, for the true filtering distribution $p(x^t | y^{1:t})$. 

An important step of our approach is the estimation of a first order Markov chain $f(x^t | y^{1:t-1})$ at each time step $t$. Basically, this involves estimating an initial distribution $f(x^t_1 | y^{1:t-1})$ and $n-1$ transition matrices $f(x^t_{i+1} | x^t_{i}, y^{1:t-1})$, $i=1, \dots, n-1$. Since each component $x^t_i$ is a binary variable, the initial distribution $f(x^t_1 | y^{1:t-1})$ can be represented by one parameter, while the transition matrices $f(x^t_{i+1} | x^t_{i}, y^{1:t-1})$ each require two parameters. In this example, we pursued a Bayesian approach for estimating the parameters. Specifically, if we let $\theta^t$ represent a vector containing all the parameters required  to specify the model $f(x^t | y^{1:t-1})$, we 
put a prior on $\theta^t$, $f(\theta^t)$, and then set the final estimator for $ \theta^t$ equal to the mean of the corresponding posterior distribution $f(\theta^t | \mathcal X^t)$. In the specification of $f(\theta^t)$ we assumed that all the parameters in the vector $\theta^t$ are independent and that each parameter follows a Beta distribution $\mathcal B (\alpha, \beta)$ with parameters $\alpha = 2, \beta = 2$. 

As well as studying the performance of our proposed approach, we are in this experiment interested in studying the results one would get without constructing $q(\tilde x^t | x^t, y^{1:t})$ at each time step $t$, instead updating the prior ensemble by simulating independent samples from the assumed Markov chain model $f(x^t | y^{1:t})$. Below, we refer to  this procedure as the \emph{assumed model approach}. As with our approach, we reran the assumed model approach  $B = 1,000$ times, yielding a total of $M B = 20,000$ posterior samples of each state vector $x^t$, $t=1, \dots, T$. These samples can then be used to construct an estimate, denoted $\hat p_a (x^t | y^{1:t})$, for the true filtering distribution $p(x^t | y^{1:t})$. 
By comparing $\hat p_a(x^t | y^{1:t})$ and $\hat p_q(x^t | y^{1:t})$ with $\hat p_c (x^t | y^{1:t})$, which essentially represents the true model $p(x^t|y^{1:t})$, we can get an understanding of how much we gain by doing our approach instead of the much simpler assumed model approach. 
In the next two sections we investigate how well $\hat p_q (x^t | y^{1:t})$ and $\hat p_a (x^t | y^{1:t})$ capture marginal and joint properties of the true distribution $ p (x^t | y^{1:t})$, for which the  MCMC estimate $\hat p_c(x^t | y^{1:t})$ works as a reference.

\subsubsection{Evaluation of marginal distributions}

\label{sec:numerical marginal}

In this section, we are interested in studying how well our approach estimates the marginal filtering distributions $p (x^t_i | y^{1:t} )$.  Following the notations introduced above, we let 
 $\hat p_q(x^t_i | y^{1:t})$ and $\hat p_a(x^t_i | y^{1:t})$ denote estimates of the marginal distribution $p(x^t_i | y^{1:t}) $ obtained with our approach and the assumed model approach, respectively. 
The values of $\hat p_q(x^t_i =1 | y^{1:t})$ and $\hat p_a(x^t_i = 1| y^{1:t})$ are in each case set equal to the mean of the corresponding set of samples of $x_i^t$. Figures \ref{fig:Example 2, Figure 1}(c) and (d)  present grayscale images of  $\left  \{ \hat p_q(x^t_i =1 | y^{1:t}) \right \}_{t=1}^{100} $ and $\left  \{ \hat p_a(x^t_i = 1| y^{1:t}) \right \}_{t=1}^{100}$, respectively. From a visual inspection, the image of $\left  \{ \hat p_q(x^t_i = 1| y^{1:t}) \right \}_{t=1}^{100}$ is more grey and noisy than that of $\left  \{ \hat p_c(x^t_i =1 | y^{1:t}) \right \}_{t=1}^{100}$, which contains more tones closer to pure black and white. This is to be expected, since $\left  \{ \hat p_c(x^t_i | y^{1:t}) \right \}_{t=1}^{100}$ essentially is the ideal solution, and we do not expect our approach to perform this well. However, the image of $\left  \{ \hat p_a(x^t_i = 1 | y^{1:t}) \right \}_{t=1}^{100}$ is even more grey and noisy than $\left  \{ \hat p_q(x^t_i =1 | y^{1:t}) \right \}_{t=1}^{100} $, so presumably we do  gain something by running our approach instead of the simpler assumed model approach. 
To investigate this further,  we compute the  Frobenius norms   of the two matrices  produced by subtracting the true marginal probabilities $ \hat p_c(x^t_i = 1| y^{1:t}) $ from the corresponding estimates $\hat p_q(x^t_i = 1| y^{1:t}) $ and $ \hat p_a(x^t_i = 1| y^{1:t}) $. 
We then obtain the numbers $35.38$ and $63.00$, respectively. That is, the Frobenius norm of the difference between the true and the estimated marginal filtering distributions is reduced to almost the half with our method compared to the assumed model approach, clearly indicating that our method overall provides much better estimates of the marginal distributions $p(x_i^t | y^{1:t})$.

\begin{figure}
\centering
\hspace{-0.95cm}
\subfigure[ ]{\includegraphics[width=4.4cm]{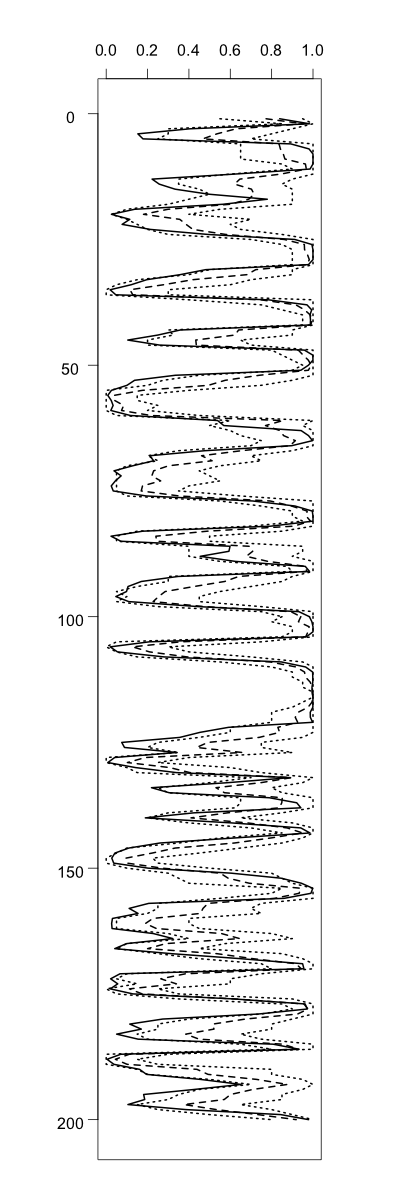}} 
\hspace{-0.95cm}
\subfigure[]{\includegraphics[width=4.4cm]{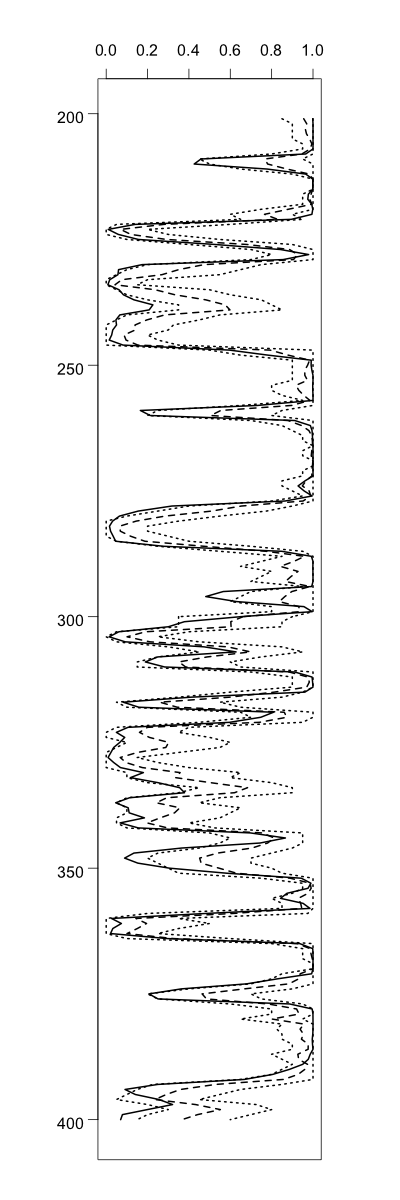}} 
\hspace{-0.95cm}
\subfigure[]{\includegraphics[width=4.4cm]{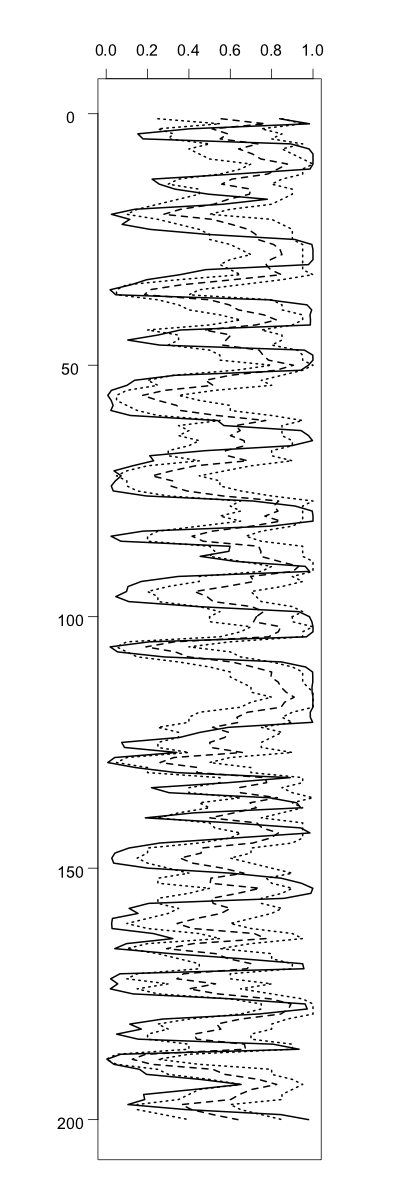}} 
\hspace{-0.95cm}
\subfigure[]{\includegraphics[width=4.4cm]{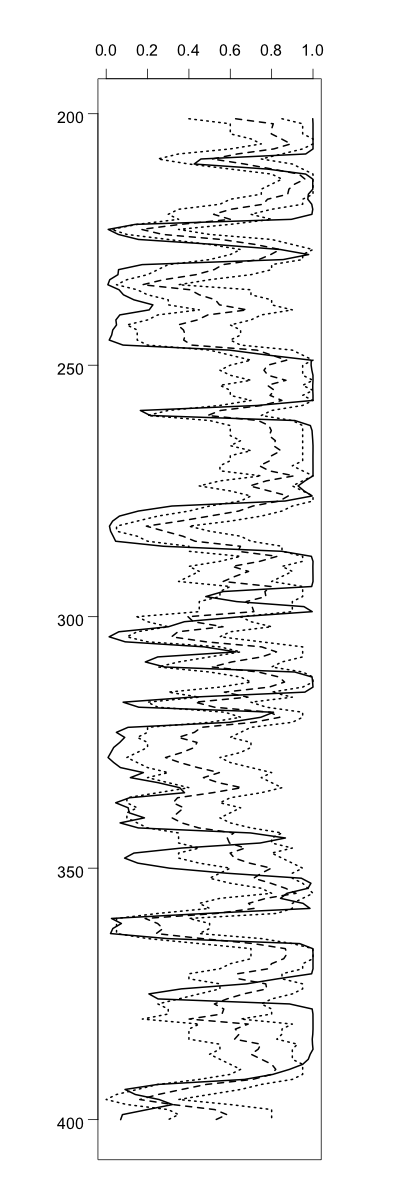}} 
\hspace{-0.95cm}
\caption{Results obtained at time step $t=60$ in the numerical experiment of Section \ref{sec:advanced example}. Figures (a) and (b) present marginal estimates $\hat p_q(x_i^t = 1 | y^{1:t})$ (dashed) and corresponding 90$\%$ quantile intervals (dotted), in (a) from $i=1$ to $i=200$, and in (b) from $i=201$ to $i=400$. Figures (c) and (d) give corresponding results for $\hat p_a(x_i^t = 1 | y^{1:t})$. The solid lines in each plot represent the MCMC estimate $\hat p_c(x^t_i | y^{1:t})$.}
\label{fig:marginals t60}
\end{figure}

To look further into the accuracy of the estimates $\hat p_q(x^t_i =1 | y^{1:t})$ and $\hat p_a(x^t_i =1 | y^{1:t})$ and to study their variability from run to run, 
we take a closer look at the results for some specific time steps. For each of these time steps we compute a 90$\%$ quantile interval for each of the estimates $\hat p_q(x_i^{t} = 1 | y^{1:t})$ and $\hat p_a(x_i^{t} = 1 | y^{1:t})$, $i=1, \dots, 400$. To compute the quantile intervals, recall that  our proposed approach and the assumed model approach were both rerun $B=1,000$ times. From run $b=1, \dots, B$ of our approach, we obtained an estimate $\hat p_q^{(b)}(x_i^t | y^{1:t})$ of $p(x_i^t | y^{1:t})$ for each $i$. Likewise, from run $b=1, \dots, B$ of the assumed model approach, we obtained an estimate $\hat p_a^{(b)}(x_i^t | y^{1:t})$ of $p(x_i^t | y^{1:t})$ for each $i$. 
Hence, for each marginal distribution $p(x_i^t | y^{1:t})$, we have $B=1,000$ estimates $\{\hat p_q^{(b)}(x_i^t | y^{1:t}) \}_{b=1}^B$  obtained with our approach  and $B=1,000$ estimates $\{ \hat p_a^{(b)}(x_i^t | y^{1:t}) \}_{b=1}^B$  obtained with the assumed model approach. From these two sets of samples, corresponding quantile intervals for $\hat p_q(x_i^{t} = 1 | y^{1:t})$ and $\hat p_a(x_i^{t} = 1 | y^{1:t})$ can be constructed. 
Figure \ref{fig:marginals t60} 
presents the computed results for time step $t=60$. For simplicity, we have not included similar figures from any of the other time steps that we studied, since they look very much the same as those obtained at time $t=60$. According to Figures \ref{fig:marginals t60}(a) and (b),  it seems that the essentially true value $\hat p_c(x^{60}_i | y^{1:60})$ typically is within the $90\%$ quantile interval corresponding to $\hat p_q(x_i^{60} | y^{1:60})$, but often closer to one of the interval boundaries rather than the estimate $\hat p_q(x_i^{60} | y^{1:60})$ itself. In particular, we note that 
$\hat p_c(x^{60}_i | y^{1:60})$ often is close to either zero or one, while $\hat p_q(x_i^{60} | y^{1:60})$ is a bit higher than zero or a bit lower than one. 
This is not unreasonable, since 
we have used approximations to construct $\hat p_q(x_i^{60} | y^{1:60})$. Thereby, we loose  information about the true quantity
$\hat p_c(x^{60}_i | y^{1:60})$ and end up with estimated values closer to 0.5. 
From Figures \ref{fig:marginals t60}(c) and (d),  we observe that this is even more the case for the estimate $\hat p_a(x_i^{60} | y^{1:60})$ whose quantile interval often not even covers $\hat p_c( x^{60}_i | y^{1:60})$.

\subsubsection{Evaluation of joint distributions}
\label{sec:numerical joint}

In this section, we want to evaluate
how well our proposed approach manages to capture properties about 
the joint distribution $p(x^t | y^{1:t})$. 
To do so, we select three specific time steps to study, namely $t=60$, $t=70$, and $t=80$. For each of these steps, we perform two  tests on our samples, both concerning a feature we call \emph{contact} between a pair of nodes of $x^t$. 
 So first, we need to explain the concept of contact between a pair of nodes. 
Consider two components $x_i^t$ and $x_j^t$ at a given time step $t$. 
Given that $x_i^t$ is equal to one, we say that there is contact between node $i$ and node $j$ in $x^t$ if all components of $x^t$ between and including node  $i$ and node $j$ are equal to one. 
That is, there is contact between node $i$ and $j$, given that $x_i^t$ is equal to one, if the function
$$
\kappa_{ij}(x^t) = 
\begin{cases} 
1(x_j^t = 1 \cap x_{j+1}^t \cap \ldots \cap x_{i}^t = 1), & \text{ if } j \leq i, \\ 
1(x_i^t = 1 \cap x_{i+1}^t \cap \ldots \cap x_{j}^t = 1), & \text{ if } j > i, 
\end{cases}
$$
is equal to one. 

Keeping $i$ fixed,  we are in our first test interested in studying the probability that there is contact between node $i$ and node $j$ for various values of $j$, given that $x_i^t$ is equal to one. 
Mathematically, that means we are interested in 
\begin{equation}
p^t(i,j) = P \left ( \kappa_{ij}(x^t) = 1 | x_i^t = 1, y^{1:t} \right ).
\label{eq:p(i,j)}
\end{equation}
It is most informative to study \eqref{eq:p(i,j)} for a node $i$ whose corresponding component $x_i^t$ has a high probability of being equal to one. 
Therefore, we concentrate on estimating \eqref{eq:p(i,j)} for three specific choices of $i$, each corresponding to a component $x_i^t$ with a relatively high probability of being equal to one. According to Figure \ref{fig:Example 2, Figure 1} this appears to be the case for the three nodes $i=115$, $i=210$ and $i=290$ at all three time steps $t=60$, $t=70$ and $t=80$. For each $i$ and $t$, we can then use our three sets of samples of $x^t$ to obtain three different estimates of \eqref{eq:p(i,j)} for all $j$. 
 Following previous notations, we let $\hat  p_c^t(i,j)$ denote the MCMC estimate of $p^t(i,j)$, while $\hat  p_q^t(i,j)$ and $\hat  p_a^t(i,j)$ denote the estimates obtained  with our approach and the assumed model approach, respectively. 
 Figure� \ref{fig: Example 2, Figure 3} presents the computed results. 
 Comparing the curves representing the estimates $\hat p_c^t(i,j)$, $\hat p_q^t(i,j)$ and $\hat p_a^t(i,j)$, we observe that $\hat p_q^t(i,j)$ and $\hat p_a^t(i,j)$ typically decrease to zero for increasing values of $j$ quicker than $\hat p_c^t(i,j)$. However, we see that $\hat p_a^t(i,j)$ decreases considerably faster than $\hat p_q^t(i,j)$. 
 This makes sense, since the posterior samples used to construct the estimate $\hat p_a^t(i,j)$ are drawn independently from the assumed model $f(x^t | y^{1:t})$, not taking the state of the prior samples  into account.

\begin{figure}
\centering
\subfigure{\includegraphics[width=4.85cm]{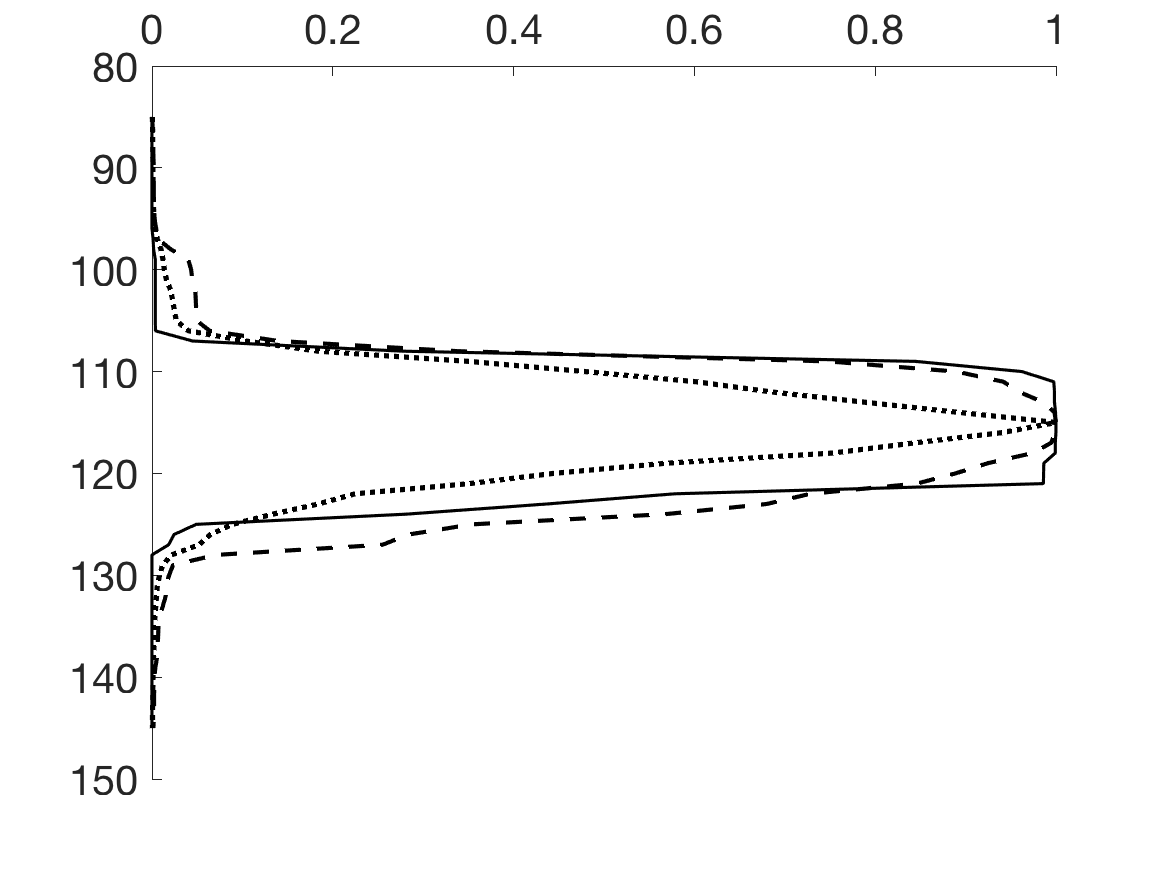}} 
\subfigure{\includegraphics[width=4.85cm]{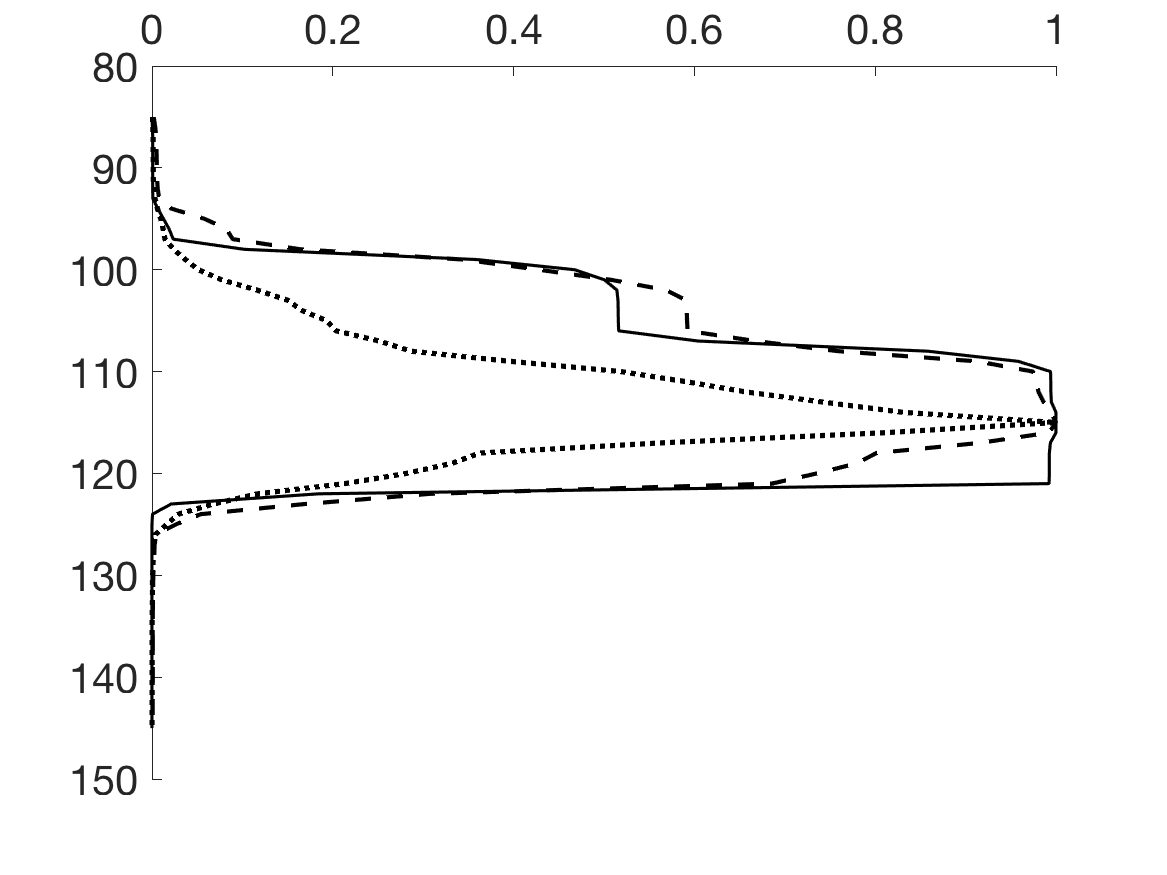}} 
\subfigure{\includegraphics[width=4.85cm]{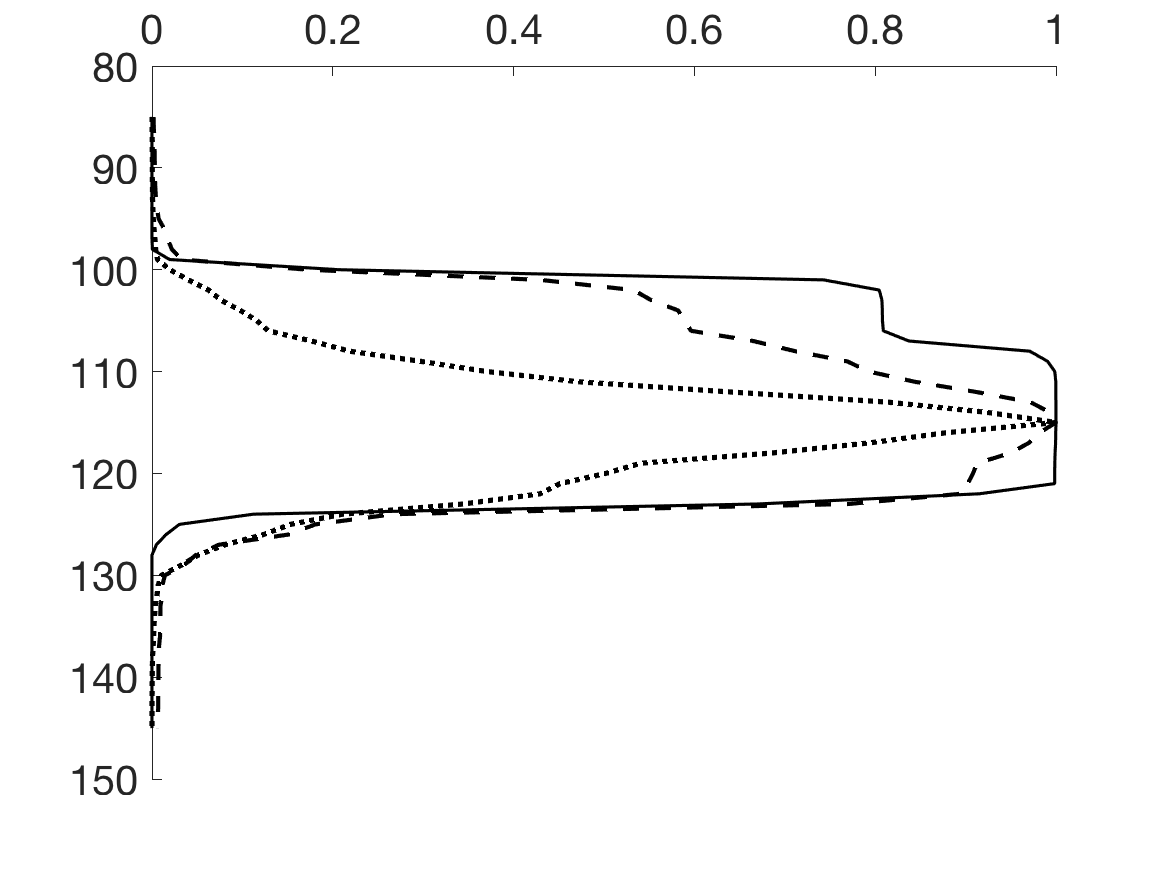}} 
\subfigure{\includegraphics[width=4.85cm]{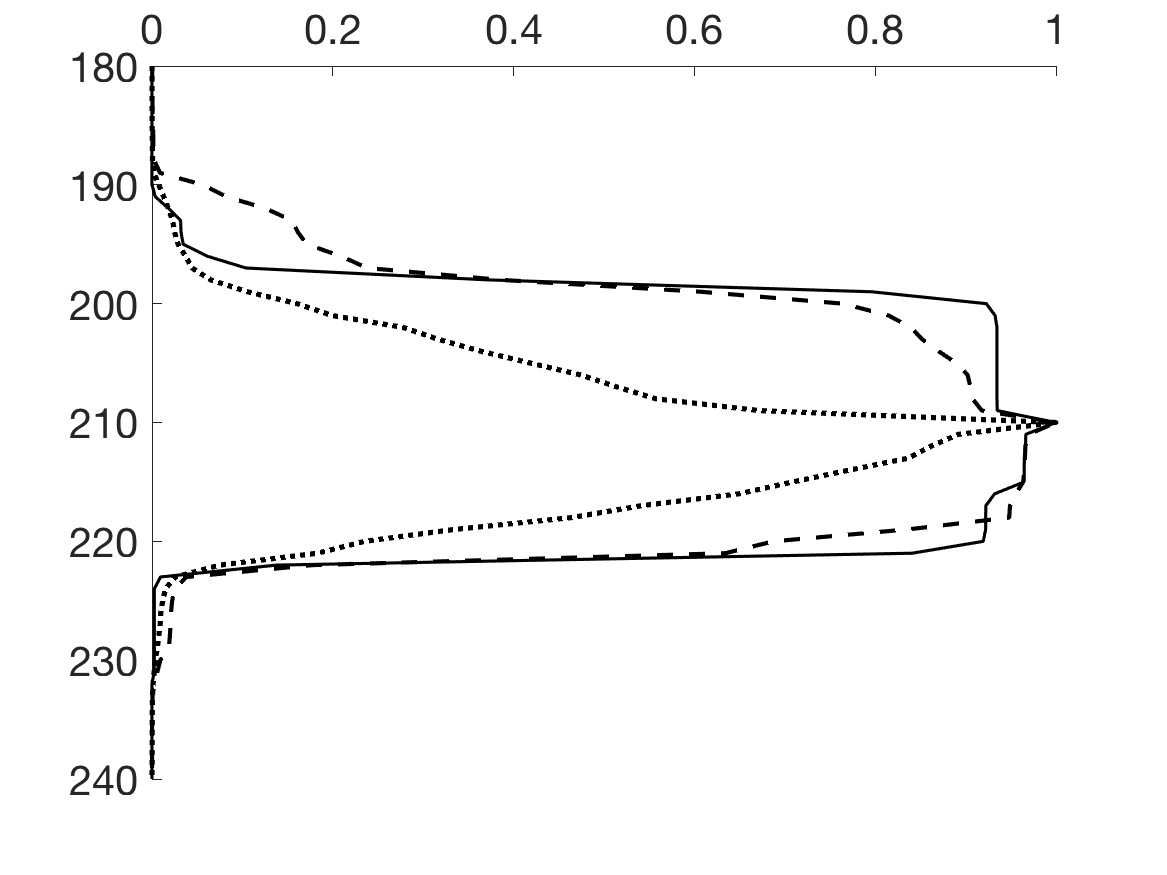}} 
\subfigure{\includegraphics[width=4.85cm]{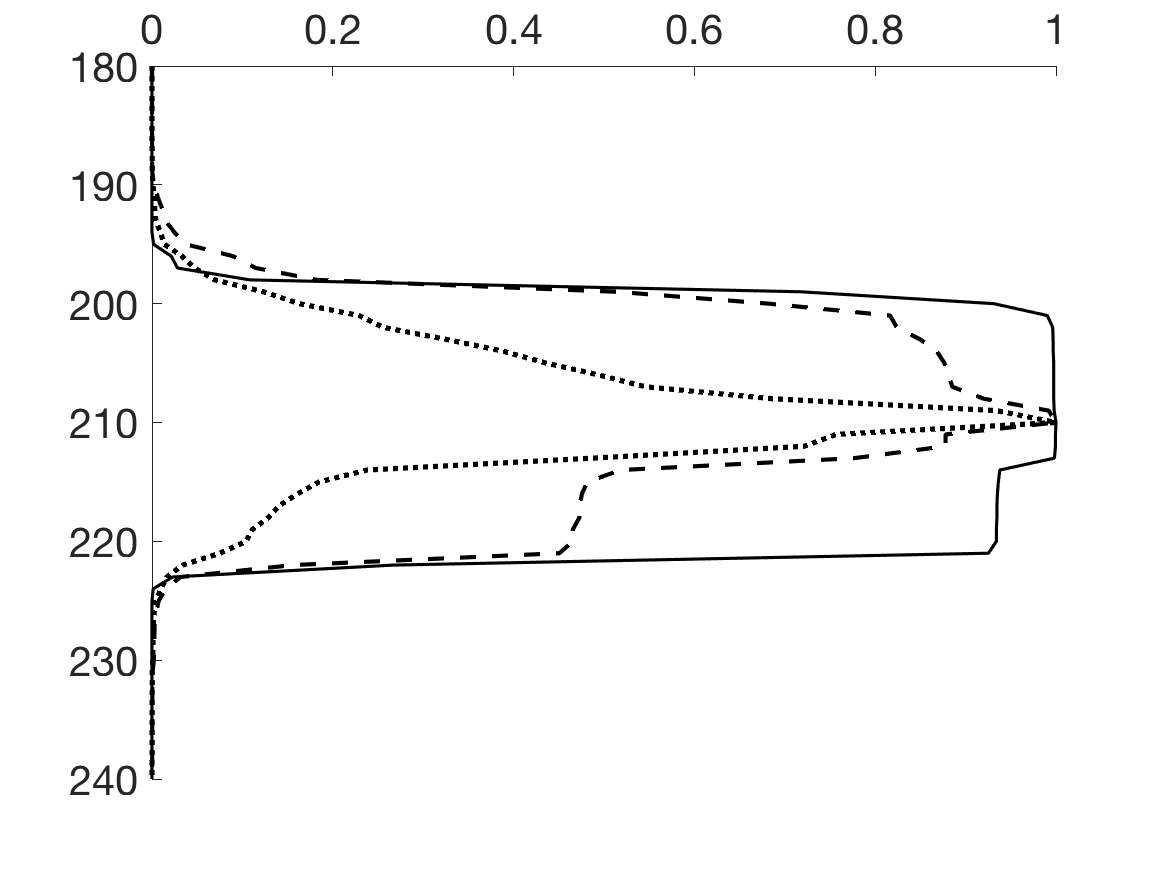}} 
\subfigure{\includegraphics[width=4.85cm]{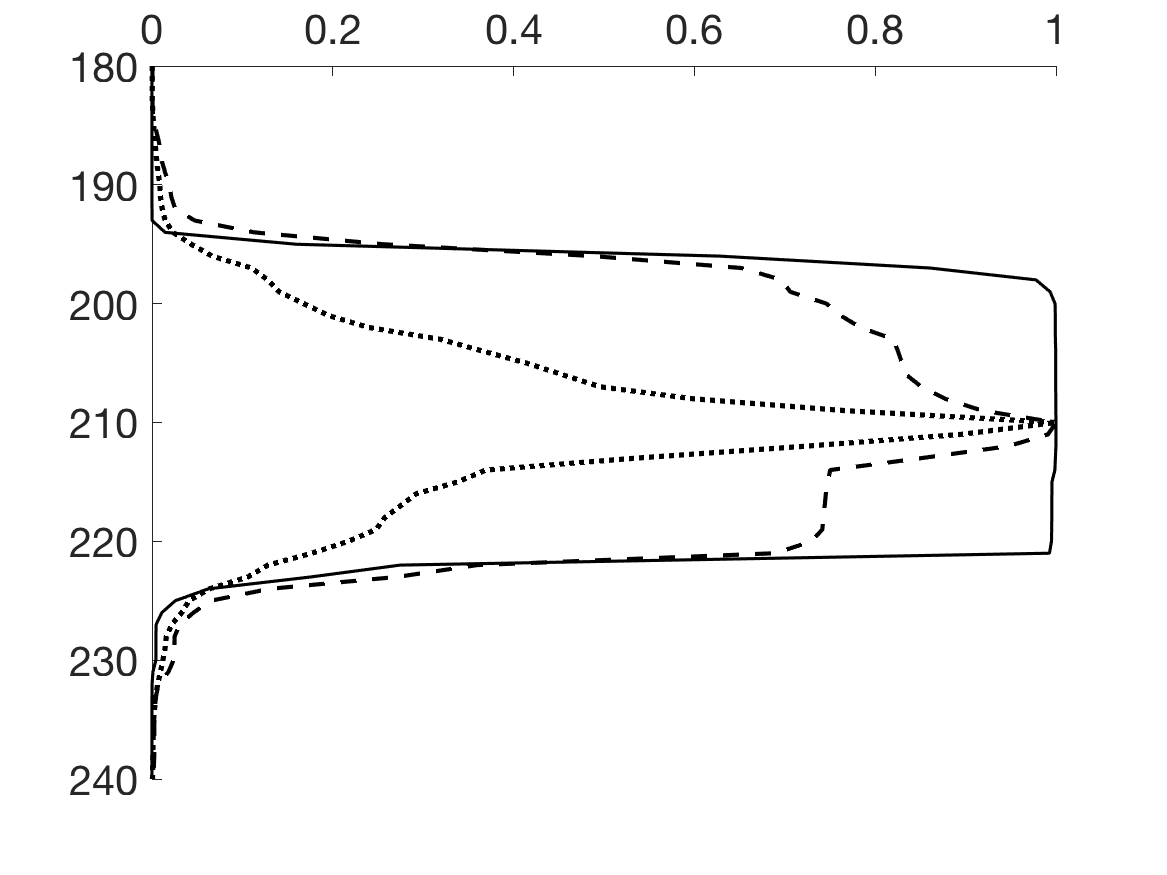}} 
\subfigure{\includegraphics[width=4.85cm]{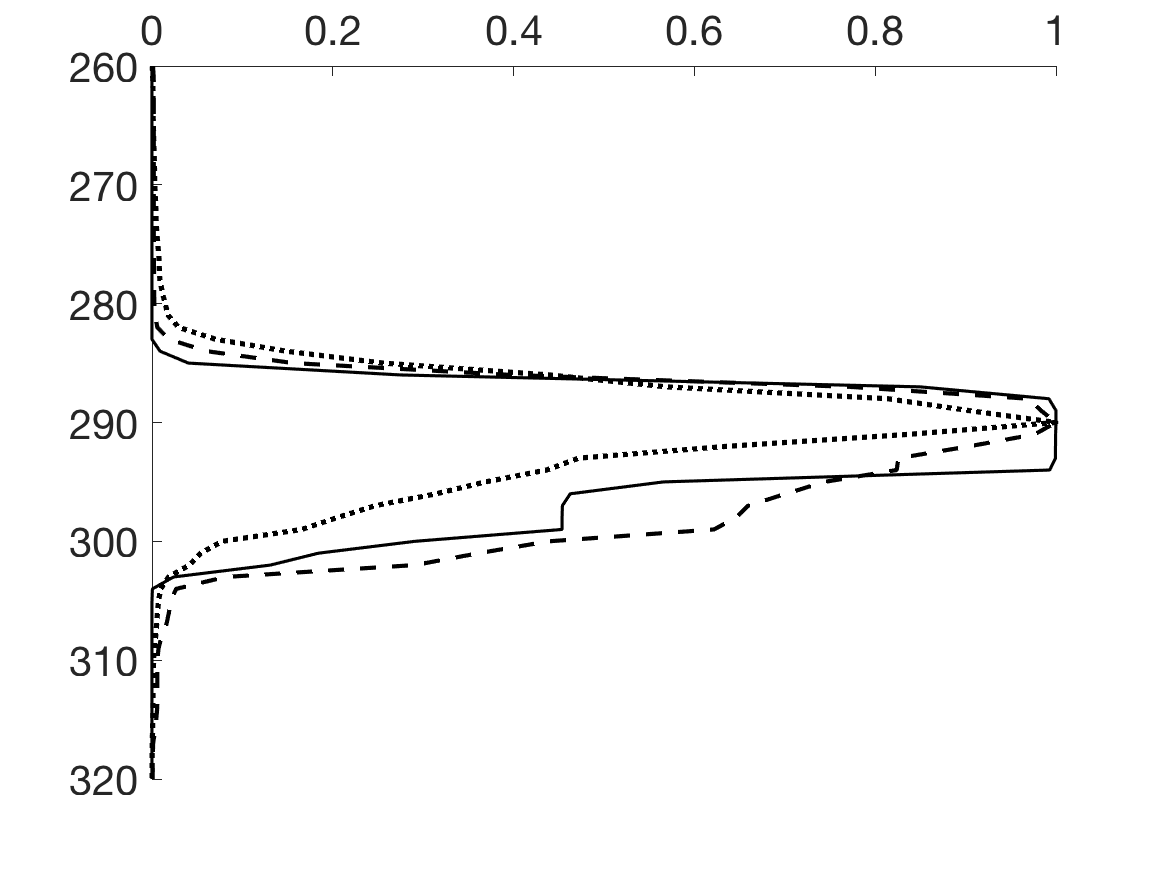}} 
\subfigure{\includegraphics[width=4.85cm]{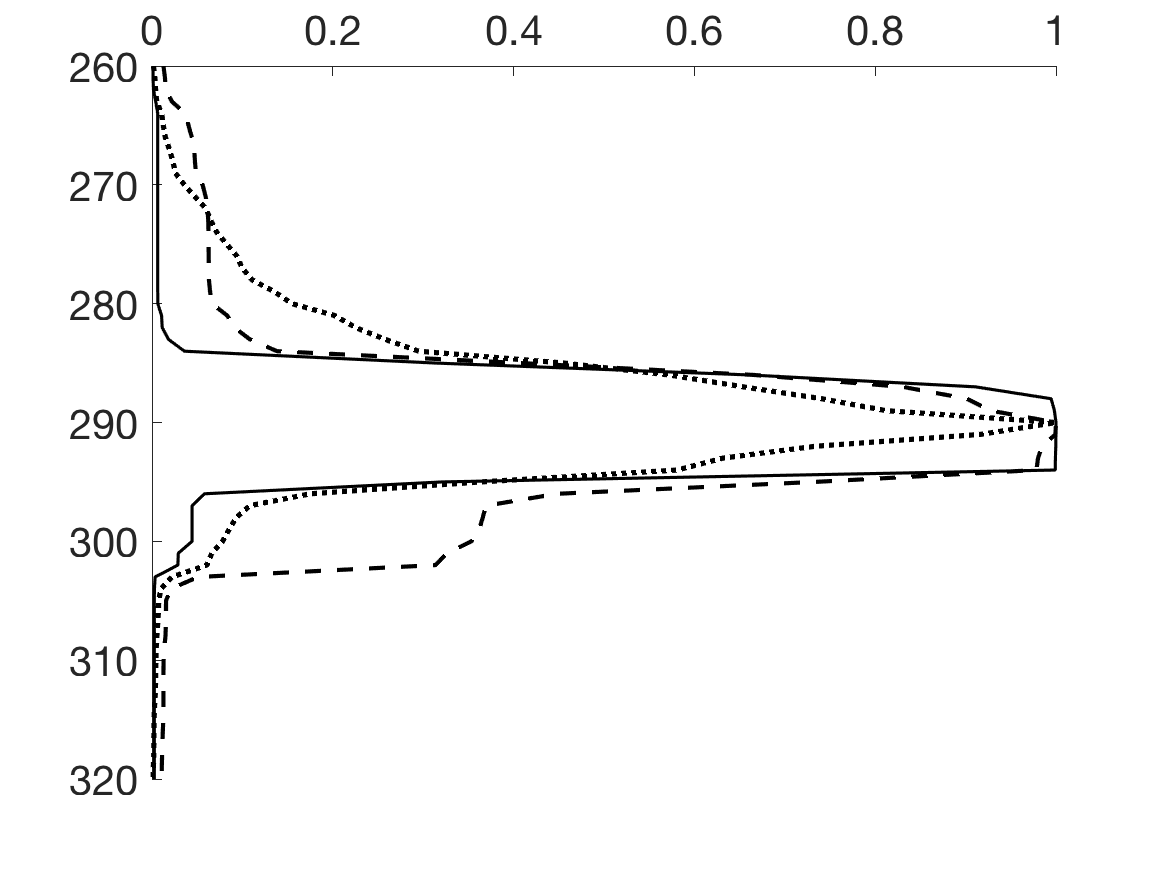}} 
\subfigure{\includegraphics[width=4.85cm]{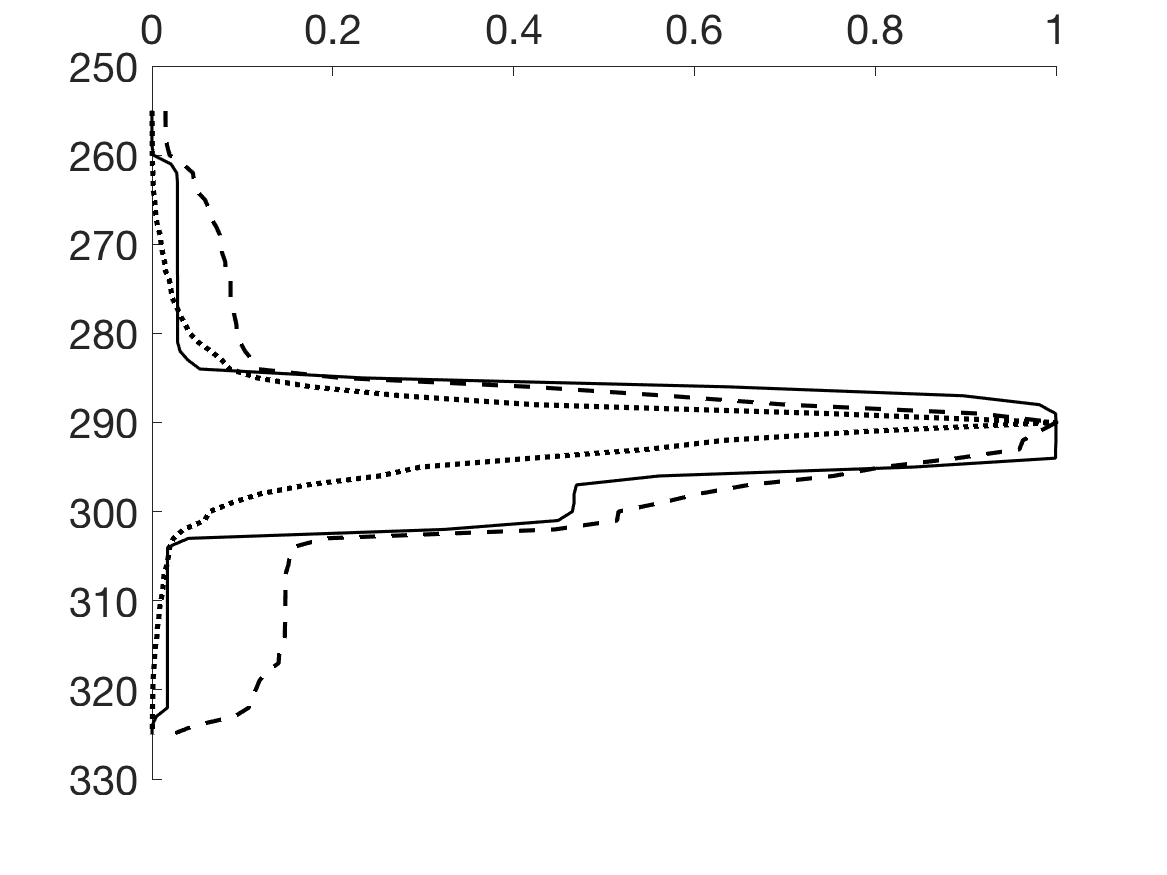}} 
\caption{Results from the numerical experiment of Section \ref{sec:advanced example}. The graphs present $\hat p^t_c(i,j )$ (solid),  $\hat p_q(i,j )$ (dashed), and  $\hat p_a(i,j )$ (dotted) for the three components $i=115$, $i=210$ and $i=290$ at time steps $t=60$ (left), $t=70$ (middle) and $t=80$ (right). }
\label{fig: Example 2, Figure 3}
\end{figure}

In our second test, we focus on the total number of nodes an arbitrary node $i$ with $x_i^t=1$ is in contact with. We denote this quantity by $L_i(x^t)$. Mathematically, $L_i(x^t)$ can be written
$$
L_i(x^t) = \max_{j \geq i} \left  \{ j; \kappa_{ij} (x^t) = 1 \right \} - \min_{j\leq i}  \left \{ j; \kappa_{ij} (x^t) = 1 \right \} + 1.
$$
For each time step $t=60, 70$ and 80, we want to study the cumulative distribution of $L_i(x^t)$, 
\begin{equation}
F ( l ) = P( L_i(x^t) \leq l | x_i^t = 1),
\label{eq:F(L<l)}
\end{equation}
when randomising over both $i$ and $x^t$, with $i \sim \text{unif}\{1, n\}$ and $x^t \sim p(x^t | y^{1:t})$. 
 Again, we can use our three sets of samples to construct three different estimates of 
\eqref{eq:F(L<l)}. That is, we can construct $\hat F_c(\cdot)$ from the MCMC samples, $\hat F_q(\cdot)$ from the samples generated with our approach, and $\hat F_a(\cdot)$ from the samples generated with the assumed model approach.
Figure \ref{fig: Example 2, Figure 4} presents the results.
 \begin{figure}
\centering
\hspace{-1cm}
\subfigure[]{\includegraphics[width=5.15cm]{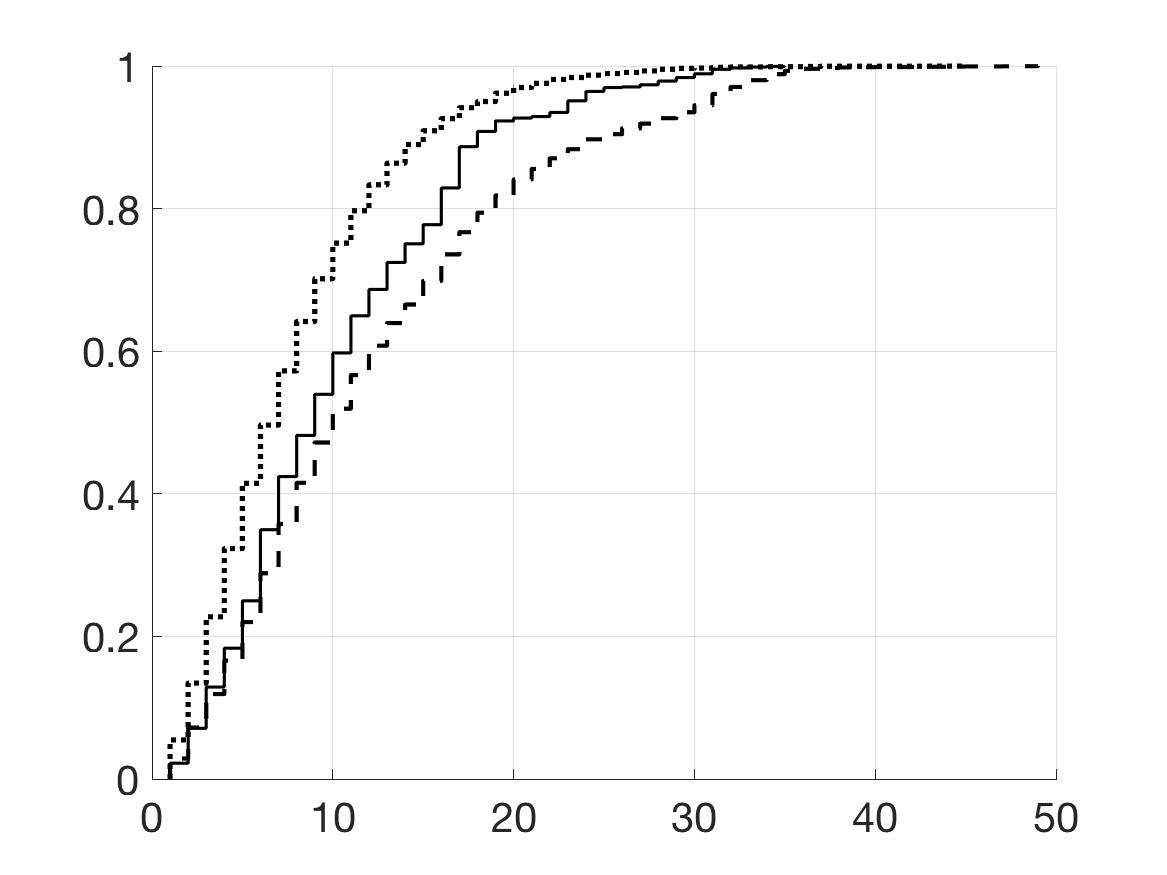}}
\subfigure[]{\includegraphics[width=5.05cm]{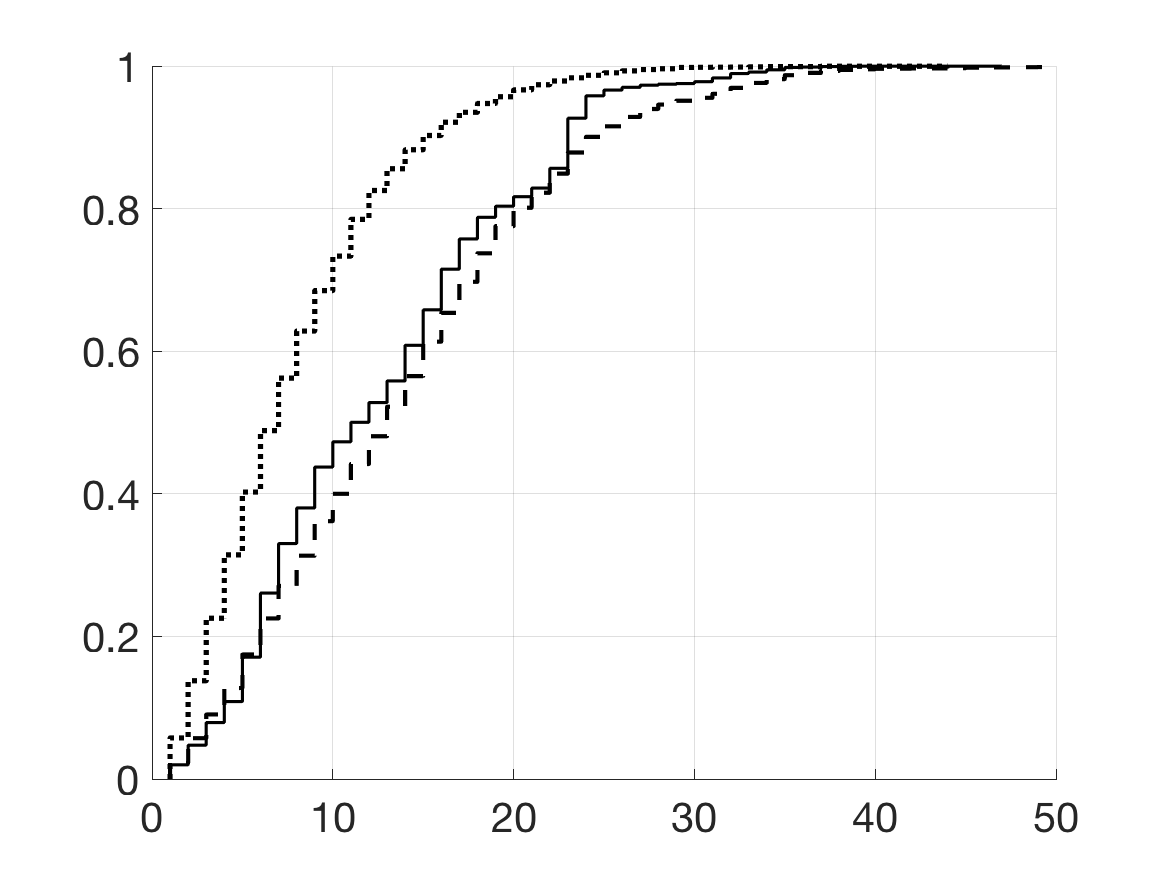}} 
\subfigure[]{\includegraphics[width=5.05cm]{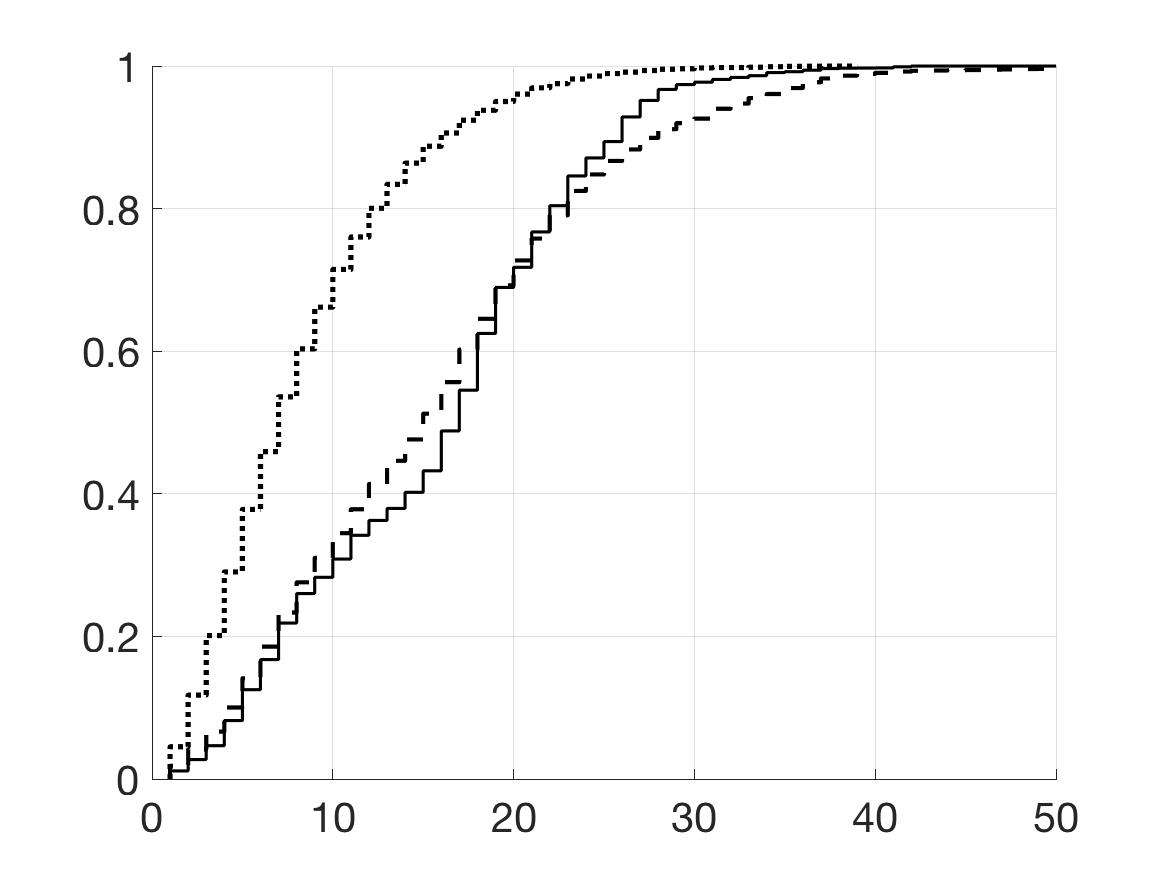}}
\caption{Results from the numerical experiment of Section \ref{sec:advanced example}. Estimates of $ P (L_i(x^t) \leq l | x_i^t  = 1)$  with $i \sim \text{unif}\{1,n\}$ and $x^t \sim p(x^t | y^{1:t})$. The graphs present $\hat F_c(l)$ (solid), $\hat F_q(l)$ (dashed) and $\hat F_a(l)$ (dotted) at time steps $t=60$ (left), $t=70$ (middle) and $t=80$ (right). }
\label{fig: Example 2, Figure 4}
\end{figure}
 Here, we can see that $\hat F_a(\cdot)$ is above $\hat F_c(\cdot)$ at all three time steps $t=60$, 70 and 80, indicating that $L_i(x^t)$ typically is too small and that  the assumed model approach seems to underestimate the level of contact between nodes. This makes sense and agrees with the performance of $\hat p^t_a(i,j)$ studied in our first test. According to Figures \ref{fig: Example 2, Figure 4}(b) and (c), our estimate $\hat F_q(\cdot)$ appears to do a better job, since it is relatively close to $\hat F_c(\cdot)$. However, this is not the case in Figure \ref{fig: Example 2, Figure 4}(a). Here, the curve for $\hat F_q(\cdot)$ is below $\hat F_c(\cdot)$, suggesting that $L_i(x^t)$ typically is too high. To investigate this further, we also examined corresponding output for other time steps $t$. Then, we observed that for smaller values of $t$, typically smaller than $60$, the curve for $\hat F_q(\cdot)$ tends to be below $\hat F_c(\cdot)$, while for larger values of $t$, it tends to be quite close to $\hat F_c(\cdot)$. This is in fact not so unreasonable, since it is for higher values of $t$ that the value one (i.e. water) is dominant in $x^t$. For smaller values of $t$, the value zero (i.e. oil) becomes more and more dominant, and the length of one-valued chains is not supposed to be very high. It appears as if our requirement of 'maximising the expected number of unchanged components' results in keeping too much information from the prior samples.

	\section{Closing remarks}
	\label{sec:Closing remarks}

In this article, we present an approximate and ensemble-based method for solving the filtering problem.
The method is particularly designed for binary state vectors and is based on a generalised view of the well-known ensemble Kalman filter. 
In the EnKF, a Gaussian approximation $f(x)$ for the true prior is constructed which combined with a linear-Gaussian likelihood yields a Gaussian approximation $f(x|y)$ to the true posterior. Next, the prior ensemble is updated with a  linear shift such that the marginal distribution of  the updated samples is equal to $f(x|y)$ provided that the distribution of the prior samples is equal to $f(x)$. 
In our approach, we instead pursue a first order Markov chain for $f(x)$ and combine this with a particular  likelihood model such that the corresponding posterior model $f(x|y)$ is also a first order Markov chain. To update the prior samples, we instead of a linear shift construct a distribution $q(\tilde x | x,y)$ and simulate the updated samples from this distribution. 
In the construction of $q(\tilde x | x,y)$, we formulate an optimality criterion and, just as in EnKF, require that  the distribution of the updated samples is equal to $f(x|y)$ provided that the distribution of the prior samples is equal to $f(x)$. To compute the optimal solution of  $q(\tilde x | x,y)$ we combine dynamic and linear programming. 
Based on results from a simulation experiment, the performance of our method seems promising.

The focus of this article is on  binary state vectors with a one-dimensional spatial arrangement. Clearly, this is a very simple situation with limited practical interest, considering that most real problems 
  involve at least two spatial dimensions and multiple classes for the state variables. 
Nevertheless, we consider the work of this article as a first step towards a more applicable method, and in the future we would like to explore  possible extensions of our method. 
Conceptually, most of the material presented in the article can easily be generalised to more complicated situations. 
Computationally, however, it is more challenging. 
A generalisation of the material in Sections 3 and 4 to a similar situation with more than two possible classes, involves a growing number of free parameters in the construction of each factor $q(\tilde x_k | \tilde x_{k-1}, x_k, y)$. 
Specifically, in the  case of three classes there will be four parameters involved, while in the case of four classes there will be nine parameters involved. 
We believe, however, that it is possible to cope with a situation with more
than one free parameter via an iterative procedure. Specifically, one can start
with some initial values for each of the free parameters and thereafter
iteratively optimise with respect to one of the parameters at a time, keeping the
other parameters fixed. By iterating until convergence we thereby obtain the
optimal solution.
How many parameters we are able to deal with using this strategy will depend on how fast convergence is reached and, of course, how much computation time one is willing to use.

Another possible extension of our method  is to pursue a higher order Markov chain for the assumed prior model $f(x)$. 
 If this is within reach, a further generalisation to two spatial dimensions  may be possible by choosing a Markov mesh model \citep{art9} for $f(x)$. 
 However, similarly to the case of multiple classes, the computational complexity grows rapidly with the order of the Markov chain. The higher the order, the higher the number of free parameters there will be in the construction of each factor $q(\tilde x_k | \tilde x_{k-1},x_k, y)$. 
Computationally we can again imagine to cope with this situation by adopting an iterative optimisation algorithm
as discussed above.

An optimality criterion needs to be specified when constructing $q(\tilde{x}|x,y)$.
In our work we choose to define the optimal solution as the one that 'maximises the
expected number of equal components'. To us this seems like an intuitively reasonable criterion,
since we want to retain as much information as possible from the prior samples. However,
there may be other criteria that are more suitable and which might improve the performance
of our procedure. What optimality criterion that give the best results may even depend on
how the true and assumed distributions differ. One may therefore imagine to construct
a procedure which at each time $t$ use the prior samples to estimate, or select, the best
optimality criterion within a specified class.

	\newpage

	\bibliographystyle{apa}

	\bibliography{main_arXiv}

	\newpage	

	\appendix

\section{Appendix}

This appendix provides an informal proof of that $E_{k:n}^*(t_k)$, $2\leq k \leq n$, is continuous piecewise linear (CPL). Every iteration of the backward recursion, except the first, relies on this result. The proof is an induction proof and consists of two main steps. First, in Section A.1, we consider the first step of the backward recursion and prove that $E_n^*(t_n)$ is CPL. This corresponds to the 'base case' of our induction proof. Next, in Section A.2, we consider the intermediate steps and prove that $E_{k:n}^*(t_k)$ is also CPL, given that $E_{k+1:n}^*(t_{k+1})$ is CPL, $2\leq k < n$. This corresponds to the 'inductive step' of our induction proof. 
In Section A.3 of the appendix, we explain how to determine the breakpoints of $E_{k:n}^*(t_k)$, $2\leq k < n$, prior to solving the corresponding parametric, piecewise linear program.
This is crucial in order to avoid a numerical computation of $E_{k:n}^*(t_k)$ on a grid of $t_k$-values. 
Throughout the appendix, we assume the reader is familiar with all notations introduced previously in the main parts of the article.

\subsection{The first iteration}
\label{sec:appendix1}

The parametric linear program of the first backward iteration can easily be computed analytically. 
Because of the equality constraints, \eqref{eq: equality constr 1} and \eqref{eq: equality constr 2}, we can reformulate the optimisation problem in terms of two variables instead of four. More specifically, we can choose either $q_n^{00}$ or $q_n^{01}$ from \eqref{eq: equality constr 1}, together with either $q_n^{10}$ or $q_n^{11}$ from \eqref{eq: equality constr 2}, and then reformulate the problem in terms of the two chosen variables. Here, we pursue $q_n^{00}$ and $q_n^{10}$.  
By rearranging terms in \eqref{eq: equality constr 1} and \eqref{eq: equality constr 2}, we can write
 \begin{eqnarray}
\pi_{n}^{01}(t_n) q_{n}^{01} = f_{n}^{00} -  \pi_{n}^{00}(t_n) q_{n}^{00},
\label{eq:01}
\\
\pi_{n}^{11}(t_n) q_{n}^{11} = f_{n}^{10} -  \pi_{n}^{10}(t_n) q_{n}^{10}.
\label{eq:11}
\end{eqnarray}
Now, if we replace the terms $\pi_{n}^{01}(t_n) q_{n}^{01}$ and $\pi_{n}^{11}(t_n) q_{n}^{11}$ in the objective function $E_n(t_n, q_n)$ cf. \eqref{eq:objective n} with the right hand side expressions in \eqref{eq:01} and \eqref{eq:11}, respectively, we can rewrite $E_n(t_n, q_n)$ in terms of $q_n^{00}$ and $q_n^{10}$ as
\begin{eqnarray}
E_n (t_n, q_n) 
&=& 2 \pi_{n}^{00} (t_n) q^{00}_{n} + 2 \pi_{n}^{10} (t_n) q^{10}_{n} + c_{n}
%- (f_n^{00} + f_{n}^{10}) + f(x_n = 1) \notag \\
%&=& 2 \pi_{n}^{00} (t_n) q^{00}_{n} + 2 \pi_{n}^{10} (t_n) q^{10}_{n} - f( x_{n} = 0 | y)  + f(x_n = 1).
\label{eq:new objective}
\end{eqnarray}
where $c_n$ is a constant given as
$$
c_{n} = f (x_n = 1) - f( x_{n} = 0 | y) .
$$
Furthermore, combining  \eqref{eq:01} and \eqref{eq:11} with the inequality constraints \eqref{eq:inequality constr} allows us to reformulate the constraints for  $q_{n}^{00}$ and $q_{n}^{10}$  as
\begin{equation}
\max \left \{ 0,   \frac{f^{00}_{n} - \pi_{n}^{01} (t_n) }{\pi_n^{00}(t_n)} \right \} \leq q_{n}^{00} \leq \min \left \{ 1, \frac{f_n^{00}}{\pi_n^{00}(t_n)}\right \}, \label{eq:new ineq 1}
\end{equation}
\begin{equation}
\max \left \{ 0,   \frac{f^{10}_{n} - \pi_{n}^{11} (t_n) }{\pi_n^{10}(t_n)} \right \} \leq q_{n}^{10} \leq \min \left \{ 1, \frac{f_n^{10}}{\pi_n^{10}(t_n)}\right \} \label{eq:new ineq 2}.
\end{equation}
To summarize, we have now obtained a linear program, in which we want to maximise the objective function in \eqref{eq:new objective} with respect to  the two variables $q_n^{00}$ and $q_n^{10}$, subject to the constraints   
\eqref{eq:new ineq 1}-\eqref{eq:new ineq 2}. 

If for some fixed $t_n \in[t_n^{\min}, t_n^{\max}]$ we consider a coordinate system with $q_{n}^{00}$ along the first axis and  $q_{n}^{10}$ along the second, the constraints \eqref{eq:new ineq 1}-\eqref{eq:new ineq 2} form a rectangular region of feasible solutions, with two edges in the $q_n^{00}$-direction and two edges in the $q_n^{10}$-direction.  The optimal solution lies in a corner point of this region.
Since $\pi_{n}^{00} (t_n)$ and $\pi_{n}^{10} (t_n)$ are non-negative for any $t_n \in [t_n^{\min}, t_n^{\max}]$, it is easily seen from \eqref{eq:new objective} that $E_n(t_n, q_n)$ is maximised with respect to $q_n$ when $q_{n}^{00}$ and $q_{n}^{10}$ are as large as possible. 
Consequently, the optimal solutions of $q_{n}^{00}$ and $q_{n}^{10}$ must equal the upper bounds in \eqref{eq:new ineq 1} and  \eqref{eq:new ineq 2}, corresponding to the upper right corner of the rectangular feasible region. That is,
\begin{eqnarray}
q^{*00}_n(t_n) = \min \left \{ 1, \frac{f_n^{00}}{\pi_n^{00}(t_n)}\right \}, \label{eq:q00 star}\\
q^{*10}_n(t_n) = \min \left \{ 1, \frac{f_n^{10}}{\pi_n^{10}(t_n)}\right \} \label{eq:q10 star}.
\end{eqnarray}

Clearly,  $q^{*00}_n(t_n)$ and  $q^{*10}_n(t_n)$ are continuous and piecewise-defined functions of $t_n$, since $\pi_n^{00}(t_n)$ and $\pi_n^{10}(t_n)$ are linear functions of $t_n$. Specifically, for $t_n$-values such that $\pi_n^{00}(t_n) > f_{n}^{00}$, we get $q^{*00}_n(t_n) = f_n^{00}/\pi_n^{00}(t_n)$, while for $t_n$-values such that $\pi_n^{00}(t_n) \leq f_{n}^{00}$, we get $q^{*00}_n(t_n) = 1$. 
Likewise,  for $t_n$-values such that $\pi_n^{10}(t_n) > f_{n}^{10}$, we get $q^{*10}_n(t_n) = f_n^{10}/\pi_n^{10}(t_n)$, while for $t_n$-values such that $\pi_n^{10}(t_n) \leq f_{n}^{10}$, we get $q^{*10}_n(t_n) = 1$. 

Inserting the optimal solutions $q^{*00}_n(t_n)$ and  $q^{*10}_n(t_n)$ into \eqref{eq:new objective}, returns $E^*_n(t_n)$. 
Doing this, it is easily seen that  $E^*_n(t_n)$ is a CPL function of $t_n$, consisting of maximally three pieces, each piece having a slope equal to either -2, 0 or 2.

\subsection{The intermediate iterations}
\label{sec:appendix2}

 At each intermediate iteration of the backward recursion, we are dealing with a parametric, \emph{piecewise} linear program, whose analytic solution is, generally,  more intricate than that of the parametric linear program of the first iteration. 
 However, proving that the resulting function $E_{k:n}^*(t_k)$ is CPL, provided that $E_{k+1:n}^*(t_{k+1})$ is CPL, is not too complicated. 
 Below, we present a proof which can be summarised as follows.
 First, for each subproblem $j \in \mathcal S_{k+1}$ corresponding to the $j$'th linear piece of the previous CPL function $E_{k+1:n}^*(t_{k+1})$, we explain that the corners (or possibly edges) of the feasible region that may represent the optimal solution yield a CPL function in $t_k$ when inserted into the objective function $E_{k:n}^{(j)}(t_k, q_k)$. Second, we argue that since the boundary of the feasible region evolves in a continuous way as a function of $t_k$ and since also $E_{k:n}^{(j)}(t_k, q_k)$ is continuous in $t_k$ and $q_k$, 
 any infinitesimal change in $t_k$ can only induce an infinitesimal change in the location of the optimal solution.
Third, we conclude from these observations  that $\tilde E_{k:n}^{(j)}(t_k)$ is CPL for each subproblem $j \in \mathcal S_{k+1}$. This means that  the final function $E^*_{k:n}(t_k)$ is the maximum of multiple CPL functions. Therefore, $E^*_{k:n}(t_k)$ itself must be piecewise linear. 
The additional fact that $E^*_{k:n}(t_k)$ is continuous is an immediate consequence of the continuity 
of the whole optimisation problem and the connection between the subproblems. 

As in the first backward step, the equality constraints \eqref{eq:constr k both} for $q_k$ allow us to reformulate the optimisation problem in terms of the two variables $q^{00}_{k}$ and $q^{10}_{k}$. 
Specifically,  for each subproblem $j \in \mathcal S_{k+1}$, we can use the equality constraints to write the objective function $E_{k:n}^{(j)}(t_k, q_k)$ cf. \eqref{eq:objective k j} in terms of 
$q^{00}_{k}$ and $q^{10}_{k}$ as 
\begin{equation}
E_{k:n}^{(j)}(t_k, q_k) = \tilde \beta_{k}^{(j)} \pi_k^{00} (t_k) q_k^{00} + \tilde \beta_{k}^{(j)} \pi_k^{10} (t_k) q_k^{10}  + \tilde \alpha_{k}^{(j)}
\label{eq:new objective k}
\end{equation}
where
$$
\tilde \beta_{k}^{(j)}  = 2 + b_{k+1}^{(j)} \left( \rho_k^{0|0} - \rho_k^{0|1} \right )  
$$
and
$$
\tilde \alpha_k^{(j)} = f (x_k = 1) - f( x_{k} = 0 | y) + a_{k+1}^{(j)} + b_{k+1}^{(j)} \left (f_k^{00} + f_k^{10} \right )\rho_k^{0|1}.
$$
The corresponding constraints for 
$q^{00}_{k}$ and $q^{10}_{k}$ read 
\begin{eqnarray}
\max \left \{ 0,   \frac{f^{00}_{k} - \pi_{k}^{01} (t_k) }{\pi_k^{00}(t_k)} \right \} \leq q_{k}^{00} \leq \min \left \{ 1, \frac{f_k^{00}}{\pi_k^{00}(t_k)}\right \}, \label{eq:new ineq k 1} \\
\max \left \{ 0,   \frac{f^{10}_{k} - \pi_{k}^{11} (t_k) }{\pi_k^{10}(t_k)} \right \} \leq q_{k}^{10} \leq \min \left \{ 1, \frac{f_k^{10}}{\pi_k^{10}(t_k)}\right \} \label{eq:new ineq k 2},
\end{eqnarray}
and
\begin{eqnarray}
t_{k+1}^{B(j)} \leq \left( \rho_k^{0|0} - \rho_k^{0|1} \right ) \pi_k^{00}(t_k) q_k^{00} + \left( \rho_k^{0|0} - \rho_k^{0|1} \right ) \pi_k^{10}(t_k) q_k^{10} + \left  (f_k^{00} + f_k^{10} \right  )  \rho_k^{0|1} \leq t_{k+1}^{B(j+1)}
\label{eq:ineq k 3}
\end{eqnarray}

\begin{figure}
\centering
\subfigure[]{\includegraphics[width=2cm]{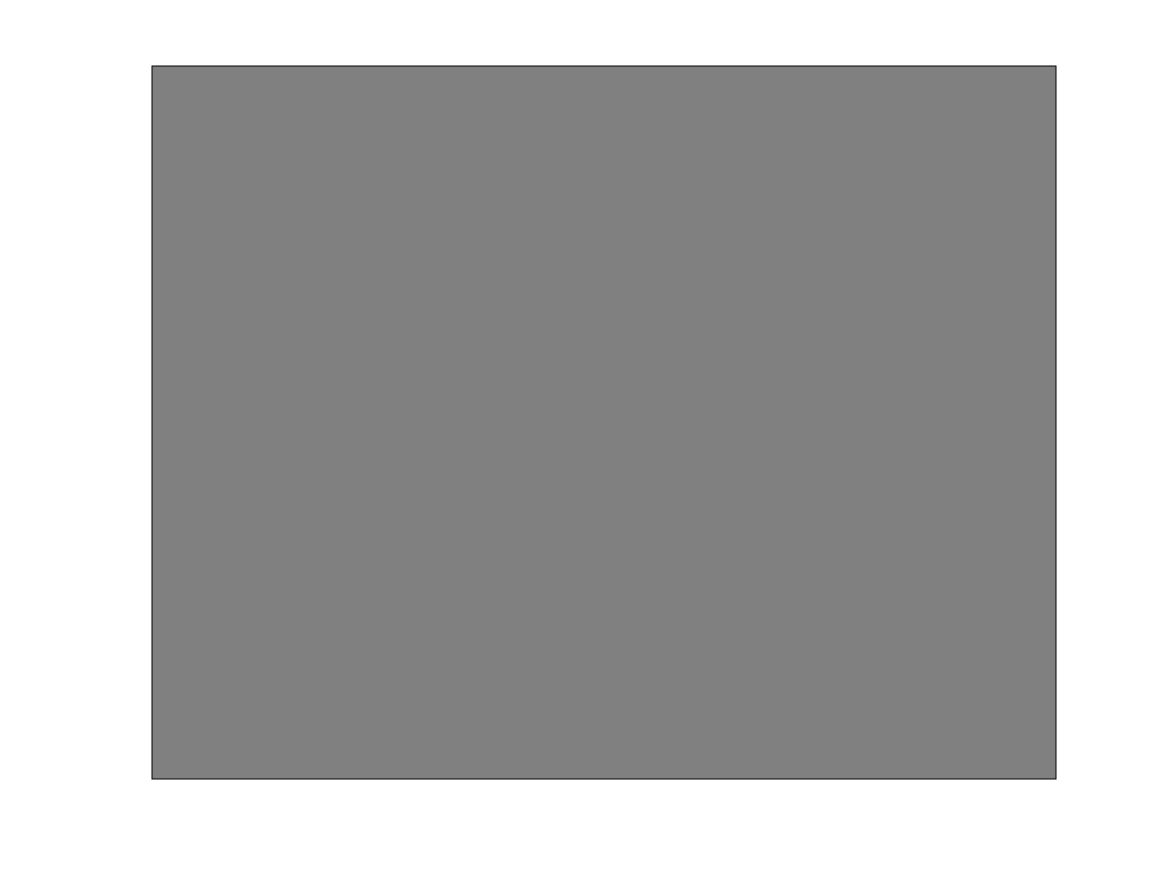}}
\subfigure[]{\includegraphics[width=2cm]{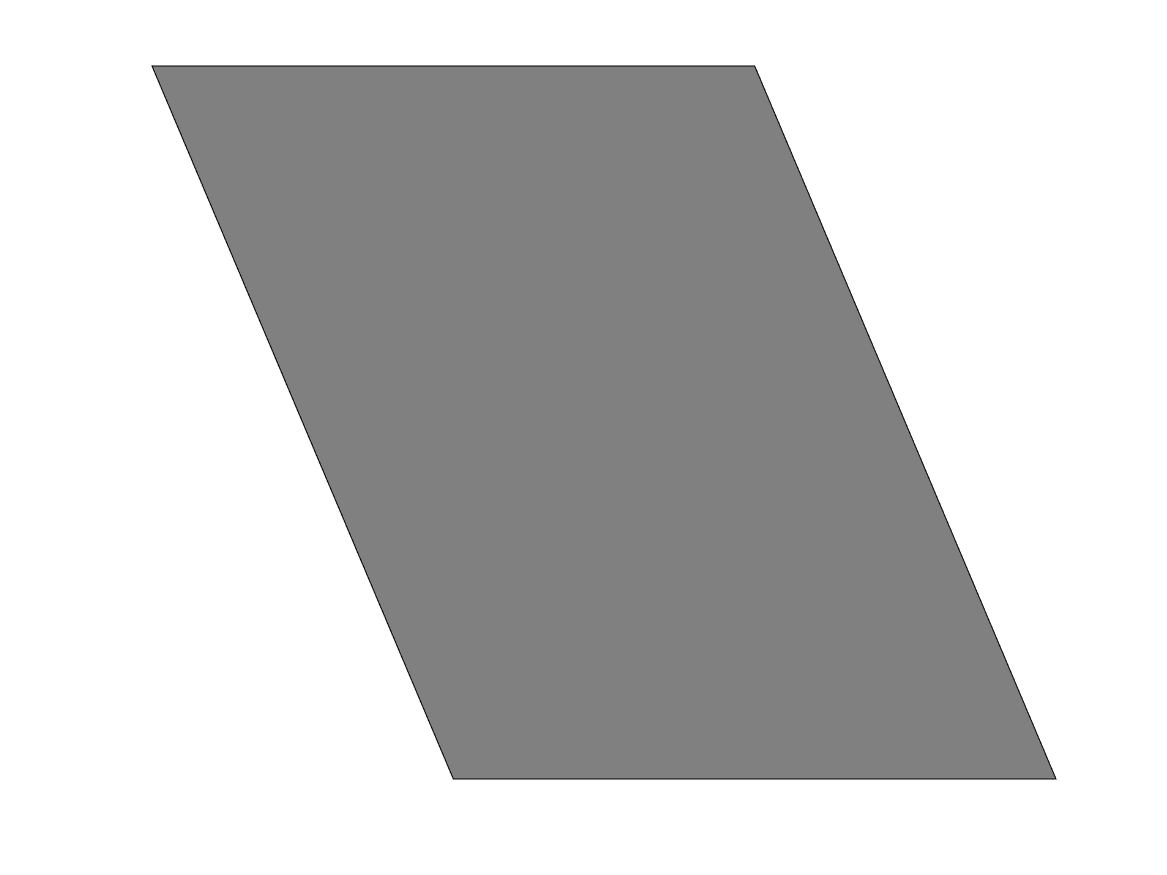}}
\subfigure[]{\includegraphics[width=2cm]{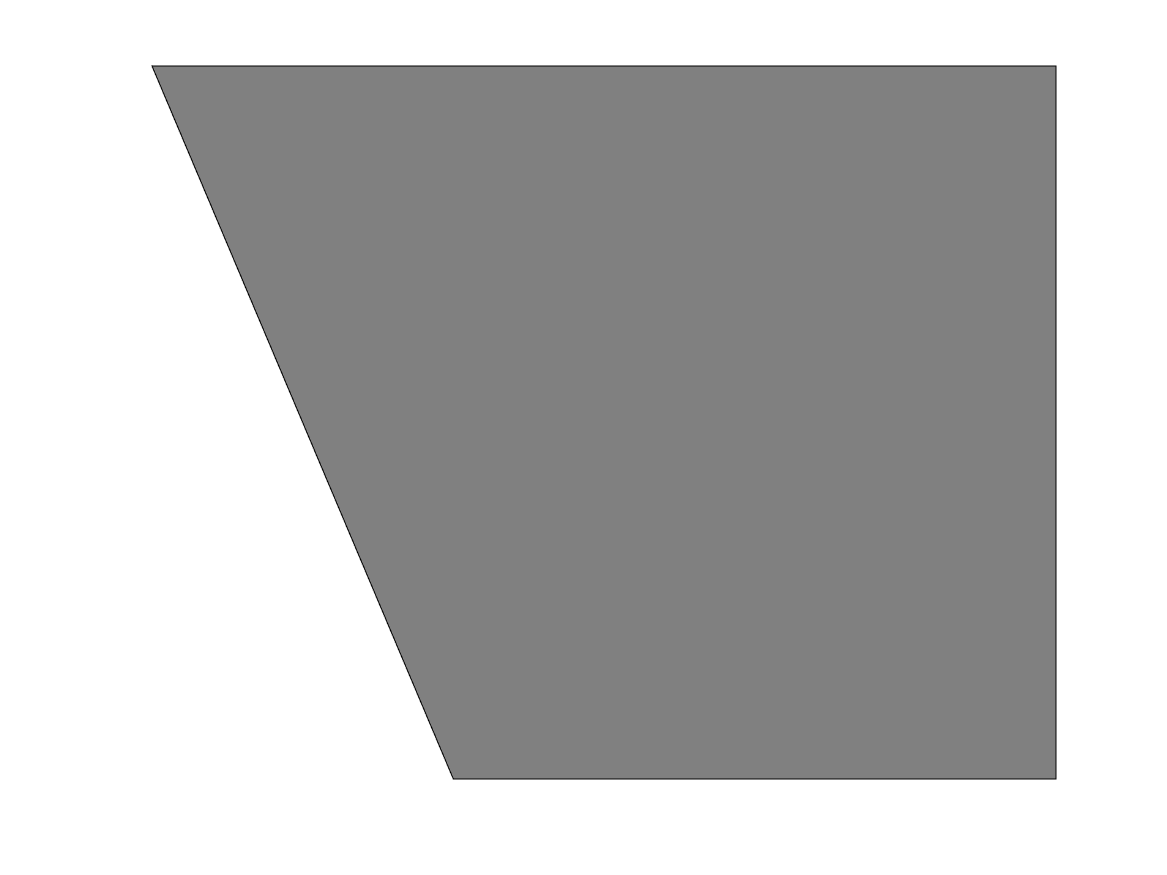}}
\subfigure[]{\includegraphics[width=2cm]{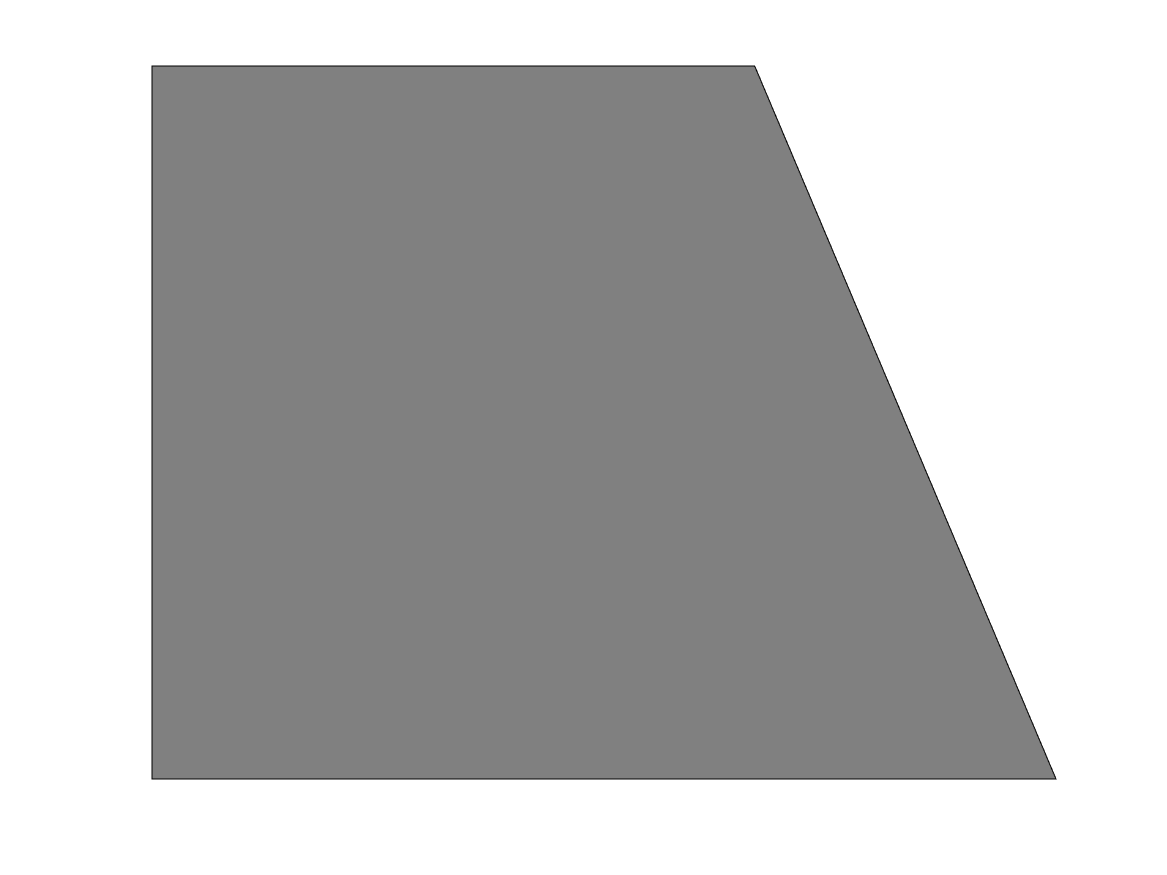}}
\subfigure[]{\includegraphics[width=2cm]{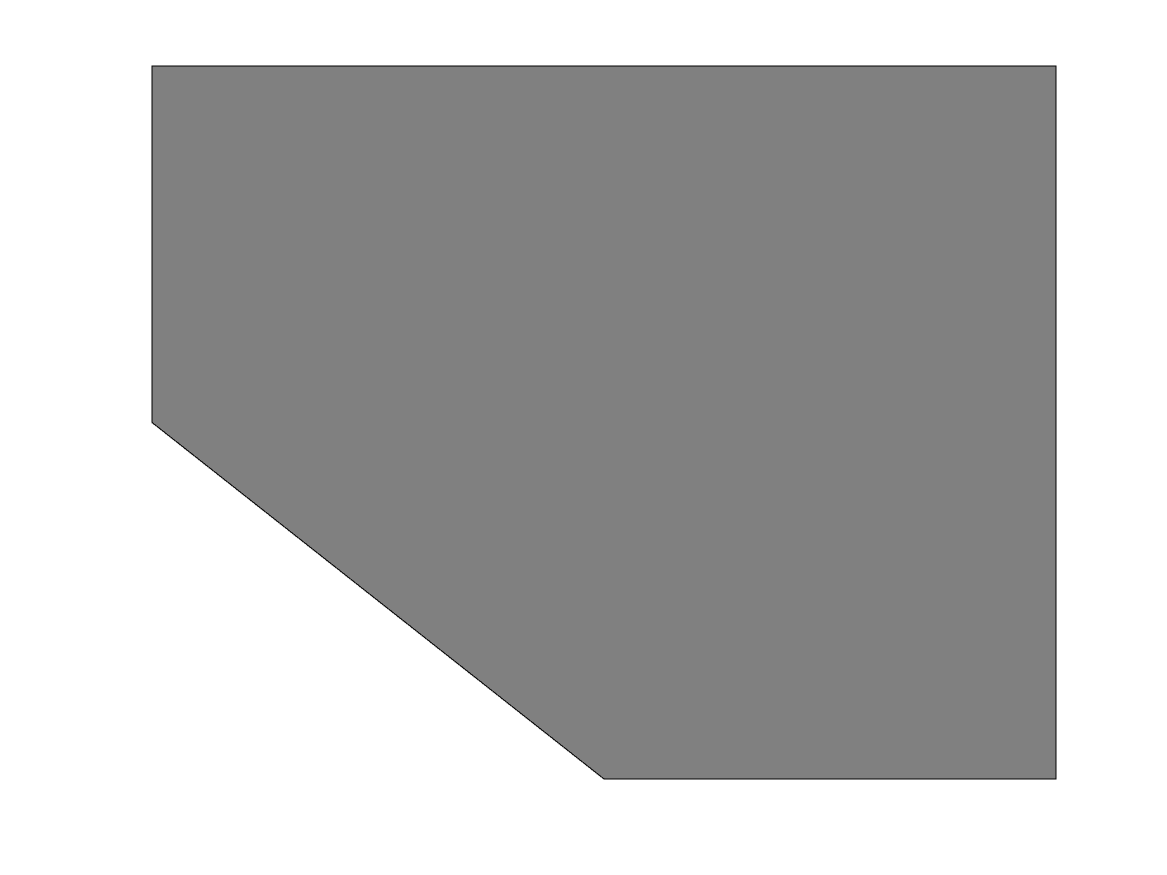}}
\subfigure[]{\includegraphics[width=2cm]{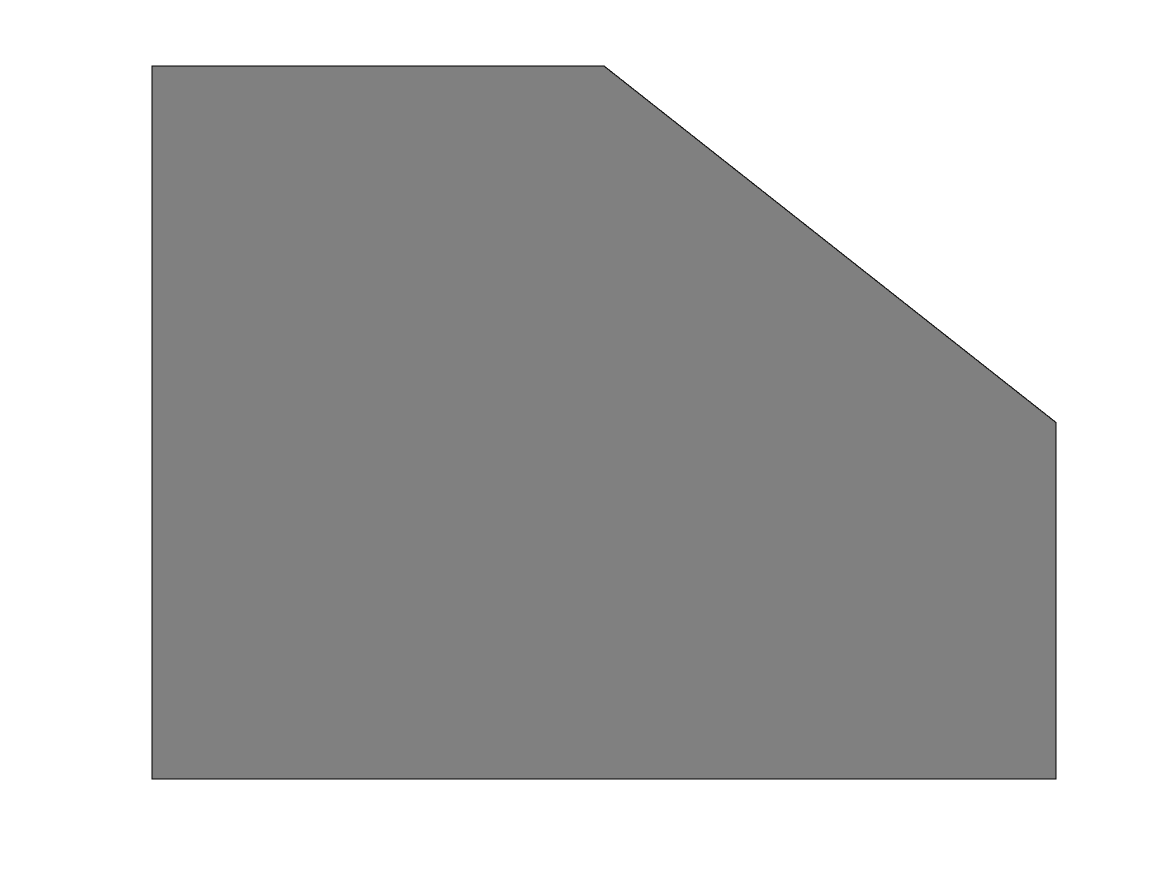}}
\subfigure[]{\includegraphics[width=2cm]{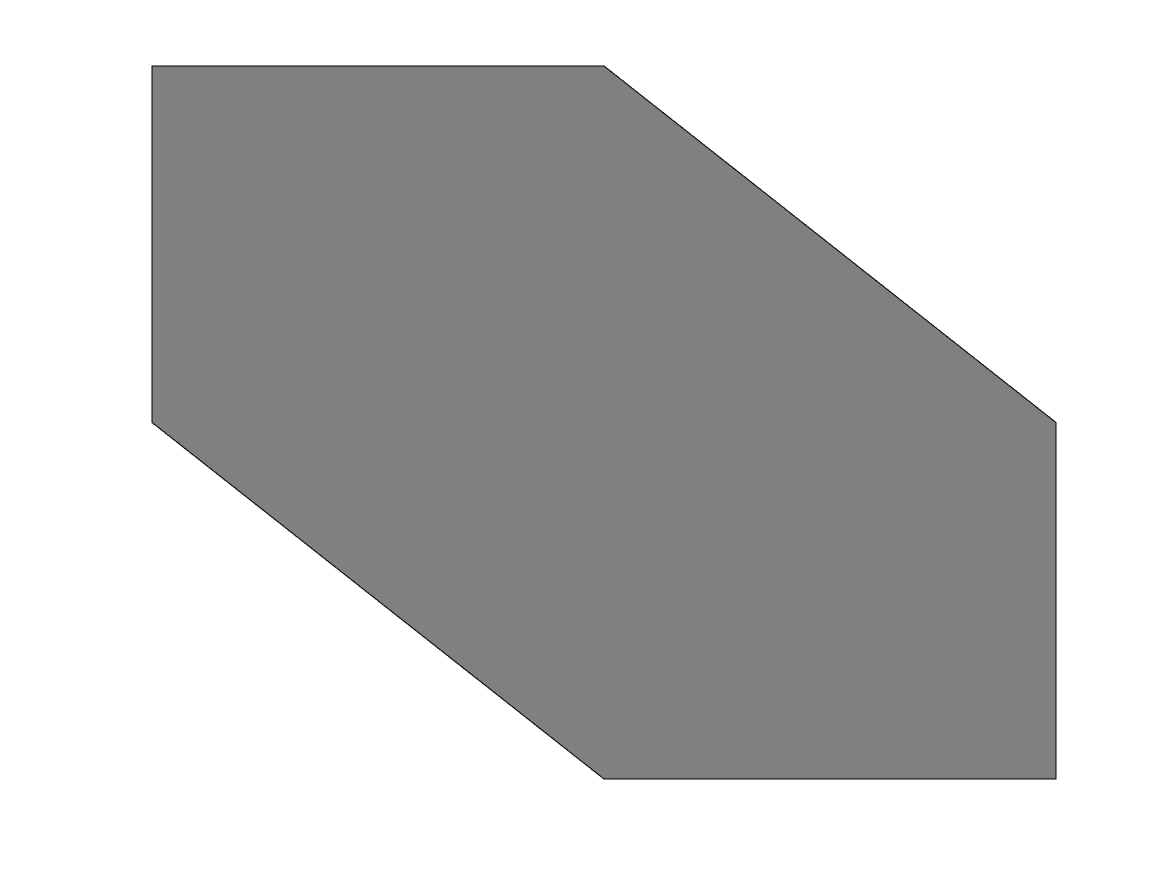}}
\caption{Illustrations of some possible shapes for the feasible regions of the linear programs at the intermediate steps of the backward recursion. The polygons are drawn in a coordinate system with $q_k^{00}$ in the horizontal direction and $q_k^{10}$ in the vertical direction. }
\label{fig:feasible region shapes}
\end{figure}

If for some fixed $t_k \in [t_k^{\min}, t_k^{\max}]$ we consider a coordinate system with $q_k^{00}$ along  one axis and $q_k^{10}$ along the other, we see that the feasible region formed by the constraints \eqref{eq:new ineq k 1}-\eqref{eq:ineq k 3} is a polygon
with maximally six corners. The region is enclosed by two lines in the $q_k^{00}$-direction cf. \eqref{eq:new ineq k 1},  two lines in the $q_k^{10}$-direction cf. \eqref{eq:new ineq k 2}, and two parallel lines with a negative slope of $-\pi_k^{00}(t_k)/{\pi_k^{10}}(t_k)$ cf. \eqref{eq:ineq k 3}.  Figure \ref{fig:feasible region shapes} illustrates some of the possible shapes that the region can take. Clearly, the optimal solution is located in a corner of the feasible region, possibly along a whole edge.

To understand where along the boundary of the feasible region the optimal solution is located, we note from  \eqref{eq:new objective k} that if $\tilde \beta_{k}^{(j)}$ is positive, then $E_{k:n}^{(j)}(t_k, q_k)$ is maximised when $q_k^{00}$ and $q_k^{10}$ are as large as possible,  while if $\tilde \beta_{k}^{(j)}$ is negative, then $E_{k:n}^{(j)}(t_k, q_k)$ is maximised when  $q_k^{00}$ and $q_k^{10}$ are as small as possible. 
For simplicity, we assume in the following that the feasible region is non-empty.
First, consider the case with $\tilde \beta_{k}^{(j)}$ positive. Then, we need to check whether or not the upper of the two lines corresponding to the two inequality constraints in \eqref{eq:ineq k 3} forms an edge of the feasible region. If this line does \emph{not} form an edge of the feasible region; see for example the shapes in Figure \ref{fig:feasible region shapes} (a), (c) and (e); we observe that the point $\left ( q_k^{00(\mathcal U)} (t_k), q_k^{10(\mathcal U)}(t_k) \right)$, where
\begin{equation}
 q_k^{00(\mathcal U)} (t_k) = \min \left \{ 1, \frac{f_k^{00}}{\pi_k^{00}(t_k)}\right \}, 
\label{eq:q00U}
\end{equation}
\begin{equation}
q_k^{10(\mathcal U)} (t_k) = \min \left \{ 1, \frac{f_k^{10}}{\pi_k^{10}(t_k)}\right \},
\label{eq:q10U}
\end{equation}
is a corner. Moreover, this corner represents the optimal solution, since $q_k^{00}$ and $q_k^{10}$ jointly take their maximal values in this point. 
Now, if we insert the functions in \eqref{eq:q00U} and \eqref{eq:q10U} into the objective function $E_{k:n}^{(j)}(t_k, q_k)$, we obtain a CPL function in $t_k$. 
Thereby, given that
\eqref{eq:q00U} and \eqref{eq:q10U} represent a corner of the feasible region for all values of $t_k$, the resulting function $\tilde  E_{k:n}^{(j)}(t_k) $ is CPL in $t_k$. 
If, on the other hand, the upper of the two lines of the constraints  \eqref{eq:ineq k 3} \emph{does} represent an edge of the feasible region; see for instance Figures \ref{fig:feasible region shapes}(b), (d), (f) and (g); then this whole edge represents the optimal solution. That is, any point along the edge is optimal. This result  is due to that  the slope of the objective function and the slope of the line for this edge are equal, from  which it follows that the objective function takes the same maximal value anywhere along the edge. 
Now, if we insert $(q_k^{00}$, $q_{k}^{10})$-coordinates located on the edge into the objective function $E_{k:n}^{(j)}(t_k, q_k)$, we get a function  which is constant, and hence CPL, in $t_k$. Thereby, given that the edge is part of the feasible region for all values of $t_k$, the resulting function $\tilde  E_{k:n}^{(j)}(t_k) $ is CPL in $t_k$.
Next, consider the case with $\tilde \beta_{k}^{(j)}$ negative.
 Then, the situation is equivalent to  the case with $\tilde \beta_{k}^{(j)}$ positive, but we need to consider the lower part of the feasible region instead of the upper. 
That is, we need to check whether or not the lower of the two lines corresponding to the constraints in \eqref{eq:ineq k 3} forms an edge of the feasible region.
If this line does \emph{not} represent an edge;  see for example Figures \ref{fig:feasible region shapes}(a), (d) and (f); the optimal solution is found in the lower left corner  point, $\left ( q_k^{00(\mathcal L)} (t_k), q_k^{10(\mathcal L)} (t_k) \right)$, where
\begin{equation}
 q_k^{00(\mathcal L)} (t_k) = \max \left \{ 0,   \frac{f^{00}_{k} - \pi_{k}^{01} (t_k) }{\pi_k^{00}(t_k)} \right \},
\label{eq:q00L}
\end{equation}
\begin{equation}
q_k^{10(\mathcal L)} (t_k) = \max \left \{ 0,   \frac{f^{10}_{k} - \pi_{k}^{11} (t_k) }{\pi_k^{10}(t_k)} \right \}.
\label{eq:q10L}
\end{equation}
Again, if we insert the functions in \eqref{eq:q00L} and \eqref{eq:q10L} into the objective function $E_{k:n}^{(j)}(t_k, q_k)$, we obtain a CPL function in $t_k$. 
Thereby, given that \eqref{eq:q00L} and \eqref{eq:q10L} represent a corner of the feasible region for all values of $t_k$, the resulting function $\tilde  E_{k:n}^{(j)}(t_k) $ is CPL in $t_k$. 
If, on the other hand, the lower of the two lines of the constraints \eqref{eq:ineq k 3} \emph{does} represent an edge of the feasible region, then this edge also represents the optimal solution since the objective function takes the same maximal value anywhere along this edge. Now, if we insert  $(q_k^{00}$, $q_{k}^{10})$-coordinates located on the optimal edge into the objective function $E_{k:n}^{(j)}(t_k, q_k)$, we obtain a function which is constant, and hence CPL, in $t_k$. Thereby, given that the edge is part of the feasible region for all values of $t_k$, the resulting function $\tilde  E_{k:n}^{(j)}(t_k) $ is CPL in $t_k$.

Because the objective function, $E_{k:n}^{(j)}(t_k, q_k)$, as well as all the constraints, \eqref{eq:new ineq k 1}-\eqref{eq:ineq k 3}, are continuous in $t_k$ and $q_k$, it follows that any infinitesimal change $\delta t_k$ in $t_k$ can only induce corresponding infinitesimal changes in the shape of the feasible region and the value of the objective function. Hence, the optimal solution at any $t_k$-value  $t_k'$
must be located in the same corner (or along the same edge) as the optimal solution at the $t_k$-value $t_k' + \delta t_k$. We note, however,  that it is possible that the infinitesimal change $\delta t_k$ may have added or deleted an edge from the region. In this case, it is possible that a single corner represented the optimal solution at  $t_k'$, while a whole edge represents the optimal solution at $t_k' + \delta t_k$, or vice versa. However, this will not cause any discontinuities in the resulting function $\tilde E_{k:n}^{(j)}(t_k)$ because of the continuity of the optimisation problem as a whole. 
We have already showed that the coordinates describing the evolution of every potentially optimal corner (or edge) as a function of $t_k$   return a CPL function in $t_k$. Hence,  we understand that $\tilde E_{k:n}^{(j)}(t_k)$ must be  CPL.

Finally, we obtain the function $E_{k:n}^*(t_k)$ by taking the maximum of the $\tilde E_{k:n}^{(j)}(t_k)$'s. 
Taking the maximum of a set of continuous piecewise linear functions necessarily produces another piecewise linear, but not necessarily a continuous, function. 
However, it is obvious without a further proof that $E_{k:n}^*(t_k)$ must be continuous, since all functions 
in the whole optimisation problem are continuous. 
Thereby, we can conclude that $E_{k:n}^*(t_k)$ is CPL.

According to numerical experiments, it seems that $q_k^{*00}(t_k)$ and $q_k^{*10}(t_k) $ are analytically given as $q_k^{*00}(t_k)  = q_k^{00(\mathcal U)} (t_k)$ and $q_k^{*10}(t_k) = q_k^{10(\mathcal U)} (t_k)$, just as in the first backward iteration.  However, we have not proved this result, since it is not really important for our application.  Yet, we note that if this result can be proved, the computation of $q(\tilde x | x, y)$ can be done particularly simple.

\subsection{Computing the breakpoints of $E_{k:n}^*(t_k)$}

This section concerns computation of the breakpoints of the CPL function $E_{k:n}^*(t_k)$ produced at each intermediate iteration $2 \leq k < n$. 
The breakpoints of $E_{k:n}^*(t_k)$ should be computed prior to solving the corresponding parametric piecewise linear program in order to
 avoid numerical computation of $E_{k:n}^*(t_k)$ on a grid of $t_k$-values. However, 
 it can in some cases be a bit cumbersome and technical to compute the explicit set  of $t_k$-values representing the breakpoints of $E_{k:n}^*(t_k)$. Fortunately, it is an easier task to compute a slightly larger set of $t_k$-values representing \emph{potential} breakpoints of $E_{k:n}^*(t_k)$, which necessarily includes all of the \emph{actual} breakpoints.
For convenience, we denote in the following the set of actual breakpoints by $A_k$ and the larger set of potential breakpoints by $A_k' \supset A_k$. 
Having computed the set $A_k'$, we can solve our parametric piecewise linear program for the $t_k$-values in this set, and afterwards go through the values of the resulting function $E_{k:n}^*(t_k)$ to check which of the elements in $A_k'$ that represent \emph{actual} breakpoints that must be stored in $A_k$, and which points that  can be omitted.

As explained in Section A.3, the function $E_n^*(t_n)$ of the first backward iteration consists of maximally three linear pieces, or equivalently, it has maximally two breakpoints in addition to its two endpoints $t_n^{\min}$ and $t_n^{\max}$. Since at each intermediate iteration we consider a more complicated parametric \emph{piecewise} linear program, additional breakpoints can occur in $E_{k:n}^*(t_k)$, with the  number of possible breakpoints for $E_{k:n}^*(t_k)$  increasing with the number of breakpoints for $E_{k+1:n}^*(t_k)$ computed at the previous step of the recursion. 
To compute the set $A_k'$ of potential breakpoints for $E_{k:n}^*(t_k)$, we need to check for which $t_k$-values the corners of the rectangular region formed by the constraints \eqref{eq:new ineq k 1} and \eqref{eq:new ineq k 2} intersect with the lines of the constraints \eqref{eq:ineq k 3} for each $j \in \mathcal S_{k+1}$.  Each $t_k$-value that causes such an intersection must be included in the set $A_k'$. 
To understand why, consider a subproblem $j \in \mathcal S_{k+1}$, and assume $\tilde \beta^{(j)}_k$ is positive. Furthermore, suppose that for all $t_k \in [t_k^{\min}, t_k^{\max}]$ the feasible region has a rectangular shape cf. Figure \ref{fig:feasible region shapes}(a), meaning that the region is only enclosed by the constraints \eqref{eq:new ineq k 1} and \eqref{eq:new ineq k 2}, while the extra constraints in \eqref{eq:ineq k 3} do not contribute to shaping the region. Then, from Section \ref{sec:appendix2}, we know that the optimal solution  lies in the upper right corner given by  \eqref{eq:q00U} and \eqref{eq:q10U} for all $t_k$. Moreover, we know that $\tilde E_{k:n}^{(j)}(t_k)$ is CPL with  breakpoints corresponding to the breakpoints of \eqref{eq:q00U} and \eqref{eq:q10U}.
Now, suppose instead that after some specific value $t_k'$  the shape of the feasible region changes from a rectangular shape as in Figure \ref{fig:feasible region shapes}(a) to a pentagon shape as in Figure \ref{fig:feasible region shapes}(f). This means that  the upper of the two lines formed by the extra constraints in \eqref{eq:ineq k 3} at the $t_k$-value $t_k'$ intersects with the upper right corner point given by \eqref{eq:q00U} and \eqref{eq:q10U}, while for $t_k > t_k'$ the constraints results in that an extra edge is added to the feasible region. From Section \ref{sec:appendix2}, we then know that for $t_k > t_k'$ this extra edge represents the optimal solution and the value of the objective function remains constant as a function of $t_k > t_k'$. Thereby, we understand that a breakpoint may occur in $\tilde E_{k:n}^{(j)}(t_k)$, and hence possibly  in $E_{k:n}^*(t_k)$, at the $t_k$-value $t_k'$. If the  feasible region were to evolve in a different way than the one considered here, similar arguments can be formulated. 
In $A_k'$, we must also include the breakpoints of the functions in \eqref{eq:q00U}-\eqref{eq:q10L}, i.e. the breakpoints of the functions describing the coordinates for the lower left and upper right corner points of the feasible region when the constraints \eqref{eq:ineq k 3} do not contribute.

\end{document}